\documentclass{aa}
\usepackage{graphicx}
\usepackage[]{hyperref}
\usepackage{lipsum, babel}
\hypersetup{colorlinks=true,citecolor=blue}
\usepackage{booktabs}
\usepackage{comment}
\usepackage{txfonts}
\usepackage{enumitem} 

\usepackage[dvipsnames]{xcolor}
\usepackage{colortbl}
\newcommand{\inputs}[1]{\textcolor{red}{\textbf{#1}}}
\def\spec{0.58}
\def\specbis{0.54}

\usepackage{adjustbox}

\usepackage{tablefootnote}
\usepackage{threeparttable}

\def\coo{CO$_2$\xspace}
\def\co{CO\xspace}

\def\hh{H$_2$\xspace}

\def\hho{H$_2$O\xspace}
\def\chhhh{CH$_4$\xspace}

\def\nn{N$_2$\xspace}
\def\na{Na\xspace}
\def\tio{TiO\xspace}
\def\vo{VO\xspace}
\def\feh{FeH\xspace}
\def\hcn{HCN\xspace}

\def\sio{SiO\xspace}
\def\nhhh{NH$_3$\xspace}

\def\dlogE{$\Delta$logE\xspace}

\begin{document}

   \title{ARES VI: Are 1D retrieval models accurate enough to characterize exo-atmospheres with transmission spectroscopy in the era of JWST and Ariel?}
   
\titlerunning{ARES VI: from 1D to 3D models}

   \author{Adam Yassin Jaziri\inst{1,2,10}, William Pluriel\inst{2}, Andrea Bocchieri\inst{3}, Emilie Panek\inst{4}, Lucas Teinturier\inst{5,6}, Anastasiia Ivanova\inst{1}, Natalia E. Rektsini\inst{4,7}, Pierre Drossart\inst{4, 5}, Jean-Philippe Beaulieu\inst{4, 7}, Aurélien Falco\inst{8, 9}, J\'{e}r\'{e}my Leconte\inst{10}, Lorenzo V. Mugnai\inst{11, 12, 3, 13} \and Olivia Venot\inst{14}}
          
\authorrunning{Yassin Jaziri et al.}

   \institute{$^1$ LATMOS/IPSL, UVSQ Universit\'{e} Paris-Saclay, Sorbonne Universit\'{e}, CNRS, Guyancourt, France \\
   $^2$ Observatoire Astronomique de l’Universit\'{e} de Gen\`{e}ve, département d’astronomie Chemin Pegasi 51, CH-1290 Versoix, Switzerland \\
   $^3$ Dipartimento di Fisica, La Sapienza Universit\`a di Roma, Piazzale Aldo Moro 2, 00185 Roma, Italy\\
   $^4$ Institut d’Astrophysique de Paris - CNRS, Sorbonne Universit\'{e}, Paris, France \\
   $^5$ LESIA, Observatoire de Paris, Universit\'{e} PSL, Sorbonne Université, Universit\'{e} Paris Cit\'{e}, CNRS, 5 place Jules Janssen, 92195 Meudon, France\\ 
   $^6$ Laboratoire de Météorologie Dynamique, IPSL, CNRS, Sorbonne Université, Ecole Normale Supérieure, Université PSL, Ecole Polytechnique, Institut Polytechnique de Paris, 75005 Paris, France \\
   $^7$ School of Natural Sciences, University of Tasmania, Private Bag 37 Hobart, Tasmania, 70001, Australia\\
   $^8$ Universit\'{e} de Paris Cit\'{e}, Institut de Physique du Globe de Paris, CNRS F-75005 Paris, 1 rue Jussieu, 75005 Paris, France \\
   $^{9}$ Laboratoire AIM, CEA, CNRS, Univ. Paris-Sud, UVSQ, Universit\'{e} Paris-Saclay, F-91191 Gif-sur-Yvette, France \\
   $^{10}$ Laboratoire d’astrophysique de Bordeaux, Univ. Bordeaux, CNRS, B18N, all\'{e}e Geoffroy Saint-Hilaire, 33615 Pessac, France\\
   $^{11}$ School of Physics and Astronomy, Cardiff University, Queens Buildings, The Parade, Cardiff, CF24 3AA, UK\\
   $^{12}$ Department of Physics and Astronomy, University College London, Gower Street, London, WC1E 6BT, UK\\
   $^{13}$ INAF – Osservatorio Astronomico di Palermo, Piazza del Parlamento 1, I-90134 Palermo, Italy\\
   $^{14}$ Universit\'{e} Paris Cit\'{e} and Univ Paris Est Creteil, CNRS, LISA, F-75013 Paris, France \\
}

   \date{Received July 6, 2023; accepted December 11, 2023}

 
  \abstract
    {The observed exoplanets transit spectra are usually retrieved using one-dimensional models to determine atmospheric composition. However, planetary atmospheres are three-dimensional. With the new state-of-the-art James Webb Space Telescope (JWST) and future space telescopes such as Ariel (Atmospheric Remote-sensing Infrared Exoplanet Large-survey), we will be able to obtain increasingly accurate transit spectra. The 3D effects on the spectra will be visible, and we can expect biases in the 1D extractions. In order to elucidate these biases, we have built theoretical observations of transit spectra, from 3D atmospheric modeling through transit modeling to instrument modeling. For that purpose, we used a Global Climate Model (GCM) to simulate the atmosphere, a 3D-radiative transfer model to calculate theoretical transmission spectra, and adapted instrument software from JWST and Ariel to reproduce telescope noise. Next, we used a 1D-radiative transfer inversion model to retrieve the known input atmosphere and disentangle any biases that might be observed. The study has been done from warm planets to ultra-hot planets to assess biases as a function of average planet temperature.
    Three-dimensional effects are observed to be strongly non-linear from the coldest to the hottest planets. These effects also depend on the planet's metallicity and gravity. Considering equilibrium chemistry, 3D effects are observed through very strong variations for certain features of the molecule, or very small variations over the whole spectrum. We conclude that we cannot rely on the uncertainty of retrievals at all pressures, and that we must be cautious about the results of retrievals at the top of the atmosphere. However the results are still fairly close to the truth at mid altitudes (those probed). We also need to be careful about the chemical models used for planetary atmosphere. If the chemistry of one molecule is not correctly described, this will bias all the others, as well as the retrieved temperature. Finally, although fitting a wider wavelength range and higher resolution has been shown to increase retrievals accuracy, we show that this could depend on the wavelength range chosen, due to the accuracy on modeling the different features. In any case, 1D retrievals are still correct for the detection of molecules, even in the event of an erroneous abundance retrieval.}


   \maketitle
%

\section{Introduction}

Over the past three decades, there has been an exponential rise in the discovery of exoplanets that have dramatically expanded our understanding of exoplanets. Their observed diversity (e.g., \citep{Gaudi2021}) has fueled the excitement and imagination of the scientific community and the general public and challenged long-held assumptions about planet formation and evolution. The focus in the exoplanet field has shifted from studying the bulk and orbital parameters to understanding the true nature of exoplanets through their compositions, atmospheres, and climates \citep{Gaudi2021,Guillot_review}. The atmospheres are shaped by the stellar environment and offer a glimpse into the planetary interior, which holds evidence of the formation process of the planet. Transiting exoplanets are very well suited for atmospheric analysis, as the stellar light filtering through their atmosphere provides a wealth of information on its composition and thermodynamic state \citep{Sing2018,tsiaras2018}. Transmission spectroscopy has therefore emerged as the most promising technique for atmospheric characterization \citep{Seager2000,Tinetti2007,Kreidberg2014,Line2016,welbanks2019b,Madhusudhan2019, Edwards2020,Skaf2020,pluriel2020,Guilluy2021,Mugnai2021}. Until very recently, the main challenges in atmospheric studies using this technique were limited sensitivity and spectral range that prevented proper sampling of molecular features. In this regard, the launch of the James Webb Space Telescope (JWST) marks a new milestone in the exploration of exoplanets and provides us with a unique opportunity to uncover their true nature and formation-evolution histories. The upcoming space mission Atmospheric Remote-Sensing Infrared Exoplanet Large-survey (Ariel) of the European Space Agency (ESA) will provide yet another important contribution, by systematically characterizing the atmospheres of entire populations of exoplanets (e.g. \citet{Charnay2023}).

The JWST \citep{Mcelwain2023} is a large, highly sensitive infrared-optimized space telescope developed in a collaboration between NASA, the European Space Agency (ESA), and the Canadian Space Agency (CSA). The JWST science goals were organized into four themes, including the Planetary Systems and the Origins of Life theme, which aims to determine the physical and chemical properties of planetary systems and investigate the potential for the origins of life in those systems \citep{Gardner2006,Beichman2018}. The overall design of the observatory includes three main components: the telescope and scientific instruments, the 5-layer sunshield, and the spacecraft bus. The telescope has a unique 6.5 m-class primary mirror comprising 18 hexagonal segments, phased together to act as a single mirror. There are four science instruments, each with several observing modes: a Near-Infrared Camera (NIRCam, 0.6–5.0 $\mu$m) (e.g., \citep{Beichman2012}), a Near-Infrared Spectrograph (NIRSpec, 0.6–5.3 $\mu$m) \citep{Jakobsen2022}, a Near-Infrared Imager and Slitless Spectrograph (NIRISS, 0.6–5.0 $\mu$m) \citep{Doyon2012}, and a Mid-Infrared Instrument (MIRI, 5–28.5 $\mu$m) \citep{Rieke2015}. The telescope and science instruments are passively cooled to below 50 K by the sunshield and thermally isolated from the spacecraft bus and solar array. The combined spectral coverage of all science instruments is from 0.6 to 28.5 $\mu$m, opening uncharted territory in atmospheric characterization, as evidenced by recent inferences of CO$_2$ and SO$_2$ in the atmosphere of the hot Jupiter WASP-39b \citep{Bean2018, Rustamkulov2023natur, Feinstein2023Natur, Ahrer2023Natur, Tsai2023Natur, Alderson2023Natur}.

The Atmospheric Remote-Sensing Infrared Exoplanet Large-survey (Ariel) \citep{Tinetti2018,Tinetti2021} is the M4 ESA space mission of the Cosmic Vision program and will operate from the L2 point starting in 2029. Ariel is a dedicated mission for the spectroscopic observation of transiting exoplanets and will conduct the first unbiased survey of a large and diverse sample of approximately 1000 planets in the optical and near-infrared \citep[e.g.,][]{Mugnai2021a}. The Ariel payload will mount an off-axis Cassegrain telescope with a 1-m class primary mirror feeding a collimated beam, split by dichroics, into two separate instruments with a coincident field of view: the Fine Guidance System (FGS) and the Ariel InfraRed Spectrometer (AIRS). FGS includes three photometric channels (VISPhot, 0.5–0.6 $\mu$m; FGS1, 0.6–0.80 $\mu$m; FGS2, 0.80–1.1 $\mu$m) and a low-resolution Near-InfraRed Spectrometer (NIRSpec, 1.1–1.95 $\mu$m and R $\geq$ 15); AIRS has two low-to-medium resolution IR channels (CH0, 1.95–3.9 $\mu$m and R $\geq$ 100; CH1, 3.9–7.8 $\mu$m and R $\geq$ 30). The payload module comprising the telescope and the science instruments will be passively cooled to $\sim$55 K with the thermal shield assembly. With this instrumentation, Ariel will provide simultaneous observations of the whole 0.5 to 7.8-$\mu$m spectral band, encompassing the emission peak of warm and hot exoplanets and well-sufficient to detect all molecular species \citep{Encrenaz2015}. Key science questions to be addressed by the Ariel mission include: "What are the physical processes shaping planetary atmospheres?", "What are exoplanets made of?" and "How do planets and planetary systems form and evolve?" \citep{Turrini2018, Changeat2020}. JWST and Ariel are complementary of each other. JWST, with its exquisite sensitivity, is already providing transformative science for the field of exoplanets and, thanks to its precision launch, will continue to do so for the next 20+ years \citep{2023PASP..135e8001M}, with about 25\% time allocated to exoplanet observations. Ariel, on the other hand, already during the four-year nominal mission, will observe a conspicuous fraction of the known planetary population, providing the statistics to interpret the observed gamut of planetary bodies and place the JWST observations into a wider context.

The exquisite sensitivity and wavelength range offered by JWST and, to a lesser extent, Ariel, enable the study of subtler effects than previously possible, opening new pathways to understanding the true nature of exoplanets. A prime example is the characterization of the 3D nature of exoplanets. In this regard, the geometry of a transit allows us to probe only a limited volume around the terminator line – the region of the atmosphere at the border between the hot day-side and the cooler night-side \citep{brown2001,kreidberg2018}. Therefore, transit spectroscopy measurements average the atmospheric features from the morning and evening sides of the planet \citep{fortney2010, caldas2019effects, Wardenier2022}. Often, a uniform terminator is assumed in interpreting transit spectroscopy data, disregarding 3D atmospheric effects. This assumption can be sufficient to interpret the data in the case of cold planets that have more homogeneous atmospheres \citep{MacDonald2020} but not for hot Jupiters and ultra-hot Jupiters. Hot and ultra hot Jupiters present strong day-night contrast in temperature due to the shorter radiative timescale compared to the dynamical timescale \citep{Sudarsky2000, Guillot2010, Bell2018, Arcangeli2018}, and also chemical heterogeneities \citep{Changeat2019, Baeyens2021}. These differences diminish as we move towards warm Jupiter (equilibrium temperature below 1400 K), where zonal circulation tends to homogenize atmospheres \citep{Venot2020, Baeyens2022}. Global Circulation Models (GCMs) are numerical models of the full 3D structures and dynamics of atmospheres that can help us explore and gain a better understanding of atmospheric processes and characteristics that cannot be captured by a 1D plane parallel approach \citep{showman2008, Leconte2013, Venot2014, parmentier2018}.

This work addresses questions on the interpretability of transmission spectra given the 3D nature of exoplanetary atmospheres. The first one is: "How does the 3D atmospheric structure affect the transmission spectra of exoplanets, from a cold planet to an ultra-hot Jupiter?". Recent works \citep{pluriel2020a, Lacy2020, Wardenier2022, Pluriel2022} concluded that, for ultra-hot Jupiters, the 3D structure plays a major role in shaping the transmission spectra. If the temperature of the atmosphere is not high enough to dissociate a molecule on both the day-side and the night-side, the amplitude of its spectral features will be larger than predicted by a 1D plane parallel approach. In addition, \citet{Falco2022} showed how the changes in planet orientation during the transit allow us to probe the horizontal variations in the atmosphere. \citet{Pluriel2022} investigated 3D effects in the transmission spectra of hot Jupiters, dividing them into three main groups: vertical effects, horizontal effects along the limb, and horizontal effects through the limb.

The second question we ask is: "Can 1D retrievals find consistent parameters (T-P profile, abundances, C/O ratio, metallicity, and clouds)?". This question is closely related to the first one: if the 3D atmospheric structure strongly affects transmission spectra – can 1D retrieval models find correct atmospheric parameters? In this regard, \citet{MacDonald2020} investigated the following conundrum: "Why are most inferred temperatures from transmission spectra far colder than expected from the equilibrium temperature?". They concluded that a 1D model can fit the transmission spectra of planets with asymmetric terminators, but retrieved atmospheric parameters may not represent true terminator averages. Also, the retrieved temperatures of planetary terminators may be biased by hundred degrees below their real value, and this bias is most extreme in the case of ultra-hot Jupiters (but see \cite{welbanks2022}). This study also mentions the biases in chemical abundances derived from 1D retrievals. \citet{pluriel2020a} concluded the same – if the temperature and the chemical composition vary across the limb, which is the case for 3D structures, 1D retrievals cannot find the correct molecular abundances. This also affects the inferred C/O ratio, which is an indirect estimate based on the abundances of all C- and O-bearing molecules. Recently \citet{Zingales2022} and \citet{welbanks2022} demonstrated that the choice of the retrieval model is critical for correctly retrieving the thermal structure of the atmosphere. Lastly, \citet{Pluriel2022} provided a "cheat sheet" of the minimum model assumptions needed to avoid biases in interpreting atmospheric properties. When optical absorbers are present, 1D models are adequate to describe transmission spectra for atmospheric equilibrium temperatures lower than 1400 K; in their absence, the 1D assumption can extend up to 2000 K. Above these temperatures, retrievals with 1D models return biased estimates of the parameters in the forward model.

The problem of terminator inhomogeneities addressed here can be solved by more sophisticated retrieval using 2D atmospheric models \citep{MacDonald2020}. As these inhomogeneities would be observable by JWST \citep{espinoza2021}, this is part of the current needs. \cite{Zingales2022} and \cite{welbanks2022} have studied hot and ultra-hot Jupiter to show that models beyond the 1D approach such as 2D models can retrieve consistent temperature profiles. Similarly, \cite{MacDonald2022} and \cite{Nixon2022} have implemented a 3D parametric pressure-temperature profile to model day-night temperature variations with improved retrieval accuracy. This study is part of this new challenge in exoplanet research, and will specifically address the issue of errors introduced by 1D retrieval models when applied to contrasting 3D atmospheres.

To address the above-mentioned issues, we need two distinct branches of action: one that can produce simulated transmission spectra for different planets, accounting for 3D structures, and one that can perform atmospheric retrievals and compare the inferred parameters with the forward models. Our goal is to assess the extent to which our retrievals can reconstruct the true atmospheric composition. This end-to-end system allows us to investigate the interpretation of our atmospheric retrievals consistently.
The first branch starts from precomputed GCMs for three planets (see Section \ref{GCM}): 
\begin{itemize}
    \item GJ~1214~b, a Neptune-like planet orbiting an M-type star; 
    \item HD~189733~b, a hot Jupiter around a K-type star; 
    \item WASP-121~b, an ultra-hot Jupiter orbiting an F-type star. 
\end{itemize}
Then, it uses the \textit{Pytmosph3R} code to produce 3D transmission spectra (see Section \ref{section: pytmosph3r}). In this step, we use two different configurations for each planet: an equilibrium chemistry model and a model with chemical profiles constant with altitude. The final products of the first branch are spectra with attached errorbars, to reproduce spectra "as observed" by JWST and Ariel. The expected errorbars are estimated using PandExo and ArielRad, two simulators of the noise performance of JWST and Ariel, respectively (see Section \ref{section: uncertainty_model}).
The second branch starts from these simulated observations and performs a Bayesian retrieval to estimate the best-fitting parameters of the model. As a retrieval tool, we use the retrieval framework \textit{TauREx} 3 (Tau Retrieval for Exoplanets), briefly described in Section \ref{subsec: taurex}. Section \ref{retproc} details the retrieval procedure, the chemical configurations, and the other atmospheric parameters. The retrieval procedure is the same for all spectra – the same set of atmospheric models is applied, to remove our a priori knowledge and compare the obtained results correctly. We discuss the results from this comparison and their implications for interpreting transmission spectra with the 1D assumption in Section \ref{section: analysis}. Our conclusions are summarized in Section \ref{section: conclu}.

\section{Transmission spectra simulations}

The following three planets (GJ~1214~b, HD~189733~b, WASP-121~b) have been chosen from warm Neptune to ultra-hot Jupiter (see Table~\ref{tab:pandexo}) in order to study the transmission spectra and retrieval biases depending on the temperature of the planets. Following \cite{Al-Refaie2022} study, we go a step further by focusing on biases arising from 1D vertical thermal variation as well as full 3D thermal structure.
We simulate JWST (NIRSpec + MIRI) and Ariel observations. We do not consider clouds to focus on general planetary properties biases.\\

\subsection{Global Climate Models}
\label{GCM}

GJ~1214~b has been simulated using the generic Planetary Climate Model. This model has been specifically developed for exoplanets and paleoclimate studies \citep{charnay2015, Leconte2013}. The dynamical core solves the primitive hydrostatic equations of meteorology on an Arakawa C grid, using a finite difference scheme. Radiative transfer is solved using the correlated-k model. Radiative effects of H$_2$, He, \hho, \chhhh, \nhhh, \co, and \coo are taken into account, assuming a 100x solar metallicity. The horizontal resolution is 64 $\times$ 48 and we use 45 vertical layers between 80 bar and 3 Pa, equally spaced in log pressure. The star is taken as a blackbody at 3026 K, and we assume an internal temperature of 60 K. The dynamical time-step is 60s and the physical/radiative time-step is 300s. The model was integrated for 1600 days. For a more complete description of the model and the simulation, we refer the reader to \citet{charnay2015}, as the model used here is from this study.

For HD~189733~b, we make use of the Met Office Unified Model \citep{Drummond2018}. The model solves the deep atmosphere, non-hydrostatic Navier–Stokes equations on an Arakawa C grid. Radiative transfer is handled through the SOCRATES\footnote{\url{https://code.metoffice.gov.uk/trac/socrates}} code adapted for hot Jupiters \citep{Amundsen2016}. A chemical relaxation scheme is used and the radiative transfer is computed in 32 wavelength bins. The chemistry includes CH$_{\rm 4}$, CO and H$_{\rm 2}$O. The simulation is integrated for 1000 Earth days. For a more complete description of the model, see \citet{Drummond2018} as the model used here is the same as in this study.

WASP-121~b has been simulated using the SPARC/MIT global circulation model \citep{showman2009}. The model solves the same primitive equations of meteorology as the generic Planetary Climate Model on a cubic-sphere grid. It has been widely used for various hot Jupiters \citep{showman2009, kataria2015,parmentier2016,parmentier2018, Parmentier2021} and has also been applied to ultra-hot Jupiters \citep{kreidberg2018, Arcangeli2019}. For this study, we use the model published in \citet{parmentier2018}. The horizontal resolution is $C32$, equivalent to 128 cells in longitude and 64 in latitude and $53$ vertical levels with pressure ranging from $200$ bar to 2 $\mathrm{\mu}$bar. Radiative transfer is handled using the two-stream approximation with 11 wavelength bins, as done in \citet{Kataria_2013}. The model assumes chemical equilibrium, taking into account the thermal dissociation of water and hydrogen. The chemical species taken into account are H$_{\rm 2}$O, H$^{-}$, CO, TiO, Na, Vo, K and H$_{\rm 2}$. However, H$_{\rm 2}$ recombination is neglected despite its non-negligible impact on the thermal and dynamical structure \citep{tan2019}. For a more complete description of the model and the simulation, we refer the reader to \citet{parmentier2018}.\\

In the three models considered here, disequilibrium chemistry is not taken into account and the models are cloud-free and haze-free.\\

For each planet, we build a pseudo-1D version of the 3D model. The temperature profiles of the whole grid are replaced by the same 1D profile. This profile is the temperature profile at the equator of the western terminator (coldest one) for each model. Thus, the pseudo-1D models are representative of each temperature condition from warm Neptune to ultra-hot Jupiter. The purpose of these pseudo-1D models is to control the correct behavior of the retrieval code. If we consider the 1D retrieval code, it should correctly retrieve the pseudo-1D models. Thus, we can confidently untangle the 3D effects.

\begin{table}[h!]
\caption{Planet-Stellar parameters used to simulate the transmission spectra and as input into \textit{PandExo}.}
\resizebox{0.5\textwidth}{!}{%
    \begin{tabular}{|l|c|c|c|}
        \hline
        Parameters   & GJ~1214~b & HD~189733~b & WASP-121~b \\
        \hline
        T$_p$ [K]    & 596       & 1209        & 2358       \\
        R$_*$ [Rsun] & 0.21      & 0.75        & 1.46       \\
        T$_*$ [K]    & 3250      & 5052        & 6459       \\ 
        Z$_*$        & 0.29      & -0.02       & 0.13       \\ 
        log(g)       & 5.03      & 4.49        & 4.24       \\ 
        R$_p$ [Rjup] & 0.243     & 1.13        & 1.91       \\
        \hline
	References & \cite{Charbonneau2009} & \cite{Stassun2017} & \cite{Delrez2016} \\
                   & \citet{Cloutier2021} & \citet{Addison2019} & \\
        \hline
    \end{tabular}
}
    \label{tab:pandexo}
\end{table}

\subsection{Pytmosph3R}\label{section: pytmosph3r}

Based on the planetary atmospheres described in Section \ref{GCM}, we used the latest version of \textit{Pytmosph3R} \citep{Falco2022} to generate the transmission spectra. It takes into account the 3D structure of the atmosphere and uses directly the line-by-line cross sections calculated by ExoMol \citep{Yurchenko2011, Tennyson2012, Barton2013, Yurchenko2014, Barton2014}, more specifically cross sections are taken from \cite{Chubb2021}. Species abundances were established in two different ways: (i) a constant chemistry model, i.e. where abundances are independent of temperature and pressure, thus constant everywhere, (ii) an equilibrium chemical model.

The abundances of the constant chemistry model have been chosen close to the equilibrium ratio for a given temperature representative of each planet, and set up over all longitudes, latitudes and altitudes. We have taken into account only the main species: H$_2$O, CO, CH$_4$, CO$_2$, HCN, C$_2$H$_2$, NH$_3$, C$_2$H$_4$, and in addition TiO, VO, K, Na, SiO, FeH for HD~189733~b and WASP-121~b. The values are given in Table \ref{tab:gj1214b_ret}, \ref{tab:hd189733b_ret}, and \ref{tab:wasp121b_ret}, input column in red. This theoretical construction, not representative of the real atmosphere, removes one degree of freedom (chemistry) to check how 1D retrieval models handle 3D atmospheric thermal structures and thus what can be the biases.

Equilibrium chemistry leads to variable chemical profiles. In the modeled atmospheres, equilibrium chemistry can be expected in the hottest and densest regions. However, we know that non-equilibrium chemistry must be accounted for, especially in the upper atmosphere \citep{Cooper_2006, Moses_2011, Moses_2012, Venot2012, Venot_2020, molaverdikhani2019, Tsai_2021, Tsai2022}. This has been very recently implemented in the retrieval code (FRECKLL code developed by \cite{alrefaie2022freckll}) but this is computationally expensive. We used ACE chemistry (see Section \ref{subsec: chemmod} for details) to model the input equilibrium chemistry of GJ~1214~b and the chemical abundances used in \cite{parmentier2016} to model the input equilibrium chemistry of HD~189733~b and WASP-121~b. This theoretical construct, relative to the constant case, focuses on the research biases that can arise from the chemical retrieval models.

\subsection{Uncertainty model}\label{section: uncertainty_model}


\subsubsection{PandExo - JWST}\label{subsec:pandexo}

To simulate JWST observables (spectra Figure \ref{fig: spectra_all}), we utilize the \textit{PandExo} 1.5 package \citep{Batalha2017}. This program is a noise simulator designed for JWST transiting observations of exoplanets. We make use of the model to simulate one transit of each planet using NIRSpec-PRISM and one transit using MIRI-LRS. For each planet, we consider a saturation limit of 80\% of the full well and a fraction of time out-of-transit to in-transit of 1. For each instrument and planet, we use the \textit{optimize} option for the number of groups per integration, which automatically defines the best settings to carry the observations. The planet and star-specific parameters used for producing the observables are summarized in Table~\ref{tab:pandexo}. \\
\textit{PandExo} neither includes the varying stellar noise nor takes into account the transit ingress and egress. However, we estimate that the produced observables are a good enough approximation to what one will observe with the JWST facility and should not change the conclusions of this work. \\
It is worth noting that the three systems studied here have a J-band stellar magnitude below 11.4, which is the estimated saturation limit of NIRSpec-PRISM. Thus, these systems are not observable with this instrument configuration. Yet, we still performed the computations and studied these systems, as the goal of this analysis is not to prepare JWST observations \textit{per se}, but to highlight possible biases introduced by retrievals on JWST-like data sets.
We did not add random noise, as explain in the following section.

\subsubsection{ArielRad - Ariel}

\paragraph{The Ariel Radiometric model} ArielRad \citep{Mugnai2020a} is the radiometric simulator of the Ariel payload, developed and maintained by the Ariel Consortium. We will briefly describe ArielRad here; for a technical description, including the detailed noise model, the reader is encouraged to read the original article. 

Given the description of the payload and a list of candidate exoplanets, ArielRad outputs the expected experimental uncertainty on their measured atmospheric transmission or emission spectra. The simulation propagates the stellar light through the payload, accounting for each transmission or dispersion by interposing optical components until reaching the focal planes. 
Then, ArielRad evaluates the noise contributions (with margins) from stationary processes, i.e., stellar photon noise, detector noise, dark current, zodiacal background, and instrument emission. Jitter noise is computed externally by ExoSim \citep{Sarkar2021b}, the end-to-end time-domain simulator of Ariel observations, and included in the final noise budget. 

Then, ArielRad returns the uncertainty estimates on a single transit or eclipse observation. Ariel defines an observation to last 2.5 times the time between the first and the last contact between the planetary and the stellar disks to obtain a sufficiently long baseline integration for the light curve fit and the transit depth estimation \citep{Mugnai2020a}.
Because the astronomical measurement is the contrast ratio with the signal from the stellar host, ArielRad uses the contrast ratio to compute the observed spectrum's expected signal-to-noise ratio (S/N). 

The Ariel mission adopts a four-tier observation strategy in which, after each observation, the resulting spectrum from each spectrometer is binned in data analysis according to specific requirements to optimize the S/N and the mission scientific return \citep{Edwards2019}. 
ArielRad, knowing the binning and spectral resolution implemented in the different tiers, computes the S/N in the spectral bin according to the tier of interest. Then, it estimates the number of observations required for each planet to reach the tier’s required S/N. 

From the number of observations, ArielRad obtains the final noise estimate for a planet by rescaling the noise on a single observation. 
These uncertainties can be attached to simulated forward models of transmission or emission spectra, binned down to the tier spectral resolution to obtain the simulated observed spectra. 

\paragraph{Computing the errorbars}

We utilize the general procedure described above to calculate the Ariel observation uncertainties for the three planetary targets of interest. 
We use an updated version of the code, which is a wrapper of ExoRad, the instrument-independent version of the radiometric simulator publicly available on GitHub\footnote{\url{https://github.com/ExObsSim/ExoRad2-public}}, and ArielRad-Payloads, the repository of configuration files for the payload maintained by the Ariel Consortium. For reproducibility, we report the code versions in Table~\ref{tab: ariel_rad}.

\begin{table}[!htbp]
    \caption{Versions of the codes used to generate the Ariel spectra.}
    \centering
    \begin{tabular}{@{}ll@{}}
    \toprule
    \textbf{Code}     & \textbf{Version} \\ \midrule
    ArielRad          & 2.4.25           \\
    ExoRad            & 2.1.111          \\
    ArielRad-Payloads & 0.0.16           \\ \bottomrule
    \end{tabular}
    \label{tab: ariel_rad}
\end{table}

For each planet, we assume a mean molecular weight of 2.3 a.m.u. to simulate \hh-He dominated atmospheres. We use this parameter to calculate the atmospheric scale height and, consequently, the contrast ratio of the transit. We acknowledge that assuming an \hh-He dominated atmosphere conflicts with the GJ~1214~b model detailed above, which uses a much higher metallicity. Assuming a lower mean molecular weight leads us to an over-optimistic estimate of the number of transits for the modeled spectrum. Therefore, the S/N (and retrievals) will reflect a worst-case scenario.

We utilize the Ariel strategy for collecting data during an observation; therefore, the ratio of observing time in and out of transit is 1/1.5. Then, for each planet, we estimate the noise per spectral bin and from there, the S/N for a transit observation in Tier 3. In this Tier, the raw spectral data from each spectrometer are binned at $R=20, \, 100, \, 30$ in FGS-NIRSpec (1.1–1.95 $\mu$m), AIRS-Ch0 (1.95–3.9 $\mu$m), and AIRS-Ch1 (3.9–7.8~$\mu$m), respectively. We find that 1, 3, and 4 observations are needed to achieve the Ariel Tier 3 required S/N for HD~189733~b, GJ~1214~b, and WASP-121~b, respectively. Then, we rescale the noise by the square root of the corresponding number of observations, assuming each observation has a Gaussian noise distribution. 
Finally, we attach the rescaled noise estimates to the respective transmission spectra, binned down at the Tier 3 wavelength grid.  It should be noted that we do not scatter the spectra according to random noise corresponding to the estimated error bars. This was done throughout the paper to avoid introducing a susceptibility to individual random noise realizations, which would defeat the purpose of characterizing retrieval biases and finding intrinsic correlations between the atmospheric parameters. While using unscattered spectra may result in unrealistically precise constraints, if the spectra contain sufficiently redundant information, discrepancies vs. using scattered spectra should not be too large~\citep{Feng2018,Changeat2020}.

Alternatively, given the noise for a single transit observation and the atmospheric spectrum, we could calculate a more realistic S/N that does not rely on the assumed atmospheric scale height \citep{Mugnai2021a}. 
However, the S/N would depend on the assumed atmospheric spectrum, changing the number of observations on a single target. 
However, (i) an observability study is outside the scope of this paper, and (ii) in the following we show that even for GJ~1214~b (lower number of transits than realistic given the mean molecular weight assumed) we can investigate the main effects of interest.

\begin{figure*}[h!]
\centering
\includegraphics[scale=\spec,trim = 0cm 0cm 0cm 0cm, clip]{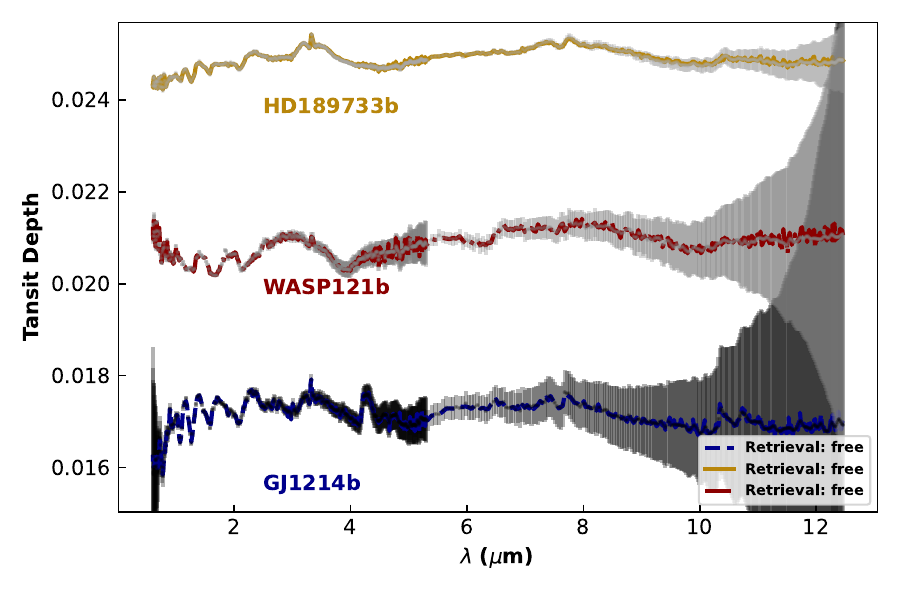}
\includegraphics[scale=\spec,trim = 0cm 0cm 0cm 0cm, clip]{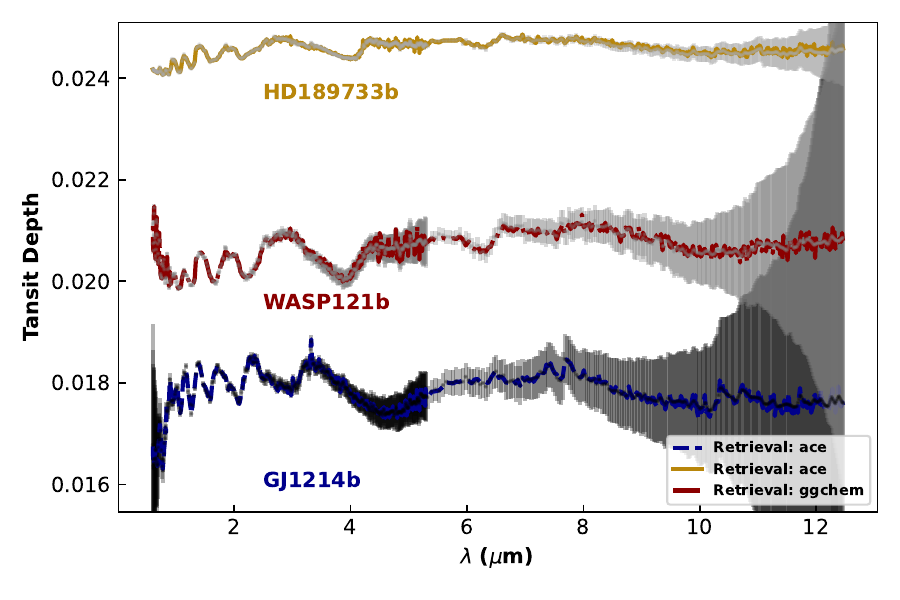}
\\
\includegraphics[scale=\spec,trim = 0cm 0cm 0cm 0cm, clip]{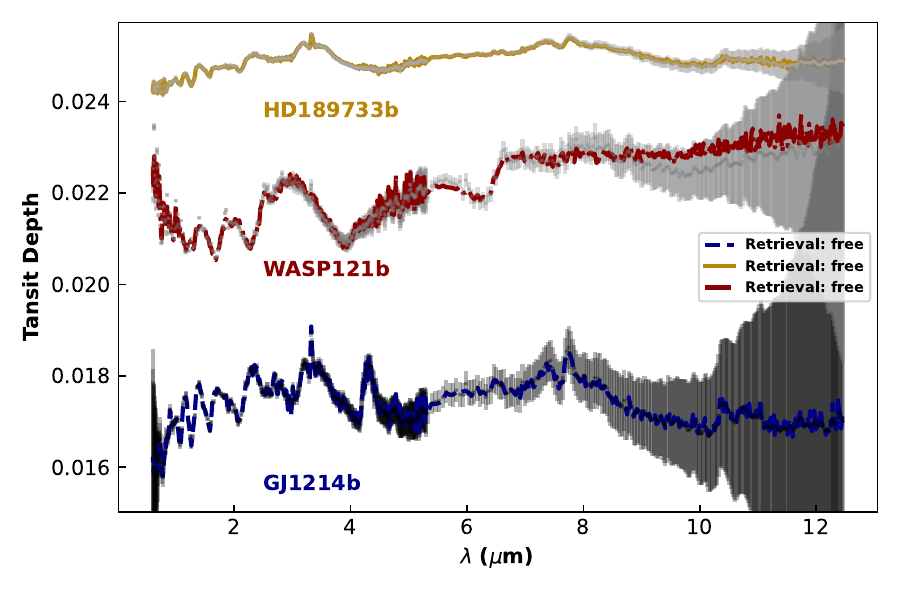}
\includegraphics[scale=\spec,trim = 0cm 0cm 0cm 0cm, clip]{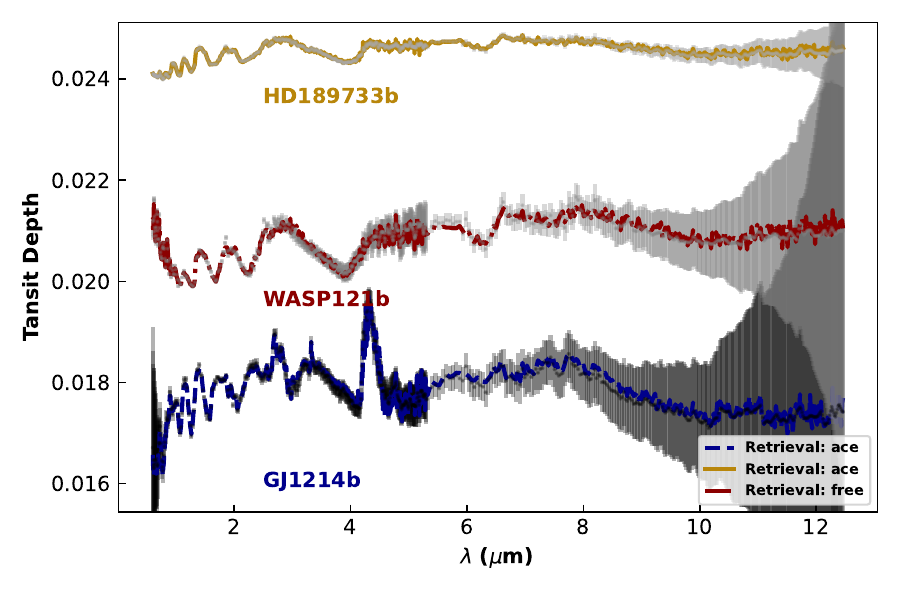}
\\
\includegraphics[scale=\spec,trim = 0cm 0cm 0cm 0cm, clip]{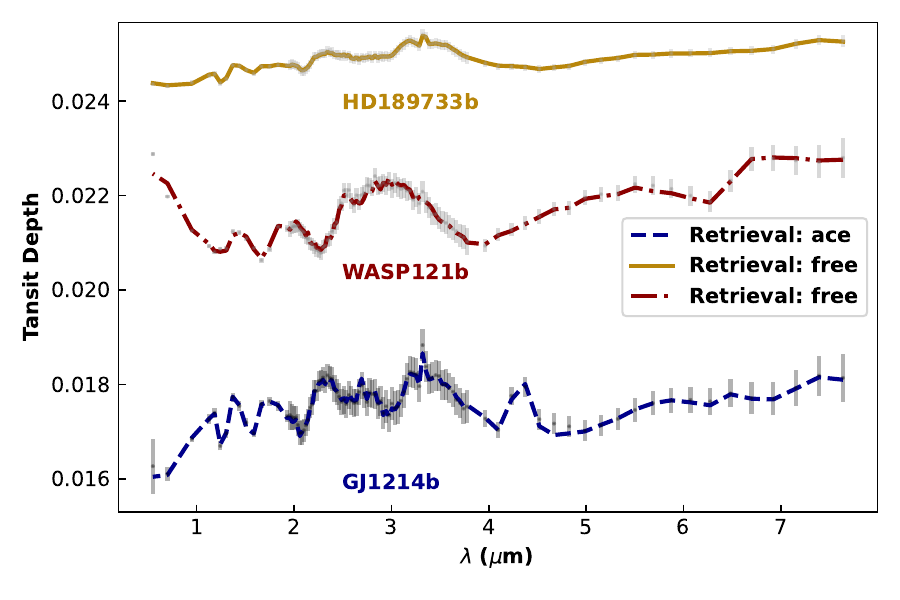}
\includegraphics[scale=\spec,trim = 0cm 0cm 0cm 0cm, clip]{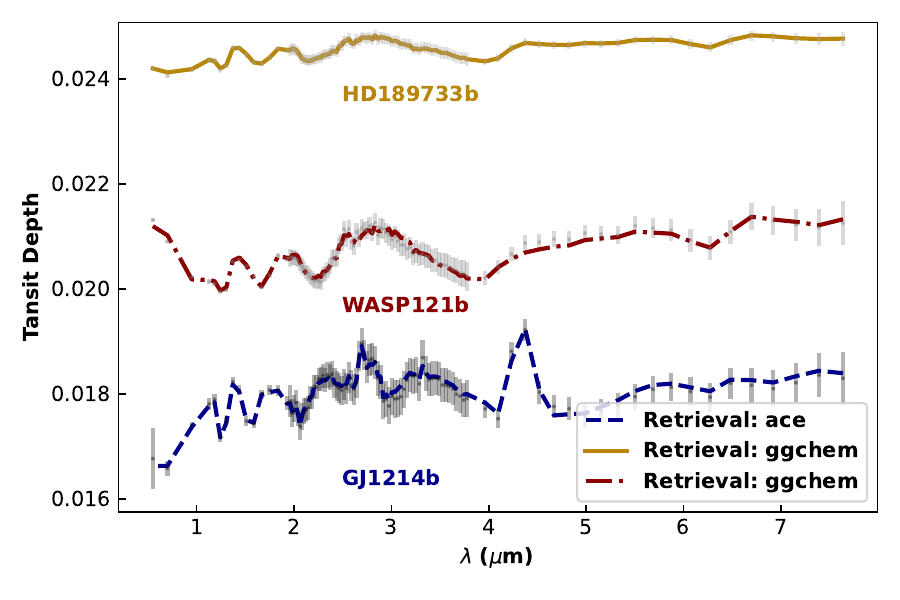}
\caption{Simulation of observations for HD~189733~b, WASP-121~b, and GJ~1214~b (grey) with the corresponding best retrieval configuration in solid yellow, red dash-dot, and dashed blue lines respectively. \textbf{Left panels:} constant input chemistry. \textbf{Right:} equilibrium input chemistry. \textbf{Top and middle:} JWST simulated observation for atmospheric with 1D and 3D assumption respectively. \textbf{Bottom:} Ariel simulated observation for a 3D atmospheric assumption. }
\label{fig: spectra_all}
\end{figure*}

All the spectra calculated with the methodology described in this section are shown in Figure \ref{fig: spectra_all}. This includes 6 input configurations for each planet (listed Table \ref{tab: config}).

\begin{table}[!htbp]
    \caption{Spectra input configurations.}
    \centering
    \resizebox{0.5\textwidth}{!}{%
    \begin{tabular}{@{}l|c|c|c|c|c|c@{}}
    \toprule
    \textbf{Instrument} & \multicolumn{4}{c|}{JWST} & \multicolumn{2}{c}{Ariel} \\ \midrule
    \textbf{Dimension}  & \multicolumn{2}{c|}{1D} & \multicolumn{4}{c}{3D} \\ \midrule
    \textbf{Chemistry}  & constant & equilibrium & constant & equilibrium & constant & equilibrium \\ \bottomrule
    \end{tabular}
    }
    \label{tab: config}
\end{table}

\section{Transmission spectra retrievals}

\subsection{TauREx} \label{subsec: taurex}

As a retrieval tool, we used \textit{TauREx 3} (Tau Retrieval for Exoplanets) \footnote{\url{https://github.com/ucl-exoplanets/TauREx3_public}} a fully Bayesian inverse atmospheric retrieval framework \citep{Al-Refaie2021}. \textit{TauREx 3} consists of two main frameworks: the Forward Model framework and the Retrieval framework. The goal of the Retrieval framework is to fit a Forward Model to an observation.
A Forward Model framework is necessary to provide information about the planet, the host star, temperature-pressure profiles as well as chemistry and contributions (e.g. collision-induced absorption (CIA), limited to H$_2$-He and H$_2$-H$_2$ in the current study, Rayleigh, gray clouds). \textit{TauREx 3} adopts the layer-by-layer approach for the temperature profile, which can be parameterized in different ways, such as isothermal, a radiative two-stream approximation, a custom profile loaded from a file, or a multi-point temperature profile. The vertical pressure profile is equally spaced in log-pressure, between a P$_{max}$ and a P$_{min}$ value specified by the user along with a number of layers N$_l$ (N$_l$=200 at current study). The cloud model provided by \textit{TauREx} 3 is discretized along P$(z)$, allowing the user to define an opacity value in square meters for layers between the pressure at the top and the pressure at the bottom of the cloud deck. \\
\textit{TauREx} 3 supports equilibrium chemistry using the ACE chemical code \citep{Agundez2012,Agundez2020}, FastChem \citep{Stock2018}, GGchem \citep{Woitke2018}, and the Free chemistry model \citep{Al-Refaie2022}.
In our study, we explore different combinations of chemistry and contribution parameters. We performed retrievals with and without clouds, using ACE, FastChem, GGchem, or Free chemistry. To perform retrievals, \textit{TauREx} 3 can use several sampling techniques PyMultiNest and MultiNest, PolyChord, or dyPolyChord. For our study, we used the nested sampling retrieval algorithm Multinest with its Python version PyMultiNest \citep{feroz2009}. The Multinest algorithm samples the parameter space by subdividing it into a set of ellipsoids according to the likelihood. This set of ellipsoids can overlap, but in cases where there are several local maxima in the parameter space, the result will be a set of multiple solutions (as can be seen in Figure \ref{fig: logE_1D3D}). For more details, see \cite{feroz2008} and \cite{feroz2019}. \textit{TauREx 3} is a full Bayesian Retrieval framework, which returns the best-fit transmission model spectrum along with all parameter posterior distributions and the Bayesian Evidence. For model comparisons, we use the Bayesian Evidence as defined by \citep{Trotta2008, waldmann2015a} to compute the logarithmic Bayes factor,

\begin{equation}
  \Delta logE = \log{\frac{E_{model_A}}{E_{model_B}}} = \log{E_{model_A}} - \log{E_{model_B}},
  \label{eq:logbayes_factor}
\end{equation}
where $E_{model_A}$ and $E_{model_B}$ are the evidences of two competing models. According to \citep{Benneke2013}, by translating these Bayes factors into a statistical significance \citep{Kass1995},  \dlogE$\leq2$ can be considered as a "weak" case of models distinguishability, while $2<$\dlogE$<5$ corresponds to "moderate" and \dlogE$\geq5$ to  "strong" cases of distinguishability. Since the Bayesian Evidences of the different retrieval models depend on the number of free parameters, and that the free chemistry model has significantly more free parameters compared the the equilibrium chemistry models, a Bayesian factor that favors the free chemistry model means a more significant improvement in goodness of fit to observations.

Finally, a molecule is detected if it has a signature greater than 3$\sigma$, which is retrieved.

\subsection{Chemical model} \label{subsec: chemmod}

We consider in this study two types of chemistry that we are going to detail here : Free chemistry and equilibrium chemistry.

Free chemistry takes into account each chemical species with constant abundances throughout the layers of atmosphere. The Free chemistry models are considering the following species: \hho, \co, \chhhh, \coo, \hcn, \nhhh, \feh, \sio, \na, K, \tio and \vo. This configuration gives the model a certain degree of freedom, as it imposes no physical or chemical constraints on what will be retrieved. In this way, it is possible to retrieve abundances corresponding to non-equilibrium chemistry, or various other species distributions since the species are not correlated with each other. However, it could also retrieve non-realistic chemical abundances. It should be remembered that this is a 1D model and we are therefore limited when faced with strong vertical variation (which could be compensated for by the two-layer method of \citealt{Changeat2019}).

Chemical equilibrium is based on temperature and pressure conditions, this assumption is a classic hypothesis when considering exoplanetary atmospheres \citep{Seager2000, Burrows_2007, Burrows_2008, Fortney2008, Madhusudhan_2011, Kataria_2014, Al-Refaie2021}. A system is at a thermodynamic equilibrium state when there is thermal, mechanical, and chemical equilibrium at the same time. This equilibrium is characterized by the minimum of a thermodynamic potential, such as the Gibbs free energy. 
It happens in exoplanets' atmospheres when the dynamical timescales can be considered longer than the chemical reaction timescales and when we suppose negligible the irradiation by a dissociating or ionizing source (photochemistry or cosmic rays induced processes). Thus, chemical abundances vary with altitude according to the retrieved TP profile, as we assume they are at chemical equilibrium. As the chemical profiles are not forced to be vertically constant, this approach should be more accurate for real atmospheres than the Free chemistry approach. For very hot planets this approximation is close to reality, on the other hand, for cooler planets, vertical mixing and photodissociation have an effect on the chemical composition and the atmospheres are no longer at a thermodynamic equilibrium state. This disequilibrium chemical composition must then be taken into account with a more complex kinetic model \citep{Cooper_2006, Moses_2011, Moses_2012, Venot2012, Venot_2020, molaverdikhani2019, Tsai_2021, Tsai2022, alrefaie2022freckll}. If the observed planet's atmosphere exhibits these non-equilibrium mechanisms, or longitudinal/latitudinal heterogeneities, the retrieved parameters will be erroneous. In such cases, it may not be possible to find a consistent fit, or the retrieval may find an adequate fit that corresponds to erroneous parameters.

Although several different algorithms have been made to calculate the chemical composition at equilibrium, we focused on three algorithms for this study: ACE \citep{Agundez2012, Agundez2020}, GGchem \citep{Woitke2018} and FastChem \citep{Stock2018}.
The basic principle of these three models is the same, they start from an initial composition made up of initial abundances for each molecule taken into account then they iterate until a convergent state. However, each model has a slightly different procedure, whether in the molecules taken into account as input, the iteration method, or the network of chemical reactions. 
We will study in more detail these differences in this part.

ACE minimizes the total Gibbs free energy by applying the algorithm first introduced by \cite{White1958}. ACE is based on an algorithm implemented in the NASA/CEA program and presented in detail in \cite{gordon1994}. For a closed system of N chemical compounds at a certain temperature and pressure, in the absence of disturbance (transport, UV radiation, etc.), the equilibrium chemical composition can be calculated theoretically, thanks to standard-state chemical potential expressed as a function of the standard-state enthalpy and entropy of the species. These thermodynamic quantities can be calculated using NASA polynomial coefficients (see e.g. \cite{McBride2002}) in databases such as NASA/CEA \citep{McBride2002} or the Third Millennium Thermochemical Database \citep{goos2016}. 
The chemical species used include 105 neutral species composed of C, H, O, and N, more specifically species up to 2 carbon atoms and the main nitrogen species (\nhhh, HCN, \nn, NO$_x$). It has been validated for temperatures as low as 300K. The reader is encouraged to consult \cite{Venot2012} for more details on the ACE code thermodynamic coefficients and calculation of thermochemical equilibrium.

Both FastChem and GGchem use a second type of method for determining the chemical composition at the equilibrium state. These two programs use the law of mass action and equilibrium constants, with some subtleties (for FastChem equilibrium constants are based on Gibbs free energy while for GGchem they are based on partition functions).
This amounts to solving a system of N algebraic equations with N unknowns, which correspond to the conservation equations for N elements. The partial pressure of each molecule is defined as the partial pressure of the constituent atoms by the atomization equilibrium constant. This system of equations can then be solved by any root-finding algorithms like the Newton–Raphson method for example \citep{Russell1934, Brinkley1947, Tsuji1973}.

The thermodynamical data used in FastChem are mainly from the NIST-JANAF database detailed in \cite{Chase1998}. The list of species used has been modified to take into account molecules that may be of interest in astrophysics with data from \cite{Tsuji1973}, \cite{Barin1995}, \cite{Burcat2005} and \cite{goos2016}. The total of species used amounts to 396 neutral and 114 charge species and the code has been validated for parameters down to 100 K and up to 1000 bar. The reader is directed to \cite{Stock2018} for more details on the list of species used. We note that the FastChem code has recently been updated with FastChem 2 \citep{Stock_2022} which is more efficient but not yet available in combination with \textit{TauREx}.

Compared to the other two codes we are using, GGchem takes condensation into account. In fact, the formation of liquids and solids in the atmosphere will have an effect on the composition at thermodynamic equilibrium. Condensed species can consume certain elements leaving a significant difference in composition between before condensation and after condensation \citep{Woitke2004, Juncher2017}. Condensation will have an effect, especially at temperatures below 2000 K so this mechanism will mainly affect GJ~1214~b and HD~189733~b in our work.
The data included in this code includes 552 molecules and 257 condensates, including 38 liquids, and GGchem has been proven robust down to 100 K. All elements from hydrogen to zirconium are included, as well as the option to add tungsten and charges. We refer the reader to \cite{Woitke2018} for more details on the list of species. 
Note that the list of active molecules that can appear as a feature in the spectra are \hho, \co, \chhhh, \coo, \hcn, \nhhh, \feh, \sio, \na, K, \tio and \vo. It depends on the opacity files loaded in the \textit{TauREx} program and is independent of the type of chemical model chosen.

\subsection{Retrieval procedure}
\label{retproc}

All spectra configurations are retrieved with the same set of retrieval models. The set of retrieval models covers all the simulated configurations. We expect that each simulated spectra will be best retrieved by the corresponding retrieval model.

Each retrieval model assumes a four points temperature-pressure (TP) profile (T$_{top}$, T$_{surface}$, T$_{1}$, and T$_{2}$) with the corresponding pressure level P$_{1}$ and P$_{2}$ free to converge between P$_{surface}$ and P$_{top}$. It has already been shown \citep{Rocchetto2016,Pluriel2022} that retrieving an isothermal temperature profile generates biases for hot planets.
We also retrieve the radius of the planet. We duplicate our retrievals by adding a gray cloud level as a parameter. Finally, we took into account two chemical configurations in the retrievals, which we call Free and equilibrium chemistry (see Section \ref{subsec: chemmod} for more details).

For the Free models, we retrieve one value of abundance for each considered species : \hho, \co, \chhhh, \coo, \hcn, \nhhh, \feh, \sio, \na, K, \tio and \vo. Note that for GJ~1214~b we did not retrieve \na, K, \tio, and \vo because these species cannot be in gaseous form under these temperature and pressure conditions.
Equilibrium chemistry is calculated using three different models included in \textit{TauREx}: ACE, FastChem, and GGchem (details in Section \ref{subsec: chemmod}). Metallicity (Z) and C/O ratio, from which abundance profiles are derived, are retrieved. ACE, FastChem and GGchem have already been shown in \cite{Al-Refaie2022} to be equivalent using the same molecules and without condensation for GGchem. Thus, we consider all the molecules of each model as well as condensation for GGchem.

Therefore, we end up with 4 retrieval models (Free, ACE, FastChem, GGchem) for each of the 3 planets (GJ~1214~b, HD~189733~b, WASP-121~b) considering each of the 6 input configurations (listed Table \ref{tab: config}). This makes 4$\times$3$\times$6 = 72 retrievals ($\times$2 with clouds). For better readability of the large number of results, we do not show GGchem results for GJ~1214~b, where the temperature is not high enough to be affected by the additional condensation considered by GGchem, and we do not show ACE results for WASP-121~b, where the missing species in ACE mean that the model is not representative of this type of planet (see Section \ref{subsec: chemmod} for more details on these chemical models). Retrievals are compared to each other considering their relative Bayes Factor as described in Sec. \ref{subsec: taurex}, equation \ref{eq:logbayes_factor}, following the same idea developed in \cite{tsiaras2018} with a baseline. However, we define here the Bayes factor (\dlogE) as the difference with the logarithmic evidence of the worse model (lower one). This allows us to compare the different models between each other as it is done in \cite{Panek2023}.

Table \ref{table_retrieval} shows free parameters and priors for the retrievals. We used a uniform sampling in log space for chemical abundances in the Free chemistry model, the metallicity, the pressure P$_1$ and P$_2$, and the pressure of the clouds, and a uniform sampling in linear space for the temperatures, the radius, and the C/O ratio.

\begin{table}[!htbp]
    \centering
    \caption{Free parameters and priors for the retrievals.}
    \resizebox{.5\textwidth}{!}{%
    \begin{tabular}{c|c|c|c}
    \hline
    \textbf{Parameters} & \multicolumn{3}{c}{\textbf{Bounds}} \\ \hline
                        & \textbf{GJ~1214~b} & \textbf{HD~189733~b} & \textbf{WASP-121~b} \\ \hline
    T$_{top}$ [K]                 & 100 to 2000 & 500 to 2500 & 450 to 3750 \\
    T$_{bot}$ [K]                 & 100 to 2000 & 500 to 2500 & 450 to 3750 \\
    T$_{1}$ [K]                   & 100 to 2000 & 500 to 2500 & 450 to 3750 \\ 
    T$_{2}$ [K]                   & 100 to 2000 & 500 to 2500 & 450 to 3750 \\
    log$_{10}$(P$_{1}$) [Pa]      & 2 to 6      & 2 to 6      & 2 to 6      \\
    log$_{10}$(P$_{2}$) [Pa]      & -1 to 6     & -1 to 6     & -1 to 6     \\
    log$_{10}$(P$_{clouds}$) [Pa] & -2 to 6     & -2 to 6     & -2 to 6     \\
    radius [R$_{jup}$]            & 0.1 to 0.5  & 0.5 to 2.0  & 1.0 to 2.5  \\
    \hline
    \multicolumn{4}{c}{\textbf{Chemistry}} \\ \hline
    \textbf{Equilibrium}          &             &             &             \\
    log(Z)                        & -1 to 3     & -1 to 3     & -1 to 3     \\
    C/O                           & 0.01 to 2   & 0.01 to 2   & 0.01 to 2   \\
    \textbf{Constant}             &             &             &             \\
    log$_{10}$(\hho)              & -12 to -1   & -12 to -1   & -12 to -1   \\
    log$_{10}$(\co)               & -12 to -1   & -12 to -1   & -12 to -1   \\
    log$_{10}$(\chhhh)            & -12 to -1   & -12 to -1   & -12 to -1   \\
    log$_{10}$(\coo)              & -12 to -1   & -12 to -1   & -12 to -1   \\
    log$_{10}$(\hcn)              & -12 to -1   & -12 to -1   & -12 to -1   \\
    log$_{10}$(\nhhh)             & -12 to -1   & -12 to -1   & -12 to -1   \\
    log$_{10}$(\feh)              & -12 to -1   & -12 to -1   & -12 to -1   \\
    log$_{10}$(\sio)              & -12 to -1   & -12 to -1   & -12 to -1   \\
    log$_{10}$(\na)               & -12 to -1   & -12 to -1   & -12 to -1   \\
    log$_{10}$(K)                 & -12 to -1   & -12 to -1   & -12 to -1   \\
    log$_{10}$(\tio)              & -12 to -1   & -12 to -1   & -12 to -1   \\
    log$_{10}$(\vo)               & -12 to -1   & -12 to -1   & -12 to -1   \\
    \end{tabular}
    }
    \label{table_retrieval}
\end{table}

\section{Results and discussion}\label{section: analysis}

\subsection{From 1D to 3D}

In this section, we present how transmission spectra are affected when considering 1D or 3D GCM models, with or without equilibrium chemistry. We have developed this study from warm to ultra-hot planets, but with some specific differences on metallicity and planet gravity at the theoretical surface. This highlights the fact that it's not just the effective temperature trend that brings biases, but that other parameters are also involved.

\begin{figure*}[]
\centering
\includegraphics[scale=\spec,trim = 0cm 0cm 0cm 0cm, clip]{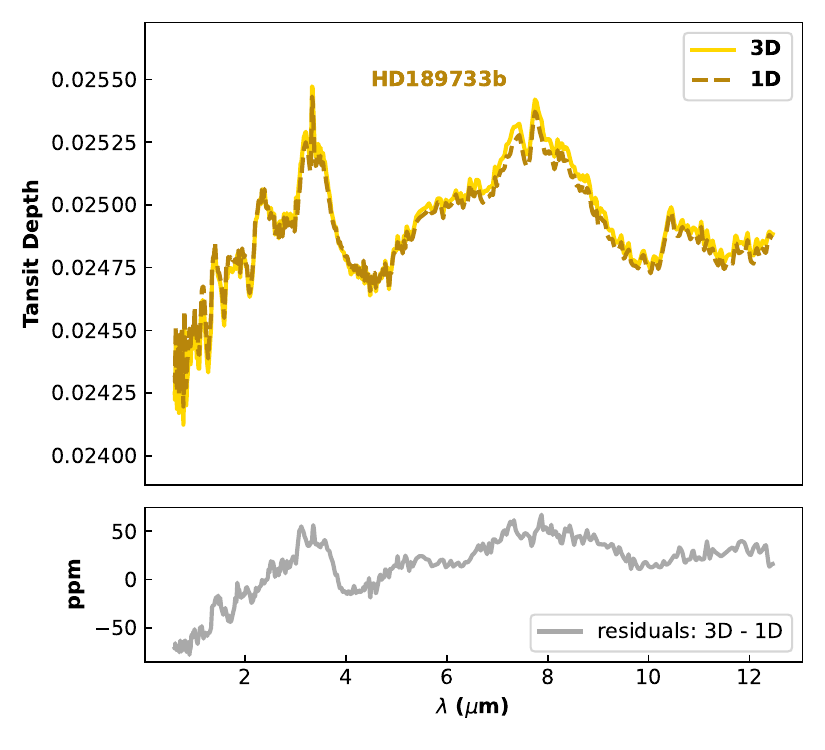}
\includegraphics[scale=\spec,trim = 0cm 0cm 0cm 0cm, clip]{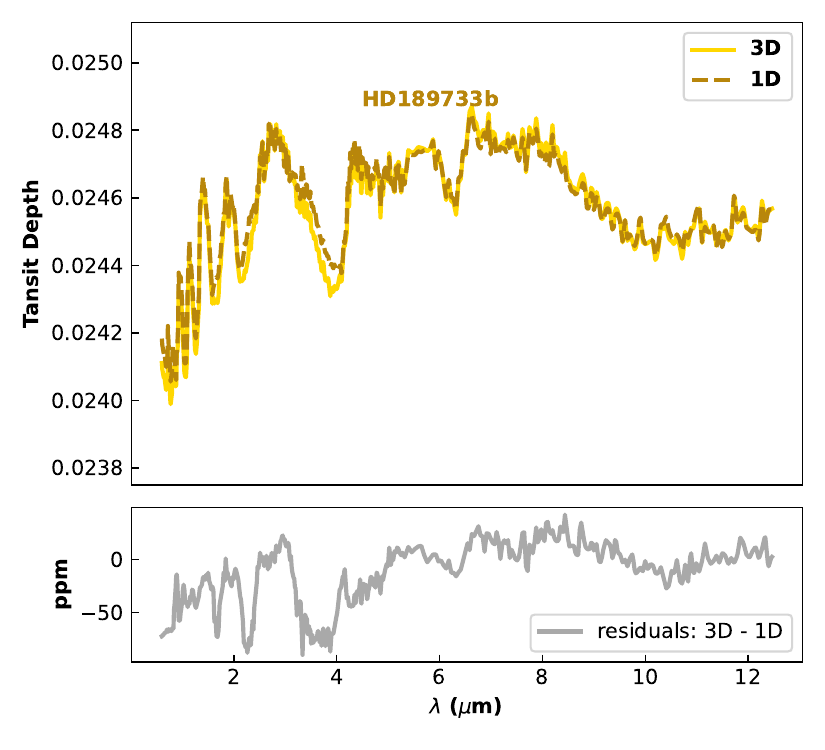}
\\
\includegraphics[scale=\spec,trim = 0cm 0cm 0cm 0cm, clip]{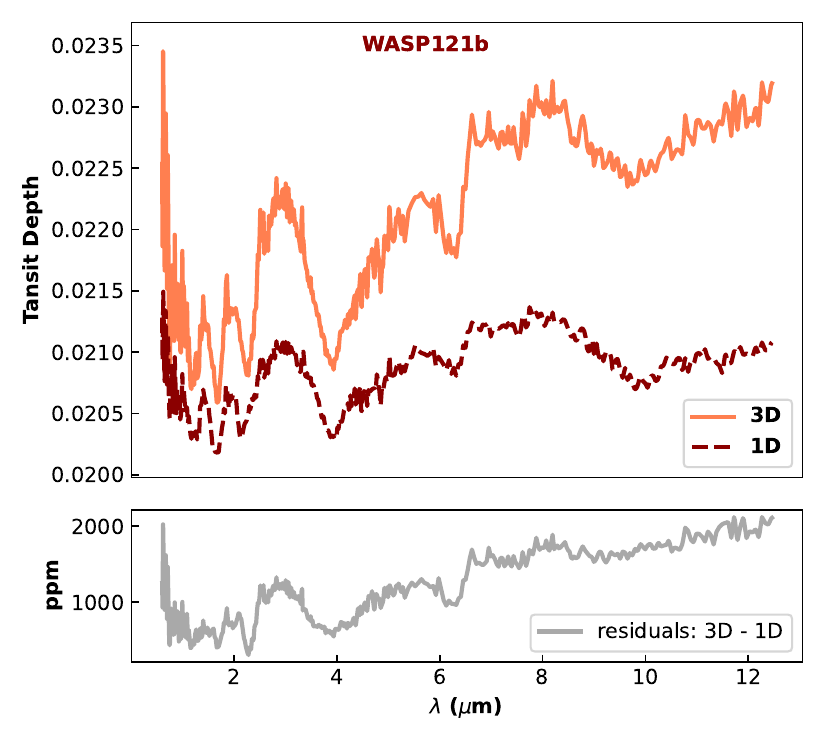}
\includegraphics[scale=\spec,trim = 0cm 0cm 0cm 0cm, clip]{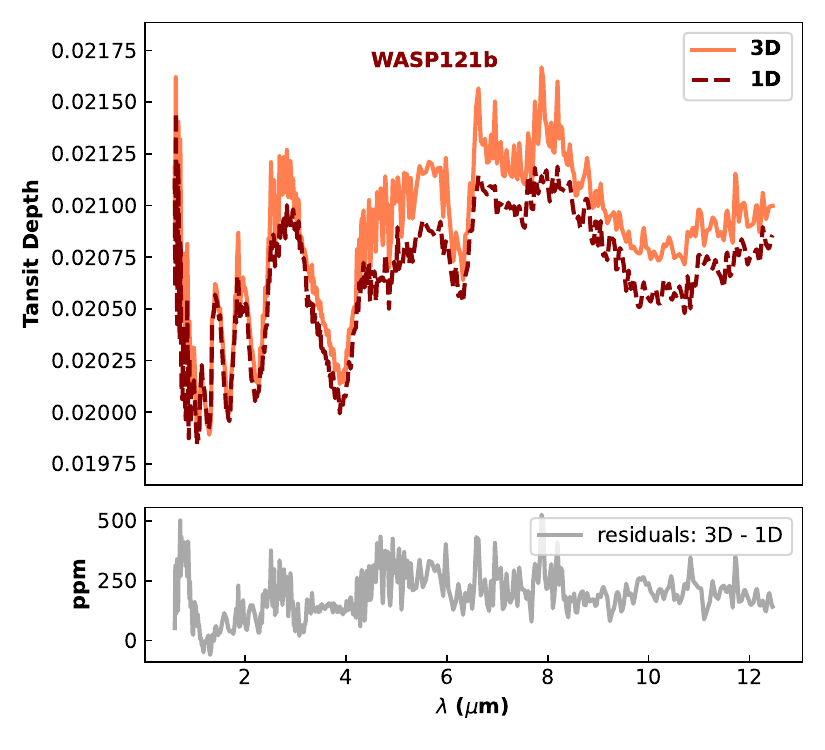}
\\
\includegraphics[scale=\spec,trim = 0cm 0cm 0cm 0cm, clip]{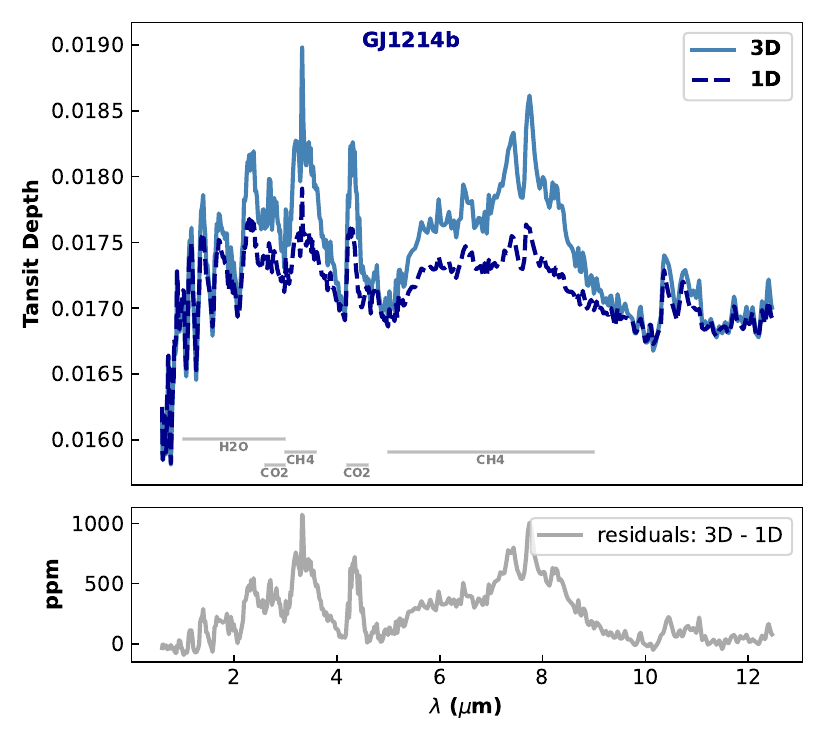}
\includegraphics[scale=\spec,trim = 0cm 0cm 0cm 0cm, clip]{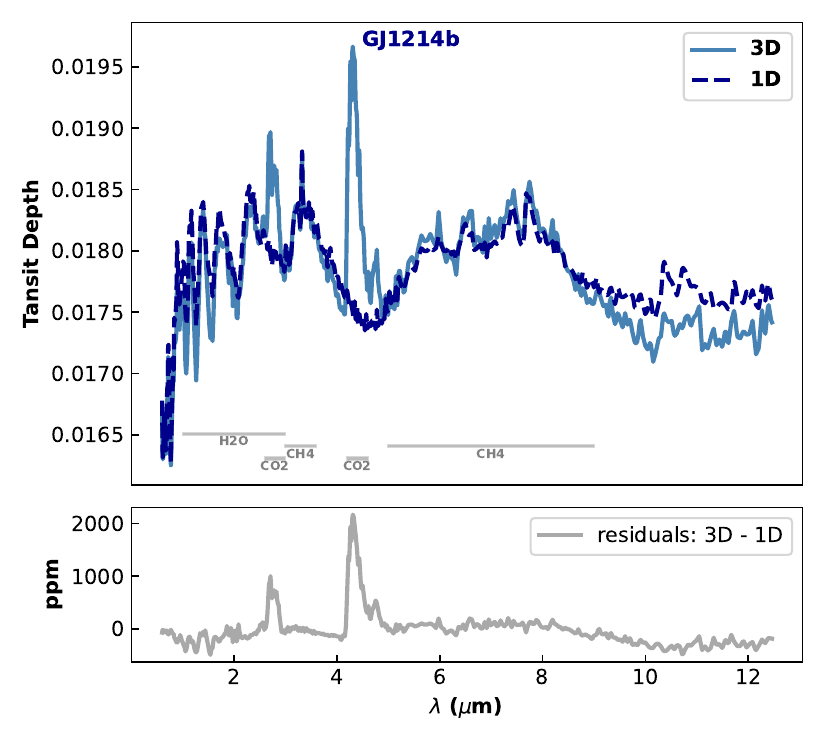}

\caption{Transmission spectra simulated with \textit{Pytmosph3R} \citep{Falco2022} for HD~189733~b (\textbf{top}), WASP-121~b (\textbf{middle}) and GJ~1214~b (\textbf{bottom}). Each panel compares two transmission spectra based on a 1D and a 3D atmosphere, respectively in dashed and solid lines. \textbf{Left panels:} constant input chemistry. \textbf{Right:} equilibrium input chemistry. The differences between the 3D and the 1D spectra are plotted below each panel in grey.}
\label{fig: spectra_1D3D}
\end{figure*}

\paragraph{\textbf{Constant chemistry}}
We will first focus on the simulated spectra assuming constant chemistry simulations which correspond to the left panels of Figure \ref{fig: spectra_1D3D}.

We consider in this case that the abundances are constant everywhere in the atmosphere no matter the temperature and pressure conditions. Thanks to this assumption, the effects on the spectra are only due to the thermal structure of the atmosphere.
For our coldest case GJ~1214~b, we see strong differences comparing 1D and 3D transmission spectra. Indeed, the 3D GCM model of GJ~1214~b \citep{charnay2015} shows large day-night asymmetries with an extended hot day side combined with a 30$^\circ$ eastward shifted hot spot. The metallicity of 100 is likely to result in a higher day-night temperature contrast compared to a solar metallicity, as the chemical composition of the atmosphere can have a significant impact on the efficiency of day-night energy redistribution \citep{Kataria_2014}. Due to this shift, the light coming from the star thus probes through the hot day side with a larger scale height compared to the limb (which is used to compute the 1D temperature profile of the pseudo-1D model). This implies hundreds of ppm differences in the transit depth in particular in the major absorbers, such as carbon dioxide and methane. However, on the water-dominated bands in the far infrared and in the visible, we see very small differences between 1D and 3D transmission spectra, in the order of 50 ppm. The altitude where the atmosphere is opaque on these bands is indeed deep in the troposphere (around 100 mbar). Thus, even when we take into account the inflated day side in 3D, the effective radius observed is very similar to that in 1D, because the region where the atmosphere is opaque remains at the same place at the limb. 
Therefore, the impact of the 3D structure of the atmosphere will depend on the wavelength and the composition.

Similar results are shown concerning the hottest study case WASP-121~b. 3D GCM models show that for highly irradiated atmospheres in tidal locking, the radiative timescale becomes substantially smaller than the dynamic timescale implying that almost no heat is transported from the day to the night side. It results in an extremely large day-side scale height compared to the night side because of the large day-night temperature contrast. Indeed, \cite{keating_2019} showed that regardless of the day-side temperature, the night-side temperature of short-period gas giants is relatively uniform, around $\sim1100$ K. This very inflated day side is thus mainly probed during the transit which explains why 3D transmission spectra are by thousands of ppm larger than in 1D. Unlike the previous case GJ~1214~b, all absorption bands are shifted. The atmosphere is so inflated for this ultra-hot Jupiter that the altitude at which the atmosphere is opaque is much higher due to the much greater scale height.

Interestingly, the results are very different in our intermediate case HD~189733~b. Here, despite a strong day-night temperature contrast, we observe less than a 50 ppm difference between 1D and 3D transmission spectra. This means that the West limb represents well the observable and that such atmospheres are more homogeneous than colder or hotter atmospheres. This is due to the high surface gravity of the planet, which mitigates the scale height differences between the different sides of the planet. Thus, using only the limb is a fair representation of the observable.

These 3D effects are independent of the differences between the east-west limbs, even though they are both related to temperature variations. Figure \ref{fig: spectra_WestEast} shows the transmission spectra of the two limbs independently compared to the whole spectra. This shows that we have similar features but with a different scale height, since the western limb is cooler than the eastern limb (see Figure \ref{fig: TP_JWST_3D}). The differences are of the order of 100 to 600 ppm. This is also observed by \citet{espinoza2021}, who shows that these differences can be observed by the JWST.

\paragraph{\textbf{Equilibrium chemistry}}
We now look at the equilibrium chemistry simulations which are shown in the right panels of Figure \ref{fig: spectra_1D3D}. Now, the simulated spectra are impacted by both the effects of the chemical and the thermal heterogeneities which is more realistic. 

For GJ~1214~b, the global picture has completely changed compared to constant chemistry. The largest difference (1000 to 2000 ppm) concerns \coo bands at 2.5 and 4.5 $\mu$m. The temperature profile of the 1D model is not hot enough to obtain abundant \coo at the equilibrium state, whereas the day side of the planet in the 3D model reaches a temperature where \coo is abundant enough to show strong signatures. Furthermore, as explained above, we probe a non-negligible part of the planet's day-side, hence the presence of broad bands of CO$_2$.
We also see differences in the water bands (between 1 and 2 $\mu$m) which weren't present in the constant chemistry model. The reason for these differences is the longitudinal variation in water abundance, which is lower on the day side of the planet. Light from the star then probes deeper regions corresponding to lower transit depth. Between 5 and 9 microns, as well as around 3.5 microns, in the region of the methane bands, the 1D and 3D spectra show fewer discrepancies than the rest of the spectra. Indeed, looking at the methane abundances in Figure \ref{fig: spe_ARIEL_3D_eq_GJ1214b}, we see that its abundance is drastically reduced in the day side above 100 mbar which is deeper than where we probe. It results that in these bands, we are not affected by the hot day side and we are mainly probing at the limb which is equivalent to the 1D spectrum.

For WASP-121~b, the 1D and 3D spectra show few differences in the whole wavelength range using equilibrium chemistry compared to constant chemistry. As shown in Figure \ref{fig: spe_ARIEL_3D_eq_WASP121b}, the abundances of almost every species drastically diminished on the day side, mainly due to thermal dissociation \citep{parmentier2018}, with the exception of carbon monoxide which is only divided by two due to its dilution in an H-dominated day side instead of a \hh-dominated atmosphere.
This implies that, on the water bands, the spectrum is not affected by the day side of the atmosphere because water is almost not present there. That's why a 1D model manages to fit the spectrum as shown in \cite{pluriel2020}. However, we can see in the residuals of Figure \ref{fig: spectra_1D3D} that in some bands, the fit is clearly less good than in the other bands. This is particularly true for the \co bands around 2.5 and 4.5 microns, as well as for the \tio and \vo bands in the visible. As we explained, \co is present on the day side of the atmosphere, where extreme temperatures induce a scale height far greater than the scale height of the limbs. Consequently, the 1D model does not represent this behavior well, resulting in a large difference (around 300-400 ppm) at these regions of the spectrum. 

The differences between 1D and 3D for HD~189733~b with equilibrium chemistry are very low, as with constant chemistry. Indeed, we see in Figure \ref{fig: spe_ARIEL_3D_eq_HD189733b} that the main species such as \hho, \chhhh, \coo, and \nhhh do not display strong longitudinal variations. In addition, we have shown before that due to similar scale height, there is no significant impact comparing 1D and 3D spectra with constant chemistry. As a consequence, a 1D model at the limb is equivalent to the 3D model meaning that for such warm atmospheres, we are probing near the limb. 

These 3D effects are independent of the differences between the east-west limbs, as explain for the constant chemistry.

\subsection{Cloud effect} \label{cloud}

Even if we have good reasons to think that cloud decks are present in many exoplanets \citep{parmentier2016, tsiaras2018}, each atmospheric model used in this study, 1D or 3D, for the three planets, is cloudless. 
It would be interesting to add clouds in the simulations, in particular, because they affect the short wavelengths observed by JWST and Ariel. However, the aim of this study is to see the impact of chemical and thermal 3D effects on the transmission spectra and how to deal with these 3D effects in the context of atmospheric characterization using 1D retrieval models. For this reason, we chose to not over-complicate our models. Clouds would require a dedicated paper. Clouds are nevertheless part of the retrieval parameters in the \textit{TauREx} model. This can break up possible degeneracies \citep{Pluriel2022,Changeat2020b}, and we verified that the model works correctly by not retrieving a cloud layer when we knew that none was implemented. For each retrieval performed, the Bayes factor has always privileged a cloudless model, and in the retrievals assuming clouds, the cloud deck was always pushed near to the surface pressure thus without impact on the spectra.
We thus decided not to present retrievals including clouds in this paper as they bring no more information compared to cloudless retrievals. 

\subsection{What if atmospheres are 1D?} \label{1datmo}

We used the theoretical 1D atmosphere (see Section \ref{GCM} for construction) to check the correct behavior of the 1D retrieval code. We remind that the 3 planets have been chosen to study the retrieval biases depending on the effective temperature of the planets, from warm Neptune to ultra-hot Jupiter. Also, the chemical construction with a constant profile or equilibrium chemistry has been considered to unravel the retrieval biases from temperature (constant chemistry) and chemistry (equilibrium chemistry). To do so, Free retrievals (constant chemistry) and equilibrium retrievals are both performed for each configuration (more details Section \ref{retproc}).

\begin{figure*}[h!]
\centering
\includegraphics[scale=\spec,trim = 0cm 0cm 0cm 0cm, clip]{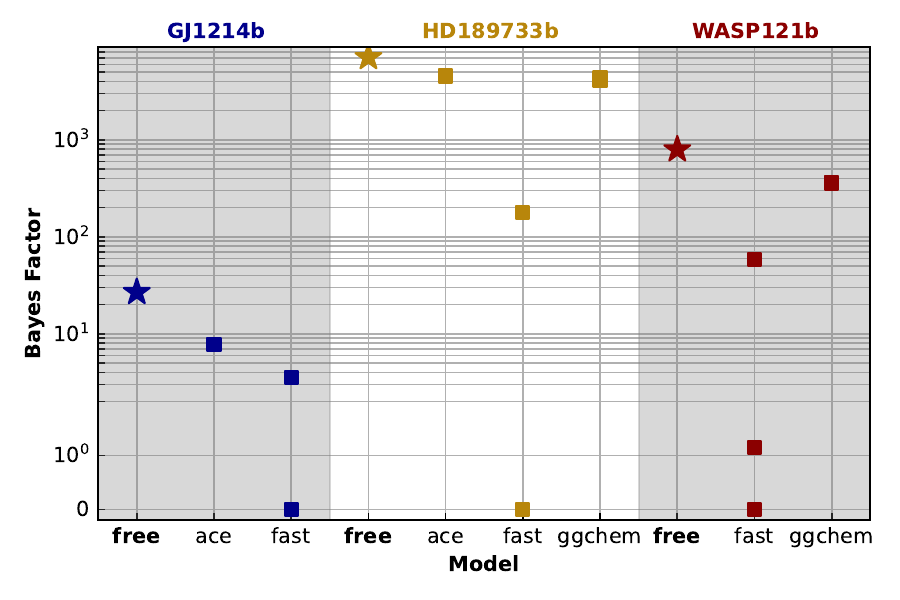}
\includegraphics[scale=\spec,trim = 0cm 0cm 0cm 0cm, clip]{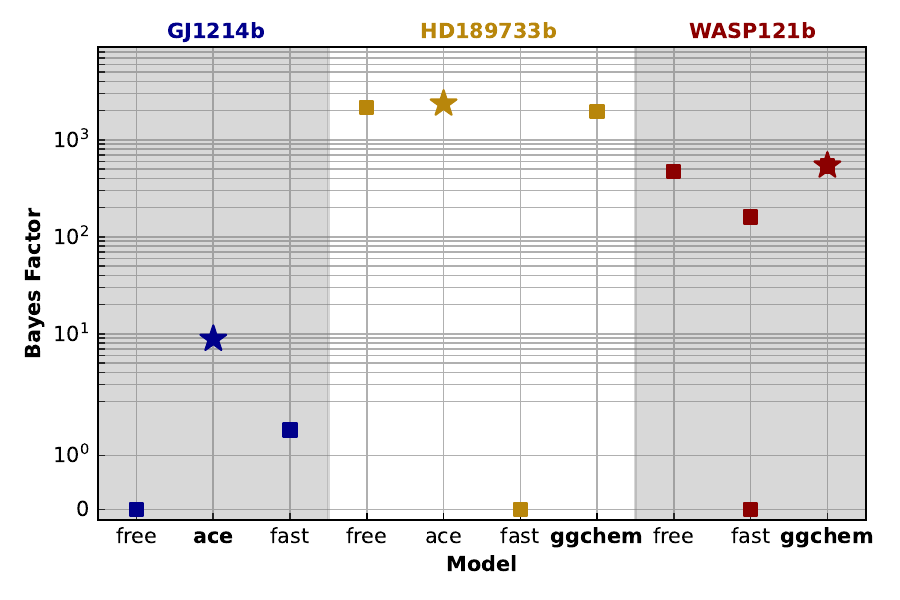}
\\
\includegraphics[scale=\spec,trim = 0cm 0cm 0cm 0cm, clip]{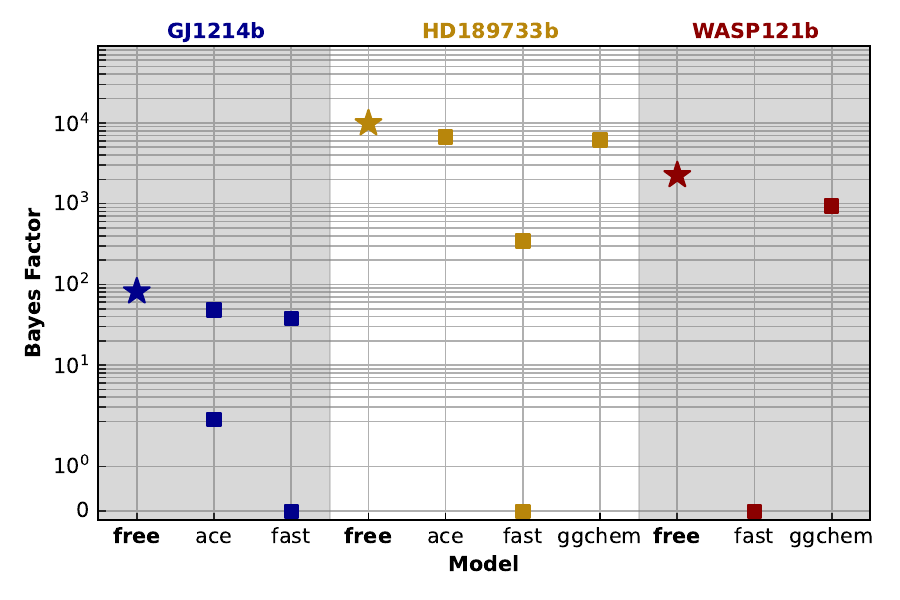}
\includegraphics[scale=\spec,trim = 0cm 0cm 0cm 0cm, clip]{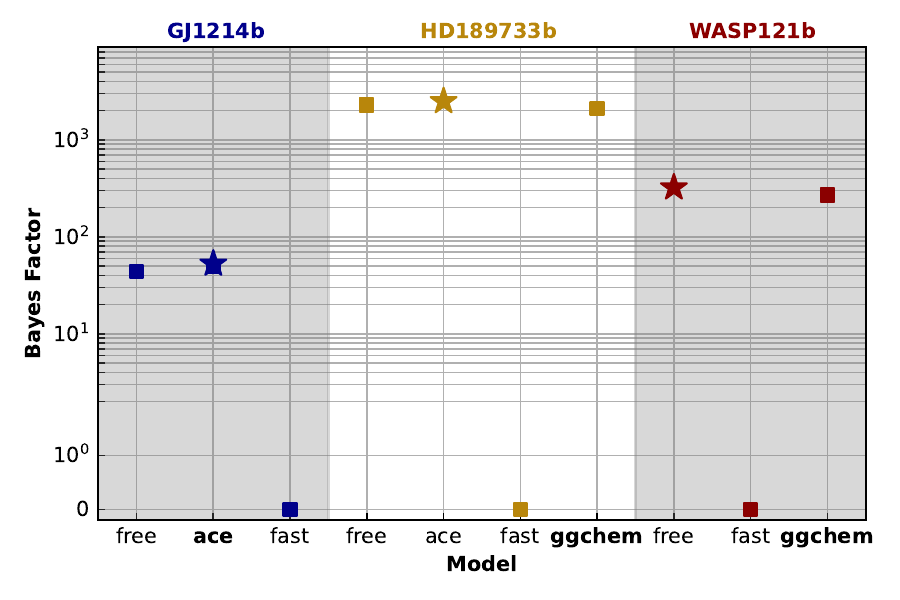}
\\
\includegraphics[scale=\spec,trim = 0cm 0cm 0cm 0cm, clip]{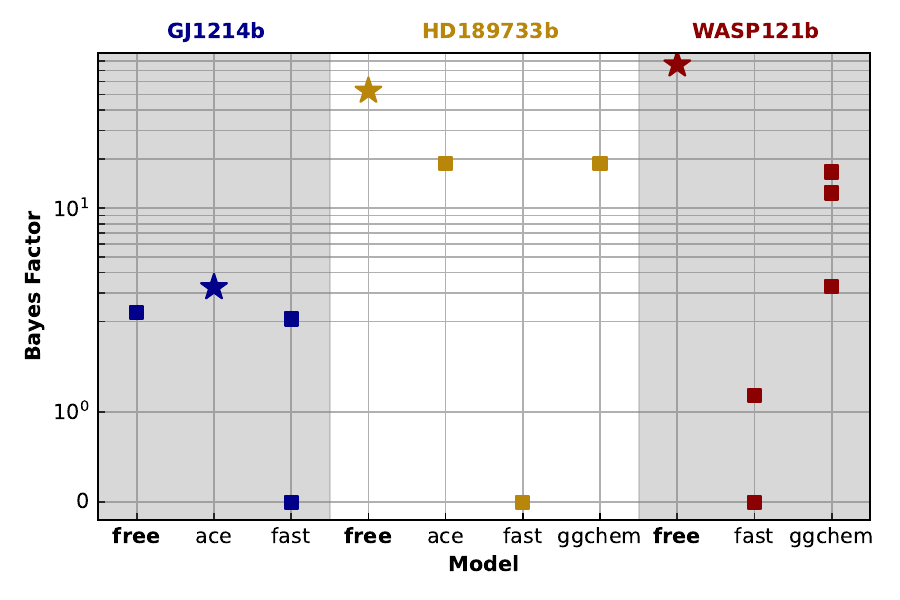}
\includegraphics[scale=\spec,trim = 0cm 0cm 0cm 0cm, clip]{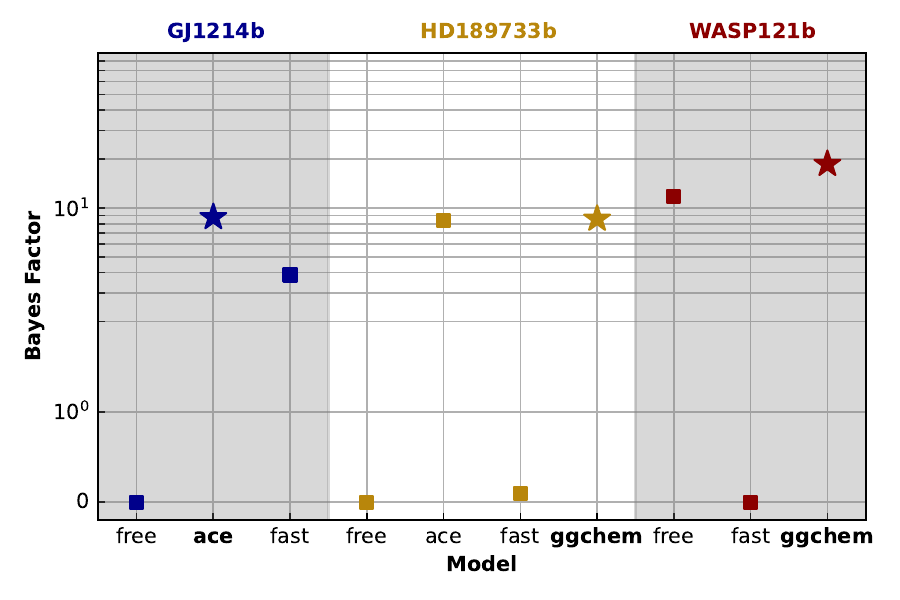}

\caption{Bayes factors for each retrieval and all their solutions\textbf{, see Section \ref{subsec: taurex}} (Free, ACE, FastChem and GGchem). By definition, we put at 0 the solution with the lowest Bayesian evidence as it is the reference of the comparison. \textbf{Left panels:} constant input chemistry. \textbf{Right:} equilibrium input chemistry. \textbf{Top and middle:} JWST simulated observation for atmospheric 1D and 3D assumption respectively. \textbf{Bottom:} Ariel simulated observation for a 3D atmospheric assumption. The star represents the highest Bayes factor. The expected best retrieval is highlighted in bold.}
\label{fig: logE_1D3D}
\end{figure*}

Figure \ref{fig: logE_1D3D} and Table \ref{tab:gj1214b_ret}, \ref{tab:hd189733b_ret}, \ref{tab:wasp121b_ret} shows that for all configurations the best retrieval is consistent with the input model configuration. This means that Free retrievals fit better input constant chemistry and equilibrium retrievals fit better input equilibrium chemistry. However, the best retrieval is not always more significant than the others and the equilibrium retrieval models are not all adapted for the different configurations. Figures \ref{fig: TP_JWST_1D}, \ref{fig: spe_JWST_1D_cst_GJ1214b}, \ref{fig: spe_JWST_1D_cst_HD189733b} and \ref{fig: spe_JWST_1D_cst_WASP121b} show that the temperature and species profiles are mostly not well retrieved below the probed altitudes (deeper than $\sim10^2$ Pa). This part of the atmosphere does not contribute to the features of the spectra, which explains why they are not well retrieved. However, the retrieved values could be well constrained while the input values is often outside the uncertainty. Thus, we cannot trust the retrieved profiles of the lower atmosphere.

\paragraph{\textbf{Temperate-Warm planet: GJ~1214~b}}
The best retrieval is consistent with the input model configuration and significantly better than the other models with \dlogE$\geq19$ for constant input chemistry and \dlogE$\geq7$ for equilibrium input chemistry (Figure \ref{fig: logE_1D3D} and Table \ref{tab:gj1214b_ret}).
\begin{itemize}
    \item Constant input chemistry: For lower pressure than $\sim10^2$ Pa (high altitudes), Free retrieval shows a significantly better retrieved temperature profile within 60 K compared to the input profile (Figure \ref{fig: TP_JWST_1D}). The main absorber \chhhh is retrieved by the Free retrieval at -0.11 dex while \hho, \coo and \nhhh giving secondary features are retrieved within -0.15 dex (Figure \ref{fig: spe_JWST_1D_cst_GJ1214b}). This allows the best model to be significantly in agreement with the input model even if not all input profiles are included in the uncertainty.
    \item Equilibrium input chemistry: For lower pressure than $\sim10^2$ Pa, all retrievals give the same temperature profile within 90 K of the input model (Figure \ref{fig: TP_JWST_1D}). All retrievals also give the same main absorber profiles (\chhhh and \hho) below $\sim10^2$ Pa, where the input equilibrium chemistry is constant at these temperature and pressure conditions (Figure \ref{fig: spe_JWST_1D_eq_GJ1214b}). However, ACE is better at retrieving \nhhh profile, which varies by 4 orders of magnitude and contributes at 3 $\mu m$, which makes ACE the best model in agreement with input ACE chemical composition. We can highlight that Free and ACE retrievals prefer to retrieve at least +0.31 dex \chhhh and down to -0.17 dex \hho than input values to better fit the spectrum, with input values out of the uncertainty. FastChem displays some discrepancies with ACE, as pointed out in \cite{Al-Refaie2022} when using an isothermal profile. ACE fits better than FastChem which is consistent with input ACE composition. However, none of the equilibrium models can reproduce the input composition. The metallicity is retrieved by ACE within 19\% and the C/O ratio only at 96\% deviation to the input value. While FastChem, which is not considered the best retrieval, has a closer value to input for C/O ratio (76\%) and metallicity (9\%) (see Table \ref{tab:gj1214b_ret} and Figure \ref{fig: corner_JWST_1D_eq_GJ1214b}). It shows that even at these "low" temperatures we already have difficulties to perfectly retrieve the input model.
\end{itemize}

To summarize for GJ~1214~b: the retrievals work well on the temperature, but not without some biases on the chemistry. Constant chemistry and 1D temperature profiles are retrieved by Free retrieval within -0.15 dex. However, even at such "low" temperatures, equilibrium input chemistry is best retrieved by equilibrium models because of species with strong vertical variations (here \nhhh). In addition, \textit{TauREx} not only has difficulty in accurately retrieving the C/O ratio and the metallicity (respectively 96\% and 19\% deviation for the best model ACE), but also the values closest to the input are not retrieved by the more significant model ACE, despite the fact that the best retrieval ACE is consistent with the input ACE equilibrium chemistry.

\paragraph{\textbf{Warm-hot planet: HD~189733~b}}
The best retrieval is consistent with the input model configuration and significantly better than the other models with \dlogE$\geq2536$ for constant input chemistry and \dlogE$\geq200$ for equilibrium input chemistry (Figure \ref{fig: logE_1D3D} and Table \ref{tab:hd189733b_ret}).
\begin{itemize}
    \item Constant input chemistry: For lower pressure than $\sim10^2$ Pa (high altitudes), Free retrieval shows a significantly better fit of the temperature profile within 150 K compared to the input profile (Figure \ref{fig: TP_JWST_1D}). However, the abundance of the main absorbers \chhhh, \hho, \nhhh, \tio, and \vo, are overestimated by +0.23 to +0.36 dex. This remains a better fit compared to equilibrium chemistry retrievals that show strong vertical variation in these temperature-pressure conditions (Figure \ref{fig: spe_JWST_1D_cst_HD189733b}).
    \item Equilibrium input chemistry: For lower pressure than $\sim10^2$ Pa, Free and FastChem retrievals overestimate the temperature profile by more than 250 K while ACE and GGchem stay mostly within 250 K deviation from the input values. However, these two chemical models underestimate the temperature in the upper atmosphere by more than 500 K (Figure \ref{fig: TP_JWST_1D}). In this peculiar configuration, with warm-hot temperature, molecules such as \tio, \vo, and K are close to their condensation temperature. Compare to the isotherm configuration of \cite{Al-Refaie2022}, condensation occurs at mid and high altitudes where the temperature is lower than in the deep atmosphere. Thus, a strong bias occurs with FastChem where it considers these molecules but not their condensates. Therefore, FastChem cannot retrieve properly this configuration where \tio, \vo, and K have strong features in the visible while they should not be because of their condensation, Figure \ref{fig: TP_JWST_1D} shows a wrong temperature profile, Table \ref{tab:hd189733b_ret} shows wrong retrieved parameters and Figure \ref{fig: spe_JWST_1D_eq_HD189733b} shows wrong species profiles. Thus, GGchem, which considers condensation of these species, should in theory solve this issue but it surprisingly does not. It shows a better fit for the temperature profile and species abundances with regards to FastChem. But still, it does not manage to give a consistent abundance of K, overestimated by a factor of 2, and shows a strong feature in the visible which bias all the spectra, since all chemical abundances are correlated. This is explained by the discrepancy between the K abundance calculated by GGchem and the input abundance calculated by \cite{parmentier2016}. We are around 1000 K, at the K condensation limit. This implies a strong variation in K abundance. \cite{Woitke2018} (GGchem) and \cite{parmentier2016,parmentier2018} use different approximations and assumptions for condensation, which leads to uncertainty, more important at the condensation limit. \cite{Al-Refaie2022} found a good agreement between ACE, FastChem, and GGchem because the same molecules are considered in the three models. However, we show here chemical discrepancies between the two models, illustrating that using imperfect chemical models with regards to what is actually observed leads to biased interpretations. ACE, which considers only C, H, O, and N atoms, ends up giving the best-fit model by getting rid of \tio, \vo, and K condensation issues. The retrieved thermal profile is similar to the one retrieved by GGchem but the main absorber \hho is now perfectly constrained under 0.01 dex (see Figure \ref{fig: spe_JWST_1D_eq_HD189733b}).
    All these biases can be seen in Table \ref{tab:hd189733b_ret} and Figure \ref{fig: corner_JWST_1D_eq_HD189733b} on the C/O ratio and the metallicity. FastChem model is far from input values while ACE has a 24\% deviation. Furthermore, even if the Free is not the more significant model, it is closest to input values.
\end{itemize}

To summarize for HD~189733~b: the retrievals work well on the temperature for constant input chemistry (within 150 K) but for equilibrium input chemistry they begin to show difficulties to retrieve the top of the atmosphere (underestimated by 500 K), as well as the bottom of the atmosphere. Only the pressures corresponding to the highest atmospheric contribution (between $\sim10^2$ and $\sim10^0$ Pa) are well retrieved (within 250 K). In addition, we encounter even more chemical abundance bias compared to GJ~1214~b. Constant chemistry profiles are best retrieved by the Free retrieval model (which is consistent) with an overestimation of +0.23 to +0.36 dex. Due to the strong vertical variation, the equilibrium configuration is better retrieved by an equilibrium model (ACE), which is not the expected one (GGchem), but which fits very well the main absorber \hho under 0.01 dex. However, the study performed on this planet shows all the limitations of the 3 chemical equilibrium models for retrievals. At this temperature, the role of condensates (such as \tio, \vo, and K here) is essential with their feature in the visible. The discrepancies between the K chemical modeling of GGchem and the K chemical modeling of \cite{parmentier2016}, bias GGchem retrieval. This is a good example of what can be encountered when fitting observations with a model that does not perfectly reproduce the true chemistry. The C/O ratio and the metallicity are well retrieved by ACE (best model) but not as well as Free. Our results show that we can retrieve values closest to input with a model statistically not favored.

\paragraph{\textbf{Ultra hot planet: WASP-121~b}}
The best retrieval is consistent with the input model configuration and significantly better than the other models with \dlogE$\geq434$ for constant input chemistry and \dlogE$\geq71$ for equilibrium input chemistry (see Figure \ref{fig: logE_1D3D} and Table \ref{tab:wasp121b_ret}). We observe a higher overall uncertainty in the retrieved temperature and chemical species profiles compared to the other two planets. The very hot temperature coupled with a sharp day-to-night gradient brings a complexity that is more difficult to retrieve with a simpler model.

\begin{itemize}
    \item Constant input chemistry: in the upper atmosphere, the retrieved thermal profiles are either overestimated or underestimated by $\sim1000$ K (Figure \ref{fig: TP_JWST_1D}). This is far from the input temperature profile ($\sim$ 50\% deviation). However, close to a pressure corresponding to the highest atmospheric contribution, between $\sim10^2$ and $\sim10^0$ Pa, retrieved profiles remain within 500 K (less than 25\% deviation), while the best model (Free) is not significantly better. The abundance of the main absorbers \hho, \chhhh and \tio, \feh in the visible, are overestimated at +0.22 to +0.29 dex for the Free retrieval (best model). Despite these high values, this is still a better fit than equilibrium chemistry. The strong vertical variation in equilibrium chemistry under these temperature and pressure conditions cannot match the constant input chemistry (see Figure \ref{fig: spe_JWST_1D_cst_WASP121b}).
    \item Equilibrium input chemistry: we observe here the same behavior as constant input chemistry for the retrieved temperature profiles (see Figure \ref{fig: TP_JWST_1D}). There is less than a 10\% deviation between $\sim10^2$ and $\sim10^0$ Pa for the best retrieval GGchem, but the retrieved profiles are mostly erroneous  outside this range (reaching more than 50\% deviation). Considering the strong temperature gradient between $\sim10^2$ and $\sim10^0$ Pa, we observe the same behavior as HD~189733~b regarding the condensation of species such as \tio, \vo, and K. This makes GGchem a better model than FastChem in this case, as shown in Figures \ref{fig: logE_1D3D}, \ref{fig: spe_JWST_1D_eq_WASP121b}. Although GGchem gives the best fit, none of the retrieval models fit all the main absorbers \hho, \tio, \coo, \co, and \vo better than the others. The Free retrieval is better for \hho and \coo, while GGchem is better for the other absorbers (despite a strong deviation for both models, reaching +0.50 dex). In addition, the best model, GGchem, retrieves an erroneous C/O ratio with a deviation of 80\% and an erroneous metallicity with a deviation of 530\%. In contrast, the Free model ends up giving values close to the input within 13\% (see Table \ref{tab:wasp121b_ret} and Figure \ref{fig: corner_JWST_1D_eq_WASP121b}). Thus, all the models retrieved different parts of the input, but none of them obtained the entire structure.
\end{itemize}

To summarize for WASP-121~b: the retrievals perform poorly outside the pressures corresponding to the highest atmospheric contribution (between $\sim10^2$ and $\sim10^0$ Pa). Due to strong large-scale vertical variation in temperature and species, 1D temperature profiles for constant and equilibrium chemistry are best retrieved at only 25\% between $\sim10^2$ and $\sim10^0$ Pa. Constant input chemistry is best retrieved by Free retrieval, overestimated by +0.22 to +0.29 dex, and equilibrium input chemistry is best retrieved by GGchem retrieval, but with a deviation reaching $\sim$+0.50 dex. The retrieval models are consistent with the input configurations, but not without bias on absolute values. GGchem retrieves the wrong C/O ratio and metallicity, while the Free model is within 13\%. The strong vertical gradient on a large scale brings a complexity that is difficult to retrieve correctly with a simpler model, which also translates into greater uncertainty in the retrievals.

\paragraph{\textbf{Summary}}
The theoretical 1D analysis validates the consistency of the 1D retrievals, but not without a few biases. This shows that below the probed altitudes, in addition to the retrieved parameters, we cannot trust the uncertainties given by the models, which are largely underestimated. This is also the case at the top of the atmosphere, as we move towards warmer planets.

Furthermore, equilibrium chemical models are not equivalent and give significantly different results:

\begin{itemize}
    \item ACE cannot retrieve ultra-hot planets whereas it might be a better approximation for cooler planets.
    \item FastChem is biased towards warm, even hot planets, where species are close to or below condensation temperature. This is never the best retrieval model.
    \item GGchem, which should be a complete model, will still be in competition with a simplified model for cold and warm planets. Apparently the best option for hot planets, but Free retrieval can give better retrieved values.
\end{itemize}

We need to be careful with equilibrium models. Our study shows that if one part of the chemistry modeling is wrong, all the chemical abundances will be biased since everything is correlated. This agrees with the conclusion of \cite{Al-Refaie2022} using an isotherm configuration.

\subsection{When atmospheres are 3D} \label{3datmo}

\subsubsection{Constant chemistry} \label{3datmocst}

Given the 3D thermal structure, we used theoretical atmospheric models with constant chemistry (see section \ref{section: pytmosph3r}) to disentangle the 3D effects of temperature without being biased by chemistry. Section \ref{1datmo} confirms the overall correct behavior of the retrieval code or highlights any biases we may encounter. So, we can be confident in this approach to focus on thermal 3D effects. To this end, Free retrievals (constant-with-depth chemistry) and equilibrium retrievals are performed for each configuration (more details in section \ref{retproc}). The aim is also to compare the biases of the JWST and Ariel instruments. \\

Figure \ref{fig: logE_1D3D} shows that the Free retrieval finds the best solution compared to equilibrium chemistry (except for GJ~1214~b in the Ariel configuration). This is consistent with the input constant chemistry. For GJ~1214~b in the Ariel configuration, ACE retrieval has a better Bayes factor but within a range of 1.5 variation compared to the others. Therefore, all models are statistically equivalent and none is preferred. We observed a stronger deviation of the Bayes factor for the JWST configuration compared to the Ariel one. This could simply be due to the higher resolution on a larger wavelength range for the JWST configuration or this may be linked to some particular wavelength band such as the lack of data points in the visible for the Ariel configuration where strong features for hot planets (such as TiO, VO, and K) are located.

\paragraph{\textbf{Temperate-Warm planet: GJ~1214~b}}
All retrieval models for the Ariel spectrum are within a Bayes factor deviation of 1.5 which makes all models equivalent (see Table \ref{tab:gj1214b_ret}). However, the JWST spectrum is better retrieved by the Free retrieval with \dlogE$=33$ compared to the second-best model. The higher resolution of the JWST spectrum, compared to the Ariel spectrum, gives more constraints on \hho and \chhhh features at low wavelength (between 1 and 2 $\mu$m), but probably also at higher wavelength (above 4 $\mu$m). Yet, this does not translate to closer-to-truth retrieved profiles, although the retrieved uncertainties are smaller. Figure \ref{fig: TP_JWST_3D} and \ref{fig: TP_ARIEL_3D} show consistent inputs and retrieved temperature profiles for GJ~1214~b. At low altitudes (deeper than $\sim$10$^2$ Pa) there is high uncertainty because these altitudes are not probed. Higher in the atmosphere (probed altitudes), the temperature day-night variation is within 300K and all models retrieve the limb profiles. \hho and \chhhh are responsible for the main features of the spectra. Both retrieved values of the Free retrieval are around -0.10 dex lower than the input. For the JWST spectrum, the uncertainty is lower but the input value is not within the uncertainty. While, for the Ariel spectrum, the \chhhh input value is within the uncertainty. Thus, even with more data point, the retrieved chemical abundances is not closer to the input and the uncertainties cannot be trusted.

\paragraph{\textbf{Warm-hot planet: HD~189733~b}}
The best retrieval (Free) is consistent with the input model configuration and significantly better than the other models with \dlogE$\geq3048$ for the JWST spectrum and \dlogE$\geq34$ for the Ariel spectrum (see Figure \ref{fig: logE_1D3D} and Table \ref{tab:hd189733b_ret}). Discrepancies between retrieval models are the same as those explained in Section \ref{1datmo}. Figure \ref{fig: TP_JWST_3D} and \ref{fig: TP_ARIEL_3D} show that the biases on the temperature profiles are the same as in Section \ref{1datmo}. Temperatures retrieved below $\sim10^2$ Pa correspond mainly to those of the limb. For the Ariel spectrum, the temperature is slightly warmer, but the solution for the JWST spectrum remains within the Ariel uncertainties, which are significantly higher. Equilibrium chemistry cannot reproduce the constant input chemistry, while Free retrieval gives consistent results but not without significant deviation (between +0.18 to +0.35 dex). Even with a variation of less than 500 K between day and night side, only temperature variation can bias the retrieval of species abundances. While \vo is largely retrieved in the JWST spectrum, it is not in the Ariel spectrum. Retrieved uncertainties on the Ariel spectrum are larger than on the JWST spectrum. However (and to the exception of \vo) the same molecules are present with both observatories. The lack of abundant \vo with Ariel is due to the coarser spectral resolution in visible light which is still sufficient to detect \tio.

\paragraph{\textbf{Ultra hot planet: WASP-121~b}}
The best retrieval (Free) is consistent with the input model configuration and significantly better than the other models with \dlogE$\geq1310$ for the JWST spectrum and \dlogE$\geq60$ for the Ariel spectrum (see Figure \ref{fig: logE_1D3D} and Table \ref{tab:wasp121b_ret}). Discrepancies between retrieval models are the same as those explained in Section \ref{1datmo}. Figure \ref{fig: TP_JWST_3D} and \ref{fig: TP_ARIEL_3D} show that the Free retrieval (best model) finds a temperature at the top of the atmosphere higher than the input model, and a temperature at the bottom of the atmosphere lower than the input model. The temperature transition occurs in the atmosphere where species absorption contributes the most, around $\sim10^2$ Pa. The temperature gradient is steep, crossing all possible temperatures from day to night side. The chemical abundances of the main absorbers \hho, \tio, \co are retrieved between -0.25 and -0.67 dex for the JWST spectrum and between -0.28 and -0.77 dex for the Ariel spectrum, with the exception of \tio with a deviation of +0.03 dex from the input value. However, \tio feature does not match the input spectrum for both JWST and Ariel configurations. The retrieved spectra are outside the uncertainties, by several sigma. The much higher JWST resolution in the visible compared with Ariel surprisingly does not provide better constraints on TiO. While we already encounter difficulties in retrieving input values with a 1D atmosphere, the huge temperature gradient between day and night brings even more biases. The end result is a temperature from both the day and night sides that does not allow the retrieval models to find input spectra and species profiles. There is not even thermal inversion retrieved by the best model.

\paragraph{\textbf{Summary}}
The conclusion in Section \ref{1datmo} remains the same in 3D. Secondly, using for retrievals a 4-point temperature profile gives good results for the cooler planets but not for hotter ones, which need at least 2D retrievals, as has already been pointed out in more detail by \cite{Pluriel2022}. The higher resolution of the JWST spectrum, particularly in the visible, reduces uncertainties but does not provide a better fit. In addition, the input values will be out of uncertainty, making them unreliable. This is probably due not only to the high resolution, but also to the good input signal-to-noise ratio, which can be improved by binning down the spectra. Thus, lower resolution would still result in extremely small error bars and over-confident retrieval results. The lower atmosphere is still poorly retrieved, especially as we move towards hotter planets. Nevertheless, it is still possible to find the presence of the input species. Finally, the low resolution in the visible wavelength range of the Ariel spectrum has missed the presence of the visible absorber \vo but never \tio with a retrieval similar to that of the JWST spectrum.

\subsubsection{Equilibrium chemistry}

Here the 3D thermal structure, as well as equilibrium chemistry, are considered as input. Considering the conclusion of Sections \ref{1datmo} and \ref{3datmocst}, this will highlight biases due to the variability of chemical abundances in the atmospheres of warm to ultra-hot planets. Section \ref{3datmocst} has already shown the temperature biases from the 3D structure. As previously done, Free retrievals (constant-with-depth chemistry) and equilibrium retrievals are both performed for each configuration (more details in Section 3.3).

Figure \ref{fig: logE_1D3D} shows that the equilibrium retrievals always find the best solution compared to Free chemistry (except for the WASP-121~b JWST configuration). This is consistent with the input equilibrium chemistry. The retrieval models of the Ariel spectrum have less deviation from each other than the JWST spectrum. This is due to the lower spectral resolution across all wavelength bands, but particularly in the visible bands. Between visible and infrared the chemical species contributing to the spectral signatures are different, \tio, \vo, and K against \hho, \chhhh, \coo, \co, and \nhhh. If we try to retrieve both parts at the same time, both will be biased, as the signatures may come from different parts of the atmosphere, at different temperatures. While a larger spectral range could benefit to interpret more accurately the observations \citep{Benneke2013,welbanks2019a}, we show that depending on the chosen range and the model used this could bias the retrieval compared to input data.

\paragraph{\textbf{Temperate-Warm planet: GJ~1214~b}} 
The ACE model provides the best fit to both JWST and Ariel spectra with \dlogE$\geq10$ and \dlogE$\geq5$ respectively (see Table \ref{tab:gj1214b_ret}), which is consistent with the ACE input chemistry modeling. Figure \ref{fig: logE_1D3D} also shows that the JWST spectrum is secondly best retrieved by the Free model, while the Ariel spectrum is secondly best retrieved by the FastChem model. FastChem's retrievals poorly fit the CO and the visible bands, because of the chemical modeling differences with the input ACE chemistry modeling. Retrievals on the Ariel spectrum circumvent this issue thanks to its low spectral resolution in visible light. The main difference from the previous input configuration is that the equilibrium chemistry results in a strong dichotomy of \coo between the day and night side (see Figure \ref{fig: spe_ARIEL_3D_eq_GJ1214b}). As a result, the retrieved temperature profiles correspond to the day side (see Figure \ref{fig: TP_JWST_3D} and \ref{fig: TP_ARIEL_3D}). Table \ref{tab:gj1214b_ret} and Figure \ref{fig: corner_ARIEL_3D_eq_GJ1214b} shows that the temperature bias on the day side still keeps a good agreement on the C/O ratio and the metallicity retrieved, within 20\%.

\paragraph{\textbf{Warm-hot planet: HD~189733~b}}
The ACE model retrieves the JWST spectrum much better than the other models with \dlogE$\geq208$, as already explained in Sections \ref{1datmo}. However, the Ariel spectrum is equivalently retrieved by ACE and GGchem (Figure \ref{fig: logE_1D3D} and Table \ref{tab:hd189733b_ret}), again due to the lack of constraint in the visible wavelength bands where discrepancies between chemical models appear. ACE is as significant as GGchem, but Table \ref{tab:hd189733b_ret} and Figure \ref{fig: corner_ARIEL_3D_eq_HD189733b} show that GGchem better retrieves the C/O ratio and metallicity. Figure \ref{fig: TP_ARIEL_3D} shows that the temperature at the top of the atmosphere is unconstrained by the huge uncertainty. This part of the atmosphere, therefore, makes no significant contribution to the features of the spectra. In contrast, Figure \ref{fig: TP_JWST_3D} shows that increasing resolution adds an erroneous constraint on the temperature of the top of the atmosphere. Only the temperature around pressures corresponding to the highest atmospheric contribution (around $\sim10^2$ Pa), is consistent between equilibrium models and the input temperature profiles. The temperature retrieved at these pressures is that of the limb. Figure \ref{fig: spe_ARIEL_3D_eq_HD189733b} shows a good agreement between the retrieved species profiles.

\paragraph{\textbf{Ultra hot planet: WASP-121~b}}
The GGchem model retrieves the Ariel spectrum better than the other models with \dlogE$\geq7$, as already explained in Section \ref{1datmo}. However, the JWST spectrum is better retrieved by the Free model with \dlogE$\geq52$ (see Figure \ref{fig: logE_1D3D} and Table \ref{tab:wasp121b_ret}). Table \ref{tab:wasp121b_ret} and Figure \ref{fig: corner_ARIEL_3D_eq_WASP121b} show that GGchem retrieves for Ariel configuration the C/O ratio very well at 5\% but not the metallicity (75\% deviation), while for JWST configuration it is higher than 44\% considering all retrieval models. The models are not suited to the high spectral resolution of JWST, which imposes severe constraints on thermal contrast and hence on chemical distribution. This shows that such a contrasted atmosphere cannot be retrieved by a 1D model with correlated chemistry. However, the higher degree of freedom of the Free retrieval allows a better match. The temperature profiles retrieved between the Free model and the GGchem model are similar (within 500 K below $10^4$ Pa, see Figure \ref{fig: TP_JWST_3D} and \ref{fig: TP_ARIEL_3D}). The conclusions on temperature biases are the same as for HD~189733~b. But Figure \ref{fig: spe_ARIEL_3D_eq_WASP121b} shows that species abundances are more difficult to retrieve.

\paragraph{\textbf{Summary}}
In addition to the previous biases from Section \ref{1datmo} and \ref{3datmocst}, the biases coming from the chemistry show that, even on a warm planet, it would make sense to fit the different molecular features separately to disentangle the temperature variation that brings chemical variability. Otherwise, using a 1D retrieval model will bias all different spectral contributions. Furthermore, only the pressure where the contribution is highest should be considered as a significantly good retrieval of the observation. The rest should be treated with caution. All models remain good at detecting input molecules.

\begin{table*}[h!]
    \caption{Overview of the ability of the best retrievals to find the input values.*}
    \centering
    \resizebox{\textwidth}{!}{%
    \begin{threeparttable}
    \begin{tabular}{l|ccc|ccc|ccc|ccc|ccc|ccc|}
        \cline{2-19}
        \rule{0pt}{4ex} & \multicolumn{12}{c|}{JWST}    & \multicolumn{6}{c|}{Ariel} \\ [0.3cm] \cline{2-19}
        \rule{0pt}{4ex} & \multicolumn{6}{c|}{1D}       & \multicolumn{12}{c|}{3D}   \\ [0.3cm] \cline{2-19}
        \rule{0pt}{4ex} & \multicolumn{3}{c|}{constant} & \multicolumn{3}{c|}{equilibrium} & \multicolumn{3}{c|}{constant} & \multicolumn{3}{c|}{equilibrium} & \multicolumn{3}{c|}{constant} & \multicolumn{3}{c|}{equilibrium} \\ [0.3cm] \cline{2-19}
        \rule{0pt}{4ex} & GJ & HD & WASP & GJ & HD & WASP & GJ & HD& WASP & GJ & HD & WASP & GJ & HD & WASP & GJ & HD & WASP \\
        \hline
        \multicolumn{1}{|l|}{Species detection  } & \cellcolor{YellowGreen}\checkmark & \cellcolor{YellowGreen}\checkmark & \cellcolor{YellowGreen}\checkmark & \cellcolor{YellowGreen}\checkmark & \cellcolor{YellowGreen}\checkmark & \cellcolor{YellowGreen}\checkmark & \cellcolor{YellowGreen}\checkmark & \cellcolor{YellowGreen}\checkmark & \cellcolor{YellowGreen}\checkmark & \cellcolor{YellowGreen}\checkmark & \cellcolor{YellowGreen}\checkmark & \cellcolor{YellowGreen}\checkmark & \cellcolor{YellowGreen}\checkmark & \cellcolor{YellowGreen}\checkmark & \cellcolor{YellowGreen}\checkmark & \cellcolor{YellowGreen}\checkmark & \cellcolor{YellowGreen}\checkmark & \cellcolor{YellowGreen}\checkmark  \\
        \multicolumn{1}{|l|}{C/O}                                        &                                   &                                   &                               & \cellcolor{RedOrange}$\times$     & \cellcolor{Dandelion}$\sim$       & \cellcolor{RedOrange}$\times$     &                                   &                               &                               & \cellcolor{YellowGreen}\checkmark & \cellcolor{Dandelion}$\sim$   & \cellcolor{RedOrange}$\times$ &                                   &                               &                               & \cellcolor{YellowGreen}\checkmark & \cellcolor{YellowGreen}\checkmark & \cellcolor{YellowGreen}\checkmark \\
        \multicolumn{1}{|l|}{Metallicity (Z)}                            &                                   &                                   &                               & \cellcolor{YellowGreen}\checkmark & \cellcolor{YellowGreen}\checkmark & \cellcolor{RedOrange}$\times$     &                                   &                               &                               & \cellcolor{YellowGreen}\checkmark & \cellcolor{Dandelion}$\sim$   & \cellcolor{RedOrange}$\times$ &                                   &                               &                               & \cellcolor{YellowGreen}\checkmark & \cellcolor{YellowGreen}\checkmark & \cellcolor{RedOrange}$\times$     \\
        \hline
        \multicolumn{19}{c}{Top of the atmosphere [$\sim10^0$ Pa - $\sim10^{-4}$ Pa]} \\
        \hline
        \multicolumn{1}{|l|}{Temperature profile}                        & \cellcolor{YellowGreen}\checkmark & \cellcolor{YellowGreen}\checkmark & \cellcolor{RedOrange}$\times$ & \cellcolor{YellowGreen}\checkmark & \cellcolor{RedOrange}$\times$     & \cellcolor{RedOrange}$\times$     & \cellcolor{YellowGreen}Limb       & \cellcolor{YellowGreen}Night  & \cellcolor{RedOrange}$\times$ & \cellcolor{YellowGreen}Day        & \cellcolor{RedOrange}$\times$ & \cellcolor{YellowGreen}Night  & \cellcolor{YellowGreen}Night      & \cellcolor{YellowGreen}Day    & \cellcolor{RedOrange}$\times$ & \cellcolor{YellowGreen}Day        & \cellcolor{YellowGreen}Day        & \cellcolor{YellowGreen}Day        \\
        \multicolumn{1}{|l|}{Chemical profile}                           & \cellcolor{YellowGreen}\checkmark & \cellcolor{Dandelion}$\sim$       & \cellcolor{Dandelion}$\sim$   & \cellcolor{YellowGreen}\checkmark & \cellcolor{Dandelion}$\sim$       & \cellcolor{RedOrange}$\times$     & \cellcolor{YellowGreen}\checkmark & \cellcolor{Dandelion}$\sim$   & \cellcolor{Dandelion}$\sim$   & \cellcolor{YellowGreen}Day        & \cellcolor{RedOrange}$\times$ & \cellcolor{RedOrange}$\times$ & \cellcolor{YellowGreen}\checkmark & \cellcolor{Dandelion}$\sim$   & \cellcolor{Dandelion}$\sim$   & \cellcolor{YellowGreen}Day        & \cellcolor{YellowGreen}Day        & \cellcolor{RedOrange}$\times$     \\
        \hline
        \multicolumn{19}{c}{Middle of the atmosphere [$\sim10^2$ Pa - $\sim10^0$ Pa]} \\
        \hline
        \multicolumn{1}{|l|}{Temperature profile}                        & \cellcolor{YellowGreen}\checkmark & \cellcolor{YellowGreen}\checkmark & \cellcolor{Dandelion}$\sim$   & \cellcolor{YellowGreen}\checkmark & \cellcolor{YellowGreen}\checkmark & \cellcolor{YellowGreen}\checkmark & \cellcolor{YellowGreen}Limb       & \cellcolor{YellowGreen}Day    & \cellcolor{YellowGreen}Limb   & \cellcolor{YellowGreen}Limb       & \cellcolor{YellowGreen}Limb   & \cellcolor{YellowGreen}Limb   & \cellcolor{YellowGreen}Limb       & \cellcolor{YellowGreen}Day    & \cellcolor{YellowGreen}Limb   & \cellcolor{YellowGreen}Limb       & \cellcolor{YellowGreen}Limb       & \cellcolor{YellowGreen}Limb       \\
        \multicolumn{1}{|l|}{Chemical profile}                           & \cellcolor{YellowGreen}\checkmark & \cellcolor{Dandelion}$\sim$       & \cellcolor{Dandelion}$\sim$   & \cellcolor{YellowGreen}\checkmark & \cellcolor{YellowGreen}\checkmark & \cellcolor{Dandelion}$\sim$       & \cellcolor{YellowGreen}\checkmark & \cellcolor{Dandelion}$\sim$   & \cellcolor{Dandelion}$\sim$   & \cellcolor{YellowGreen}Limb       & \cellcolor{YellowGreen}Limb   & \cellcolor{RedOrange}$\times$ & \cellcolor{YellowGreen}\checkmark & \cellcolor{Dandelion}$\sim$   & \cellcolor{Dandelion}$\sim$   & \cellcolor{YellowGreen}Limb       & \cellcolor{YellowGreen}Limb       & \cellcolor{RedOrange}$\times$     \\
        \hline
        \multicolumn{19}{c}{Bottom of the atmosphere [$\sim10^6$ Pa - $\sim10^2$ Pa]} \\
        \hline
        \multicolumn{1}{|l|}{Temperature profile}                        & \cellcolor{Dandelion}$\sim$       & \cellcolor{RedOrange}$\times$     & \cellcolor{RedOrange}$\times$ & \cellcolor{RedOrange}$\times$     & \cellcolor{YellowGreen}\checkmark & \cellcolor{RedOrange}$\times$     & \cellcolor{YellowGreen}\checkmark & \cellcolor{RedOrange}$\times$ & \cellcolor{RedOrange}$\times$ & \cellcolor{YellowGreen}\checkmark & \cellcolor{YellowGreen}Night  & \cellcolor{YellowGreen}Night  & \cellcolor{YellowGreen}\checkmark & \cellcolor{RedOrange}$\times$ & \cellcolor{RedOrange}$\times$ & \cellcolor{YellowGreen}\checkmark & \cellcolor{RedOrange}$\times$     & \cellcolor{RedOrange}$\times$     \\
        \multicolumn{1}{|l|}{Chemical profile}                           & \cellcolor{YellowGreen}\checkmark & \cellcolor{Dandelion}$\sim$       & \cellcolor{Dandelion}$\sim$   & \cellcolor{RedOrange}$\times$     & \cellcolor{YellowGreen}\checkmark & \cellcolor{Dandelion}$\sim$       & \cellcolor{YellowGreen}\checkmark & \cellcolor{Dandelion}$\sim$   & \cellcolor{Dandelion}$\sim$   & \cellcolor{RedOrange}$\times$     & \cellcolor{Dandelion}$\sim$   & \cellcolor{RedOrange}$\times$ & \cellcolor{YellowGreen}\checkmark & \cellcolor{Dandelion}$\sim$   & \cellcolor{Dandelion}$\sim$   & \cellcolor{YellowGreen}\checkmark & \cellcolor{RedOrange}$\times$     & \cellcolor{RedOrange}$\times$     \\
        \hline
    \end{tabular}
    \begin{tablenotes}\footnotesize
        \item[*] For species detection, C/O ratio, metallicity (Z), and chemical profiles, it focuses only on main absorbers. The temperature and chemical profiles are split depending on the region on the atmosphere, where around the highest atmospheric contribution (between $\sim10^2$ Pa and $\sim10^0$ Pa) the atmosphere in globally well retrieved, contrary to the bottom of the atmosphere (between $\sim10^6$ Pa and $\sim10^2$ Pa). See Section \ref{1datmo} and \ref{3datmo} for more details on the specific biases.
    \end{tablenotes}
    \end{threeparttable}
    }
    \label{tab:results}
\end{table*}



\section{Conclusions}\label{section: conclu}
We present in Table \ref{tab:results} an overview of the main results obtained in this study. 
A limitation to the approach described in this paper has been the use of simplified equilibrium chemistry models: recent re-analysis work on transit retrievals from HST has shown the necessity to go towards non-equilibrium chemical models at least for temperate planets \citep{Panek2023}. This additional complexity is still beyond current 3D modeling for retrievals, but will certainly in the future be an important aspect to develop.\\
The three-dimensional effects that are presented above will be an improvement in future retrievals of the JWST observations, like transit spectroscopy at JWST/NIRSpec resolution on WASP-39~b which presently use 1D models \citep{Alderson2023}. Phase curve observations as observed by HST, Spitzer, and JWST on WASP-43~b \citep{Stevenson2014,Stevenson2017,Murphy23,Bell2023} give access to preliminary constraints on the 3-dimensional composition, cloud coverage, and temperature structure of the planet's atmosphere thanks to JWST/MIRI/LRS sensitivity. A limitation of the parameter retrieval from the observations is today reached by the complexity of the models: combining GCM, radiative transfer codes, thermochemistry codes, and non-equilibrium chemistry is a formidable task that involves a multidisciplinary effort from various communities of molecular spectroscopists, chemists, meteorologists, and astronomers.\\
Even with the limitations described above, this paper provides warning of the approaches allowing future investigators to address properly these questions.

\begin{acknowledgements}
This work has been launched following the summer school ARES II in 2021 (https://www.ariel-mission.fr/ares-ii-2021-en/). All school participants are thanked for enlightening discussions.
We would like to thank the anonymous referee for his excellent comments, which improved the presentation of our results.
This work was granted access to the HPC resources of MesoPSL financed by the Region Ile de France and the project Equip@Meso (reference ANR-10-EQPX-29-01) of the programme Investissements d’Avenir supervised by the Agence Nationale pour la Recherche. This project has been carried out in the frame of the National Centre for Competence in Research PlanetS supported by the Swiss National Science Foundation (SNSF). WP acknowledges financial support from the SNSF for project 200021\_200726. This project has received funding from the ERC under the European Union’s Horizon 2020 research and innovation programme (project SPICE DUNE; grant agreement No 947634). This project has received funding from the European Research Council (ERC) under the ERC OxyPlanets projects (grant agreement No. 101053033).
E. Panek, J-P. Beaulieu and P. Drossart have been supported by CNES convention n$^{0}$6512/7493. O. Venot acknowledges funding from the ANR project `EXACT' (ANR-21-CE49-0008-01) and from the Centre National d'\'{E}tudes Spatiales (CNES).
\end{acknowledgements}

\bibliographystyle{aa}
\bibliography{biblio}

\begin{appendix}

\onecolumn

\section{\textbf{Transmission spectra}}

Transmission spectra simulated with \textit{Pytmosph3R} \citep{Falco2022} for HD~189733~b (\textbf{top}), WASP-121~b (\textbf{middle}) and GJ~1214~b (\textbf{bottom}). We compare here the differences between the East limb, the West limb and the all planet spectra.

\begin{figure*}[h!]
\centering
\includegraphics[scale=\specbis,trim = 0cm 0cm 0cm 0cm, clip]{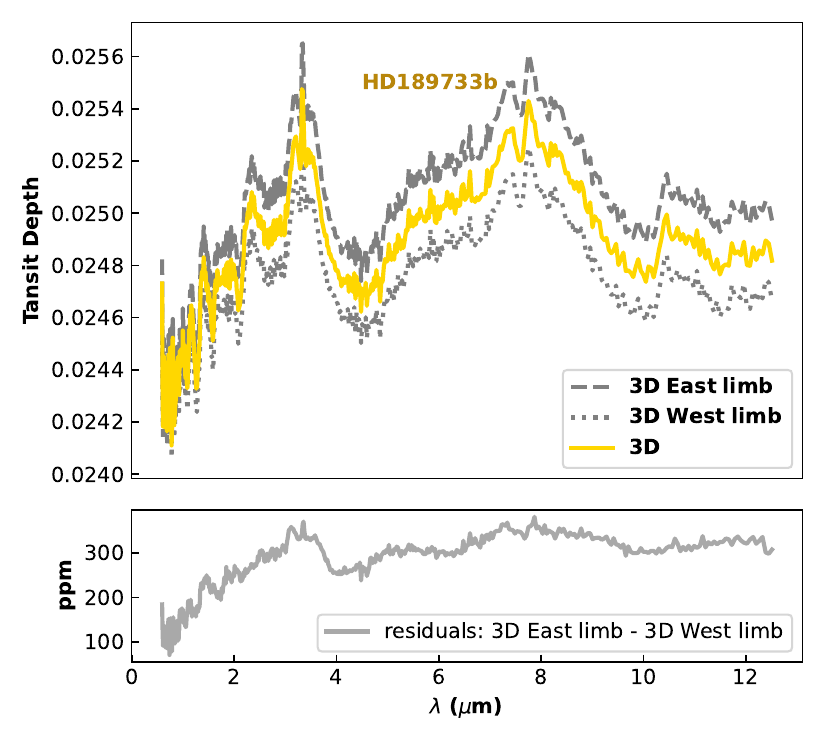}
\includegraphics[scale=\specbis,trim = 0cm 0cm 0cm 0cm, clip]{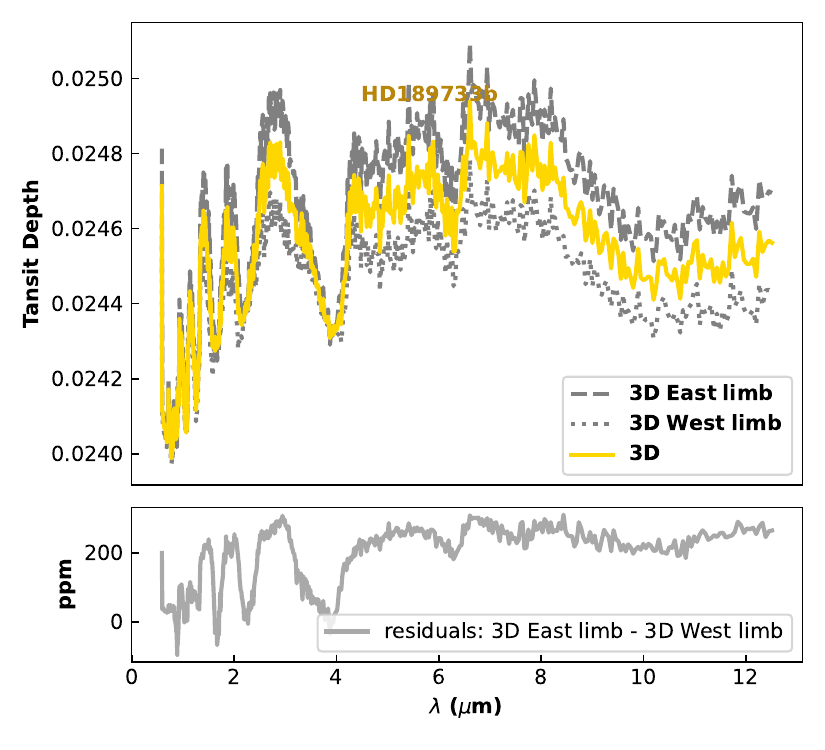}
\\
\includegraphics[scale=\specbis,trim = 0cm 0cm 0cm 0cm, clip]{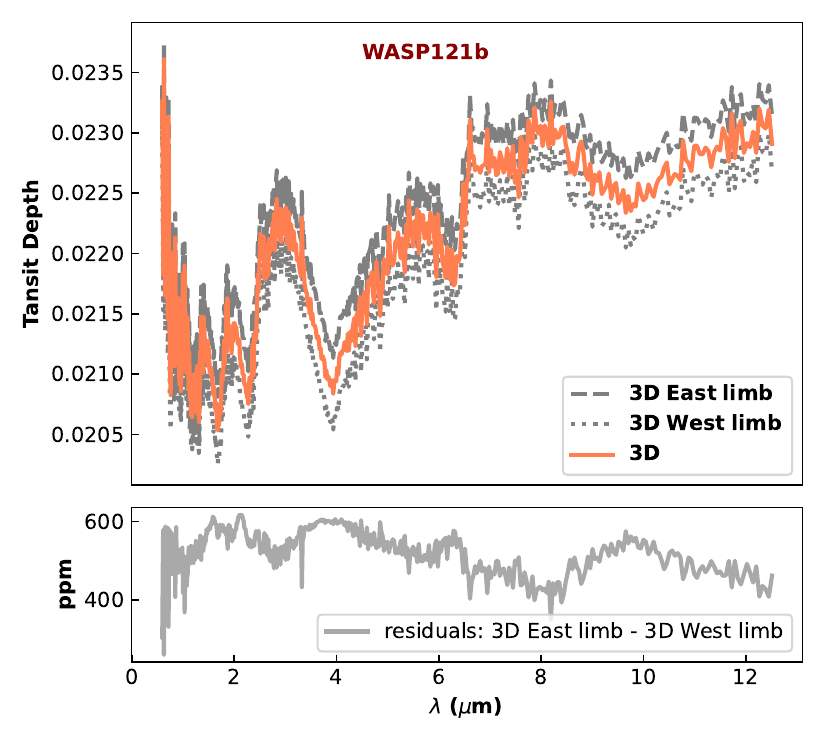}
\includegraphics[scale=\specbis,trim = 0cm 0cm 0cm 0cm, clip]{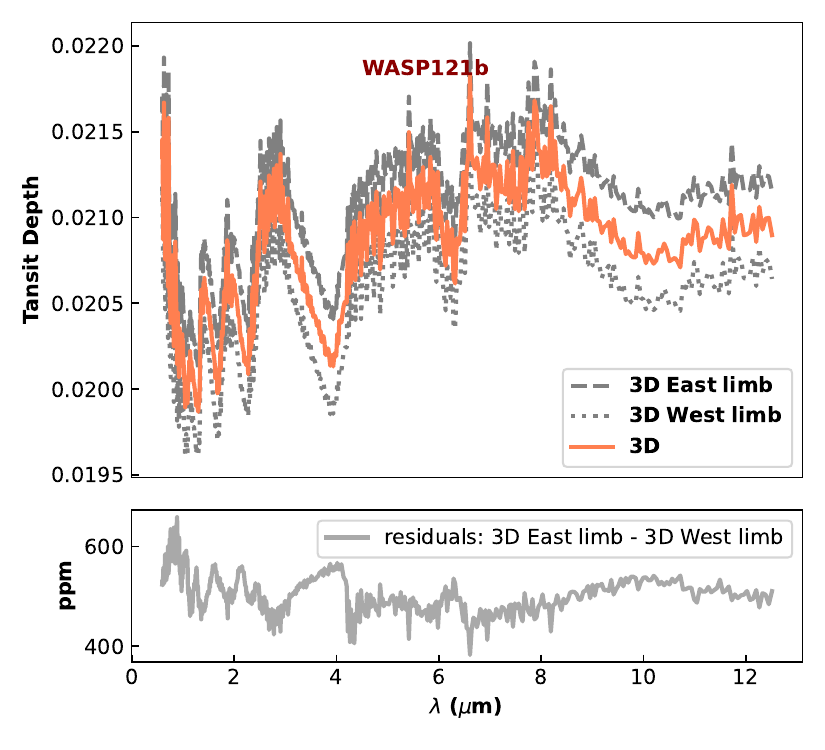}
\\
\includegraphics[scale=\specbis,trim = 0cm 0cm 0cm 0cm, clip]{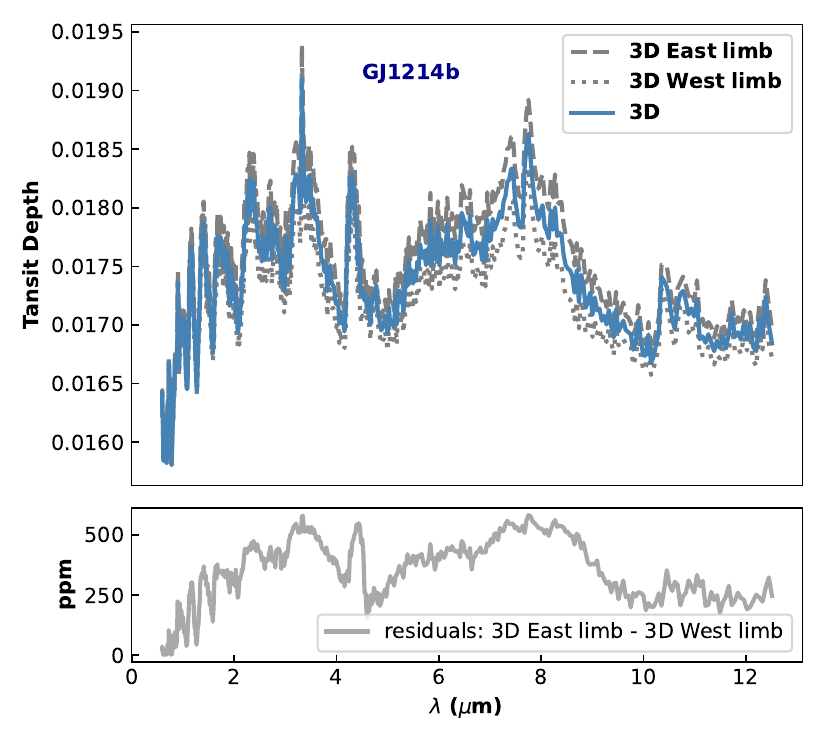}
\includegraphics[scale=\specbis,trim = 0cm 0cm 0cm 0cm, clip]{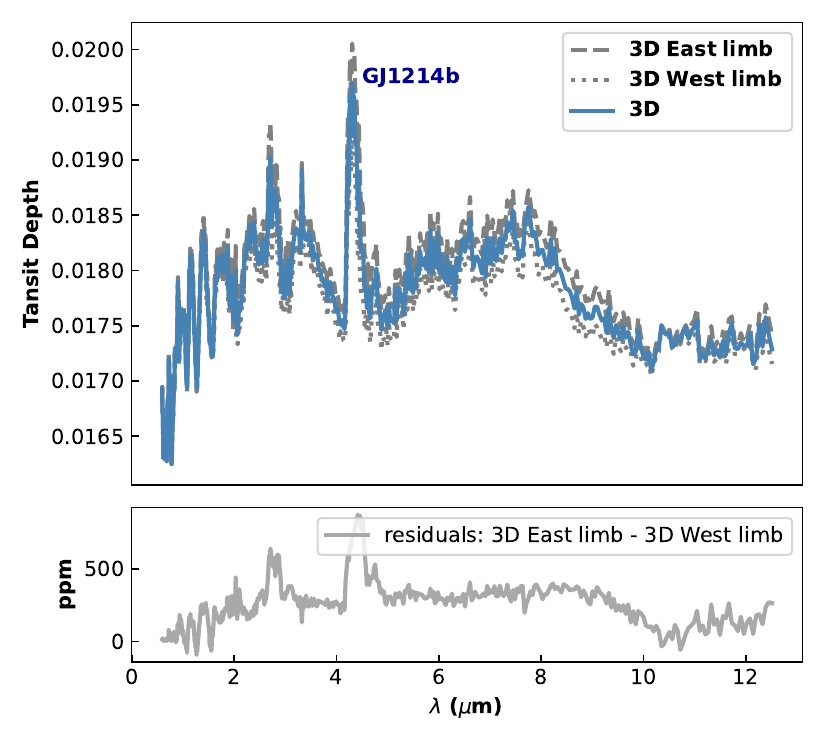}

\caption{Transmission spectra simulated with \textit{Pytmosph3R} \citep{Falco2022} for HD~189733~b (\textbf{top}), WASP-121~b (\textbf{middle}) and GJ~1214~b (\textbf{bottom}). Each panel compares three transmission spectra based on a 3D atmosphere for the East limb, the West limb and the all planet, respectively in dashed grey, dot grey and solid blue lines. \textbf{Left panels:} constant input chemistry. \textbf{Right:} equilibrium input chemistry. The differences between the East limb and the West limb spectra are plotted below each panel in grey.}
\label{fig: spectra_WestEast}
\end{figure*}

\newpage

\section{Retrievals results}

Tables of values retrieved by all retrieval models for all input configurations considering the best solution for each retrieval are shown here. For Free retrievals, we calculate the metallicity using \textit{TauREx 3} which includes it in the available derived parameters, while we compute the C/O ratio using formula~2 from \cite{Lee2013} and the marginalized posterior distributions of CO, CH$_4$, CO$_2$, and H$_2$O from each retrieval (best solution).

\begin{table*}[h!]
    \caption{Retrieval results of \textbf{GJ~1214~b}. Best retrieval of each configuration is highlighted in bold.}
    \centering
    \resizebox{\textwidth}{!}{%
    \begin{tabular}{lc|ccc|ccc|ccc|ccc|ccc|ccc|}
        \cline{3-20}
        \rule{0pt}{4ex} & &\multicolumn{12}{c|}{JWST} & \multicolumn{6}{c|}{Ariel} \\ [0.3cm] \cline{3-20}
        \rule{0pt}{4ex} & &\multicolumn{6}{c|}{1D} & \multicolumn{12}{c|}{3D} \\ [0.3cm] \cline{3-20}
        \rule{0pt}{4ex} & &\multicolumn{3}{c|}{constant} &\multicolumn{3}{c|}{equilibrium} &\multicolumn{3}{c|}{constant} &\multicolumn{3}{c|}{equilibrium} &\multicolumn{3}{c|}{constant} &\multicolumn{3}{c|}{equilibrium} \\ [0.3cm] \cline{1-20}
        \multicolumn{1}{|l}{Parameters} & \inputs{Input} & \textbf{Free} & ACE & Fast & Free & \textbf{ACE} & Fast & \textbf{Free} & ACE & Fast & Free & \textbf{ACE} & Fast & Free & \textbf{ACE} & Fast & Free & \textbf{ACE} & Fast \\
        \hline
        \multicolumn{1}{|l}{T$_{surf}$ [K]}         & \inputs{}      & \textbf{970  } & 348   & 330    & 770   & \textbf{1037 } & 1097  & \textbf{1092 } & 1303  & 950   & 1055   & \textbf{1165 } & 1422  & 918   & \textbf{1084 } & 1162  & 1064  & \textbf{1086 } & 1128  \\
        \multicolumn{1}{|l}{T$_1$ [K]}              & \inputs{}      & \textbf{639  } & 745   & 875    & 1035  & \textbf{1345 } & 1341  & \textbf{756  } & 305   & 306   & 407    & \textbf{391  } & 317   & 684   & \textbf{610  } & 325   & 621   & \textbf{673  } & 533   \\
        \multicolumn{1}{|l}{T$_2$ [K]}              & \inputs{}      & \textbf{215  } & 236   & 189    & 331   & \textbf{378  } & 334   & \textbf{426  } & 560   & 602   & 306    & \textbf{580  } & 674   & 443   & \textbf{569  } & 597   & 331   & \textbf{470  } & 579   \\
        \multicolumn{1}{|l}{T$_{top}$ [K]}          & \inputs{}      & \textbf{198  } & 628   & 627    & 185   & \textbf{212  } & 334   & \textbf{536  } & 306   & 335   & 1234   & \textbf{619  } & 532   & 524   & \textbf{350  } & 367   & 983   & \textbf{814  } & 833   \\
        \multicolumn{1}{|l}{log$_{10}$(P$_1$) [Pa]} & \inputs{}      & \textbf{4.67 } & 2.25  & 2.63   & 3.95  & \textbf{4.38 } & 4.24  & \textbf{4.98 } & 2.15  & 2.15  & 2.44   & \textbf{2.00 } & 2.09  & 4.72  & \textbf{4.55 } & 2.44  & 4.59  & \textbf{4.61 } & 4.06  \\
        \multicolumn{1}{|l}{log$_{10}$(P$_2$) [Pa]} & \inputs{}      & \textbf{0.131} & 1.58  & 2.08   & 2.24  & \textbf{3.13 } & 2.91  & \textbf{3.26 } & 1.20  & 1.27  & -0.648 & \textbf{1.31 } & 1.37  & 2.60  & \textbf{2.61 } & 1.06  & 0.32  & \textbf{2.52 } & 1.68  \\
        \multicolumn{1}{|l}{log$_{10}$(\hho)}       & \inputs{-1.00} & \textbf{-1.15} &       &        & -1.17 & \textbf{     } &       & \textbf{-1.09} &       &       & -1.24  & \textbf{     } &       & -1.13 & \textbf{     } &       & -1.26 & \textbf{     } &       \\
        \multicolumn{1}{|l}{log$_{10}$(\co)}        & \inputs{-2.00} & \textbf{-2.24} &       &        & -8.62 & \textbf{     } &       & \textbf{-2.11} &       &       & -4.86  & \textbf{     } &       & -5.59 & \textbf{     } &       & -7.48 & \textbf{     } &       \\
        \multicolumn{1}{|l}{log$_{10}$(\chhhh)}     & \inputs{-1.52} & \textbf{-1.63} &       &        & -1.21 & \textbf{     } &       & \textbf{-1.64} &       &       & -2.27  & \textbf{     } &       & -1.61 & \textbf{     } &       & -2.27 & \textbf{     } &       \\
        \multicolumn{1}{|l}{log$_{10}$(\coo)}       & \inputs{-2.00} & \textbf{-2.10} &       &        & -9.34 & \textbf{     } &       & \textbf{-2.06} &       &       & -1.57  & \textbf{     } &       & -1.82 & \textbf{     } &       & -1.46 & \textbf{     } &       \\
        \multicolumn{1}{|l}{log$_{10}$(\hcn)}       & \inputs{-6.22} & \textbf{-8.16} &       &        & -8.35 & \textbf{     } &       & \textbf{-8.37} &       &       & -8.21  & \textbf{     } &       & -6.93 & \textbf{     } &       & -7.29 & \textbf{     } &       \\
        \multicolumn{1}{|l}{log$_{10}$(\nhhh)}      & \inputs{-3.30} & \textbf{-3.45} &       &        & -3.19 & \textbf{     } &       & \textbf{-3.45} &       &       & -9.23  & \textbf{     } &       & -3.57 & \textbf{     } &       & -8.52 & \textbf{     } &       \\
        \multicolumn{1}{|l}{Radius [R$_{jup}$]}     & \inputs{0.243} & \textbf{0.242} & 0.248 & 0.247  & 0.241 & \textbf{0.240} & 0.238 & \textbf{0.242} & 0.241 & 0.244 & 0.245  & \textbf{0.240} & 0.240 & 0.242 & \textbf{0.240} & 0.241 & 0.247 & \textbf{0.245} & 0.247 \\
        \multicolumn{1}{|l}{C/O}                    & \inputs{0.458} & \textbf{     } & 0.459 & 0.442  & 0.90  & \textbf{0.898} & 0.807 & \textbf{     } & 0.255 & 0.145 & 0.29   & \textbf{0.441} & 0.313 &       & \textbf{0.309} & 0.169 & 0.33  & \textbf{0.461} & 0.371 \\
        \multicolumn{1}{|l}{log$_{10}$(Z)}          & \inputs{2.00}  & \textbf{     } & 2.14  & 2.20   & 1.53  & \textbf{1.91 } & 1.96  & \textbf{     } & 2.02  & 1.82  & 1.58   & \textbf{2.01 } & 1.83  &       & \textbf{2.06 } & 1.79  & 1.61  & \textbf{2.06 } & 1.93  \\
        \multicolumn{1}{|l}{logE}                   & \inputs{-}     & \textbf{4642 } & 4623  & 4619   & 4631  & \textbf{4640 } & 4633  & \textbf{4619 } & 4585  & 4575  & 4547   & \textbf{4557 } & 4504  & 736   & \textbf{737  } & 736   & 731   & \textbf{740  } & 735   \\
        \hline
    \end{tabular}
    }
    \label{tab:gj1214b_ret}
\end{table*}

\begin{table*}[h!]
    \caption{Retrieval results of \textbf{HD~189733~b}. Best retrieval of each configuration is highlighted in bold.}
    \centering
    \resizebox{\textwidth}{!}{%
    \begin{tabular}{lc|cccc|cccc|cccc|cccc|cccc|cccc|}
        \cline{3-26}
        \rule{0pt}{4ex} & &\multicolumn{16}{c|}{JWST} & \multicolumn{8}{c|}{Ariel} \\ [0.3cm] \cline{3-26}
        \rule{0pt}{4ex} & &\multicolumn{8}{c|}{1D}    & \multicolumn{16}{c|}{3D}   \\ [0.3cm] \cline{3-26}
        \rule{0pt}{4ex} & &\multicolumn{4}{c|}{constant} &\multicolumn{4}{c|}{equilibrium} &\multicolumn{4}{c|}{constant} &\multicolumn{4}{c|}{equilibrium} &\multicolumn{4}{c|}{constant} &\multicolumn{4}{c|}{equilibrium} \\ [0.3cm] \cline{1-26}
        \multicolumn{1}{|l}{Parameters} & \inputs{Input} & \textbf{Free} & ACE & Fast & GGchem & Free & \textbf{ACE} & Fast & GGchem & \textbf{Free} & ACE & Fast & GGchem & Free & \textbf{ACE} & Fast & GGchem & \textbf{Free} & ACE & Fast & GGchem & Free & ACE & Fast & \textbf{GGchem} \\
        \hline
        \multicolumn{1}{|l}{T$_{surf}$ [K]}         & \inputs{     } & \textbf{1352 } & 1554    & 508   & 503    & 626   & \textbf{1760  } & 836    & 1610  & \textbf{1313 } & 503    & 980    & 899   & 705   & \textbf{1612 } & 729    & 1371  & \textbf{1400 } & 1610   & 1038   & 1451   & 1641  & 1509   & 1103   & \textbf{1433  } \\
        \multicolumn{1}{|l}{T$_1$ [K]}              & \inputs{     } & \textbf{731  } & 517     & 969   & 1643   & 1382  & \textbf{834   } & 694    & 732   & \textbf{745  } & 1898   & 1864   & 1541  & 1509  & \textbf{879  } & 735    & 761   & \textbf{1194 } & 1531   & 2265   & 1431   & 1557  & 1098   & 667    & \textbf{1158  } \\
        \multicolumn{1}{|l}{T$_2$ [K]}              & \inputs{     } & \textbf{1140 } & 507     & 536   & 677    & 659   & \textbf{1267  } & 1251   & 1322  & \textbf{1308 } & 720    & 688    & 664   & 1069  & \textbf{1367 } & 1175   & 1426  & \textbf{1239 } & 555    & 587    & 578    & 1048  & 1175   & 1179   & \textbf{1182  } \\
        \multicolumn{1}{|l}{T$_{top}$ [K]}          & \inputs{     } & \textbf{931  } & 760     & 501   & 501    & 2299  & \textbf{535   } & 1264   & 516   & \textbf{1020 } & 501    & 502    & 501   & 1411  & \textbf{548  } & 1748   & 522   & \textbf{1311 } & 681    & 1653   & 655    & 1440  & 1509   & 1569   & \textbf{1242  } \\
        \multicolumn{1}{|l}{log$_{10}$(P$_1$) [Pa]} & \inputs{     } & \textbf{4.31 } & 0.228   & 3.53  & 3.06   & 2.65  & \textbf{2.68  } & 2.00   & 2.28  & \textbf{4.18 } & 3.12   & 2.69   & 2.90  & 3.63  & \textbf{2.87 } & 2.01   & 2.35  & \textbf{4.46 } & 4.12   & 3.00   & 4.20   & 4.49  & 4.18   & 3.29   & \textbf{4.58  } \\
        \multicolumn{1}{|l}{log$_{10}$(P$_2$) [Pa]} & \inputs{     } & \textbf{2.52 } & -2.63   & 0.552 & 2.63   & 2.42  & \textbf{2.39  } & 1.59   & 2.12  & \textbf{2.13 } & 2.85   & 2.47   & 2.20  & 3.27  & \textbf{2.60 } & 1.91   & 2.22  & \textbf{1.61 } & 1.03   & 2.56   & 1.22   & 1.33  & 1.51   & 1.21   & \textbf{1.64  } \\
        \multicolumn{1}{|l}{log$_{10}$(\hho)}       & \inputs{-3.15} & \textbf{-2.87} &         &       &        & -3.19 & \textbf{      } &        &       & \textbf{-2.90} &        &        &       & -3.44 & \textbf{     } &        &       & \textbf{-3.05} &        &        &        & -3.71 &        &        & \textbf{      } \\
        \multicolumn{1}{|l}{log$_{10}$(\co)}        & \inputs{-4.00} & \textbf{-7.86} &         &       &        & -3.12 & \textbf{      } &        &       & \textbf{-7.92} &        &        &       & -3.30 & \textbf{     } &        &       & \textbf{-7.92} &        &        &        & -3.92 &        &        & \textbf{      } \\
        \multicolumn{1}{|l}{log$_{10}$(\chhhh)}     & \inputs{-3.40} & \textbf{-3.04} &         &       &        & -5.34 & \textbf{      } &        &       & \textbf{-3.05} &        &        &       & -6.17 & \textbf{     } &        &       & \textbf{-3.28} &        &        &        & -7.77 &        &        & \textbf{      } \\
        \multicolumn{1}{|l}{log$_{10}$(\coo)}       & \inputs{-7.70} & \textbf{-9.16} &         &       &        & -5.89 & \textbf{      } &        &       & \textbf{-9.15} &        &        &       & -6.58 & \textbf{     } &        &       & \textbf{-9.37} &        &        &        & -7.05 &        &        & \textbf{      } \\
        \multicolumn{1}{|l}{log$_{10}$(\hcn)}       & \inputs{-7.00} & \textbf{-9.19} &         &       &        & -9.32 & \textbf{      } &        &       & \textbf{-9.09} &        &        &       & -7.99 & \textbf{     } &        &       & \textbf{-8.41} &        &        &        & -9.11 &        &        & \textbf{      } \\
        \multicolumn{1}{|l}{log$_{10}$(\nhhh)}      & \inputs{-4.52} & \textbf{-4.29} &         &       &        & -6.53 & \textbf{      } &        &       & \textbf{-4.34} &        &        &       & -6.30 & \textbf{     } &        &       & \textbf{-5.49} &        &        &        & -8.89 &        &        & \textbf{      } \\
        \multicolumn{1}{|l}{log$_{10}$(\feh)}       & \inputs{-8.05} & \textbf{-7.93} &         &       &        & -11.8 & \textbf{      } &        &       & \textbf{-7.91} &        &        &       & -10.8 & \textbf{     } &        &       & \textbf{-9.43} &        &        &        & -10.2 &        &        & \textbf{      } \\
        \multicolumn{1}{|l}{log$_{10}$(\sio)}       & \inputs{-4.70} & \textbf{-8.55} &         &       &        & -9.52 & \textbf{      } &        &       & \textbf{-8.44} &        &        &       & -9.37 & \textbf{     } &        &       & \textbf{-7.59} &        &        &        & -8.22 &        &        & \textbf{      } \\
        \multicolumn{1}{|l}{log$_{10}$(\na)}        & \inputs{-5.52} & \textbf{-8.00} &         &       &        & -8.73 & \textbf{      } &        &       & \textbf{-11.7} &        &        &       & -9.18 & \textbf{     } &        &       & \textbf{-7.06} &        &        &        & -6.80 &        &        & \textbf{      } \\
        \multicolumn{1}{|l}{log$_{10}$(K)}          & \inputs{-6.70} & \textbf{-11.7} &         &       &        & -11.9 & \textbf{      } &        &       & \textbf{-7.75} &        &        &       & -11.9 & \textbf{     } &        &       & \textbf{-7.36} &        &        &        & -8.28 &        &        & \textbf{      } \\
        \multicolumn{1}{|l}{log$_{10}$(\tio)}       & \inputs{-10.0} & \textbf{-9.75} &         &       &        & -11.9 & \textbf{      } &        &       & \textbf{-9.77} &        &        &       & -12.0 & \textbf{     } &        &       & \textbf{-9.63} &        &        &        & -11.4 &        &        & \textbf{      } \\
        \multicolumn{1}{|l}{log$_{10}$(\vo)}        & \inputs{-9.00} & \textbf{-8.68} &         &       &        & -11.9 & \textbf{      } &        &       & \textbf{-8.70} &        &        &       & -11.9 & \textbf{     } &        &       & \textbf{-9.97} &        &        &        & -11.2 &        &        & \textbf{      } \\
        \multicolumn{1}{|l}{Radius [R$_{jup}$]}     & \inputs{1.13 } & \textbf{1.13 } & 1.13    & 1.13  & 1.13   & 1.16  & \textbf{1.12  } & 1.13   & 1.12  & \textbf{1.12 } & 1.13   & 1.13   & 1.13  & 1.12  & \textbf{1.12 } & 1.13   & 1.12  & \textbf{1.12 } & 1.13   & 1.12   & 1.13   & 1.12  & 1.12   & 1.12   & \textbf{1.12  } \\
        \multicolumn{1}{|l}{C/O}                    & \inputs{0.550} & \textbf{     } & 0.962   & 0.726 & 0.797  & 0.54  & \textbf{0.677 } & 0.0377 & 0.418 & \textbf{     } & 0.921  & 2.00   & 0.874 & 0.58  & \textbf{0.614} & 0.0368 & 0.308 & \textbf{     } & 0.954  & 1.87   & 0.869  & 0.38  & 0.676  & 0.143  & \textbf{0.544 } \\
        \multicolumn{1}{|l}{log$_{10}$(Z)}          & \inputs{0.00 } & \textbf{     } & 0.00139 & 0.417 & -0.047 & 0.02  & \textbf{0.0864} & 1.22   & 0.359 & \textbf{     } & -0.444 & -0.989 & 0.263 & -0.18 & \textbf{0.144} & 1.34   & 0.541 & \textbf{     } & 0.0480 & -0.858 & -0.163 & -0.44 & 0.150  & 1.03   & \textbf{0.0782} \\
        \multicolumn{1}{|l}{logE}                   & \inputs{-    } & \textbf{5818 } & 3282    & -1089 & 3146   & 5545  & \textbf{5747  } & 3390   & 5365  & \textbf{5792 } & 2744   & -3708  & 2102  & 5541  & \textbf{5749 } & 3253   & 5354  & \textbf{843  } & 809    & 791    & 809    & 840   & 848    & 840    & \textbf{848   } \\
        \hline
    \end{tabular}
    }
    \label{tab:hd189733b_ret}
\end{table*}

\begin{table*}[h!]
    \caption{Retrieval results of \textbf{WASP-121~b}. Best retrieval of each configuration is highlighted in bold.}
    \centering
    \resizebox{\textwidth}{!}{%
    \begin{tabular}{lc|ccc|ccc|ccc|ccc|ccc|ccc|}
        \cline{3-20}
        \rule{0pt}{4ex} & &\multicolumn{12}{c|}{JWST} & \multicolumn{6}{c|}{Ariel} \\ [0.3cm] \cline{3-20}
        \rule{0pt}{4ex} & &\multicolumn{6}{c|}{1D}    & \multicolumn{12}{c|}{3D}   \\ [0.3cm] \cline{3-20}
        \rule{0pt}{4ex} & &\multicolumn{3}{c|}{constant} &\multicolumn{3}{c|}{equilibrium} &\multicolumn{3}{c|}{constant} &\multicolumn{3}{c|}{equilibrium} &\multicolumn{3}{c|}{constant} &\multicolumn{3}{c|}{equilibrium} \\ [0.3cm] \cline{1-20}
        \multicolumn{1}{|l}{Parameters} & \inputs{Input} & \textbf{Free} & Fast & GGchem & Free & Fast & \textbf{GGchem} & \textbf{Free} & Fast & GGchem & \textbf{Free} & Fast & GGchem & \textbf{Free} & Fast & GGchem & Free & Fast & \textbf{GGchem} \\
        \hline
        \multicolumn{1}{|l}{T$_{surf}$ [K]}         & \inputs{     } & \textbf{2045 } & 3206   & 2122  & 2422  & 2590   & \textbf{1830 } & \textbf{880  } & 3342   & 475    & \textbf{2736 } & 928    & 3277   & \textbf{1805 } & 1273  & 3473   & 2372  & 1604   & \textbf{1558  } \\
        \multicolumn{1}{|l}{T$_1$ [K]}              & \inputs{     } & \textbf{1519 } & 641    & 1652  & 3446  & 586    & \textbf{2828 } & \textbf{908  } & 709    & 465    & \textbf{1282 } & 521    & 1081   & \textbf{614  } & 622   & 779    & 1265  & 836    & \textbf{1151  } \\
        \multicolumn{1}{|l}{T$_2$ [K]}              & \inputs{     } & \textbf{975  } & 2290   & 776   & 978   & 2658   & \textbf{766  } & \textbf{3745 } & 2704   & 3016   & \textbf{2195 } & 2162   & 2394   & \textbf{3671 } & 2799  & 2554   & 2048  & 2186   & \textbf{2143  } \\
        \multicolumn{1}{|l}{T$_{top}$ [K]}          & \inputs{     } & \textbf{2877 } & 596    & 2714  & 2737  & 468    & \textbf{3596 } & \textbf{3727 } & 784    & 466    & \textbf{705  } & 1269   & 710    & \textbf{3003 } & 558   & 837    & 2373  & 844    & \textbf{2664  } \\
        \multicolumn{1}{|l}{log$_{10}$(P$_1$) [Pa]} & \inputs{     } & \textbf{4.78 } & 2.01   & 2.53  & 3.63  & 2.02   & \textbf{3.10 } & \textbf{2.05 } & 2.51   & 3.52   & \textbf{2.40 } & 3.81   & 2.10   & \textbf{2.18 } & 2.84  & 2.46   & 3.78  & 4.30   & \textbf{3.11  } \\
        \multicolumn{1}{|l}{log$_{10}$(P$_2$) [Pa]} & \inputs{     } & \textbf{3.22 } & 0.31   & 2.43  & 3.21  & -0.616 & \textbf{2.37 } & \textbf{1.68 } & 2.40   & 3.47   & \textbf{1.56 } & 3.28   & 1.57   & \textbf{1.33 } & 2.67  & 1.98   & 1.30  & 3.45   & \textbf{0.989 } \\
        \multicolumn{1}{|l}{log$_{10}$(\hho)}       & \inputs{-3.30} & \textbf{-3.08} &        &       & -3.16 &        & \textbf{     } & \textbf{-3.62} &        &        & \textbf{-4.12} &        &        & \textbf{-3.58} &       &        & -4.05 &        & \textbf{      } \\
        \multicolumn{1}{|l}{log$_{10}$(\co)}        & \inputs{-3.30} & \textbf{-3.24} &        &       & -3.18 &        & \textbf{     } & \textbf{-3.97} &        &        & \textbf{-3.14} &        &        & \textbf{-4.07} &       &        & -3.16 &        & \textbf{      } \\
        \multicolumn{1}{|l}{log$_{10}$(\chhhh)}     & \inputs{-5.00} & \textbf{-4.71} &        &       & -9.31 &        & \textbf{     } & \textbf{-9.60} &        &        & \textbf{-9.81} &        &        & \textbf{-9.24} &       &        & -9.50 &        & \textbf{      } \\
        \multicolumn{1}{|l}{log$_{10}$(\coo)}       & \inputs{-7.40} & \textbf{-7.98} &        &       & -6.73 &        & \textbf{     } & \textbf{-9.36} &        &        & \textbf{-7.15} &        &        & \textbf{-9.37} &       &        & -7.69 &        & \textbf{      } \\
        \multicolumn{1}{|l}{log$_{10}$(\hcn)}       & \inputs{-6.40} & \textbf{-8.52} &        &       & -9.29 &        & \textbf{     } & \textbf{-9.23} &        &        & \textbf{-9.86} &        &        & \textbf{-8.87} &       &        & -9.35 &        & \textbf{      } \\
        \multicolumn{1}{|l}{log$_{10}$(\nhhh)}      & \inputs{-5.22} & \textbf{-4.99} &        &       & -6.89 &        & \textbf{     } & \textbf{-9.09} &        &        & \textbf{-9.64} &        &        & \textbf{-9.23} &       &        & -9.29 &        & \textbf{      } \\
        \multicolumn{1}{|l}{log$_{10}$(\feh)}       & \inputs{-5.70} & \textbf{-5.46} &        &       & -9.55 &        & \textbf{     } & \textbf{-6.28} &        &        & \textbf{-9.42} &        &        & \textbf{-6.82} &       &        & -10.6 &        & \textbf{      } \\
        \multicolumn{1}{|l}{log$_{10}$(\sio)}       & \inputs{-4.22} & \textbf{-4.22} &        &       & -5.48 &        & \textbf{     } & \textbf{-6.12} &        &        & \textbf{-5.53} &        &        & \textbf{-8.11} &       &        & -7.41 &        & \textbf{      } \\
        \multicolumn{1}{|l}{log$_{10}$(\na)}        & \inputs{-5.52} & \textbf{-1.80} &        &       & -8.96 &        & \textbf{     } & \textbf{-5.28} &        &        & \textbf{-7.22} &        &        & \textbf{-3.64} &       &        & -3.66 &        & \textbf{      } \\
        \multicolumn{1}{|l}{log$_{10}$(K)}          & \inputs{-6.70} & \textbf{-9.58} &        &       & -10.9 &        & \textbf{     } & \textbf{-11.2} &        &        & \textbf{-10.6} &        &        & \textbf{-10.1} &       &        & -9.39 &        & \textbf{      } \\
        \multicolumn{1}{|l}{log$_{10}$(\tio)}       & \inputs{-7.00} & \textbf{-6.75} &        &       & -7.41 &        & \textbf{     } & \textbf{-7.25} &        &        & \textbf{-7.93} &        &        & \textbf{-6.97} &       &        & -7.62 &        & \textbf{      } \\
        \multicolumn{1}{|l}{log$_{10}$(\vo)}        & \inputs{-9.00} & \textbf{-8.90} &        &       & -11.2 &        & \textbf{     } & \textbf{-11.5} &        &        & \textbf{-9.36} &        &        & \textbf{-11.0} &       &        & -10.9 &        & \textbf{      } \\
        \multicolumn{1}{|l}{Radius [R$_{jup}$]}     & \inputs{1.91 } & \textbf{1.95 } & 1.92   & 1.94  & 1.89  & 1.93   & \textbf{1.89 } & \textbf{1.98 } & 1.95   & 2.03   & \textbf{1.92 } & 1.97   & 1.91   & \textbf{1.98 } & 1.99  & 1.94   & 1.94  & 1.96   & \textbf{1.96  } \\
        \multicolumn{1}{|l}{C/O}                    & \inputs{0.550} & \textbf{     } & 0.772  & 0.897 & 0.48  & 0.0365 & \textbf{0.114} & \textbf{     } & 0.862  & 0.883  & \textbf{0.90 } & 0.792  & 0.266  & \textbf{     } & 0.924 & 0.669  & 0.88  & 0.883  & \textbf{0.523 } \\
        \multicolumn{1}{|l}{log$_{10}$(Z)}          & \inputs{0.00 } & \textbf{     } & -0.222 & 0.132 & 0.02  & 1.23   & \textbf{0.800} & \textbf{     } & -0.214 & -0.804 & \textbf{-0.14} & -0.708 & -0.992 & \textbf{     } & 0.169 & -0.308 & 0.20  & -0.678 & \textbf{-0.614} \\
        \multicolumn{1}{|l}{logE}                   & \inputs{-    } & \textbf{4938 } & 4201   & 4504  & 4851  & 4541   & \textbf{4922 } & \textbf{4396 } & 2152   & 3086   & \textbf{4872 } & 4550   & 4820   & \textbf{641  } & 566   & 581    & 761   & 749    & \textbf{768   } \\
        \hline
    \end{tabular}
    }
    \label{tab:wasp121b_ret}
\end{table*}

\clearpage

\section{Temperature profiles}

The input temperature profiles at the equator using the 1D and 3D thermal structure, over-plot with the temperature profiles of all the retrieval models are shown here. For the homogeneous 1D thermal structure model, longitude profiles are on top of each other since it has been build homogeneously. For the 3D models, limb profiles (90$^\circ$ and 270$^\circ$ longitude) are in bold with dot markers.

\begin{figure*}[h!]
\centering
\includegraphics[scale=\spec,trim = 0cm 0cm 0cm 0cm, clip]{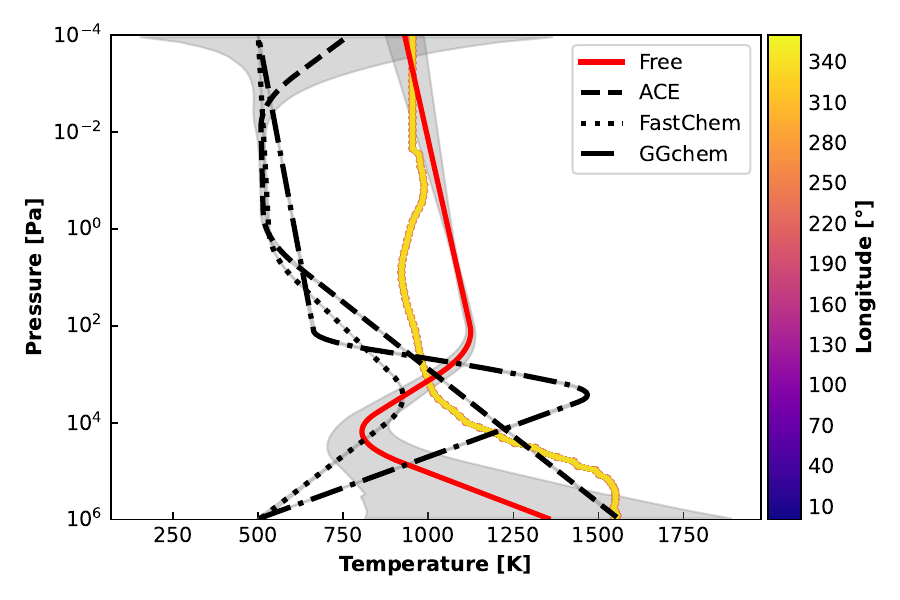}
\includegraphics[scale=\spec,trim = 0cm 0cm 0cm 0cm, clip]{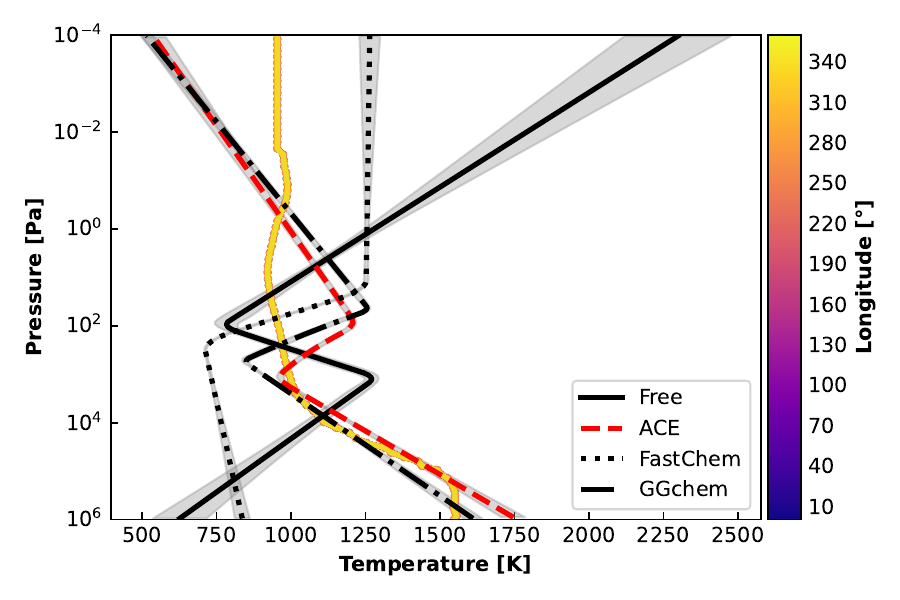}
\\
\includegraphics[scale=\spec,trim = 0cm 0cm 0cm 0cm, clip]{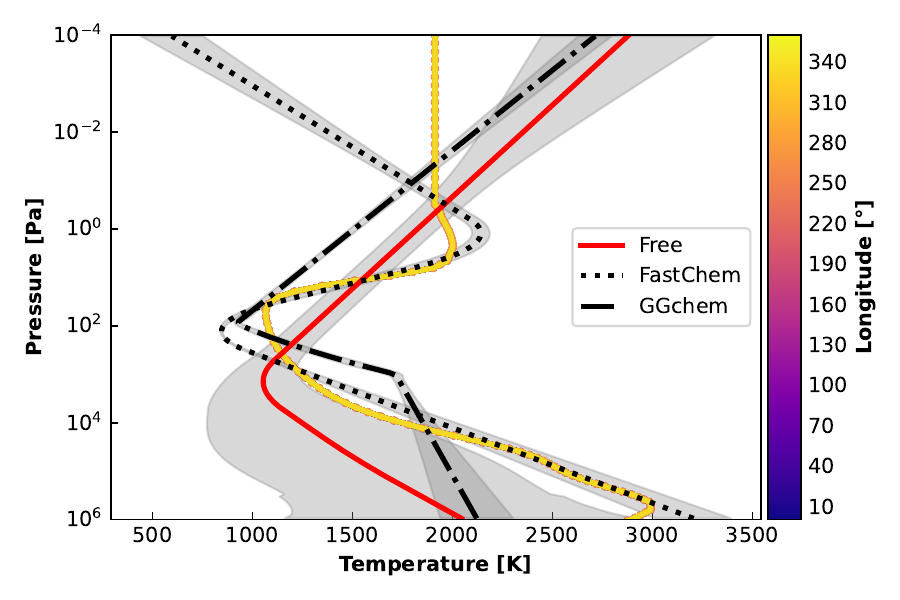}
\includegraphics[scale=\spec,trim = 0cm 0cm 0cm 0cm, clip]{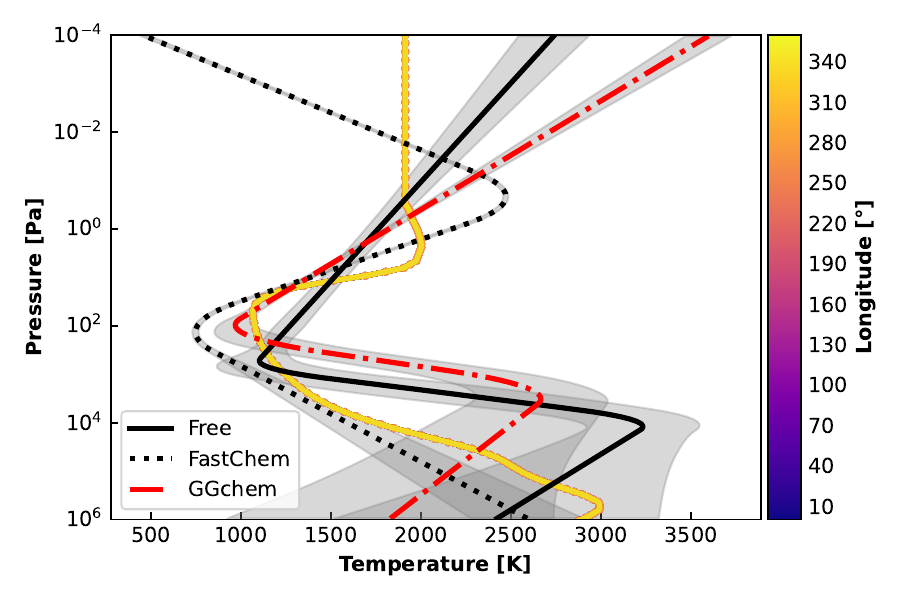}
\\
\includegraphics[scale=\spec,trim = 0cm 0cm 0cm 0cm, clip]{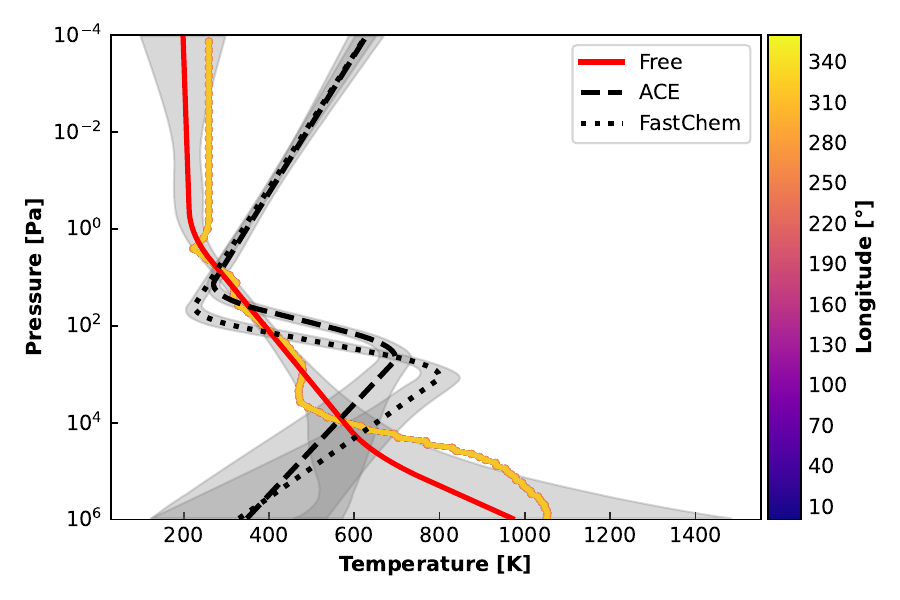}
\includegraphics[scale=\spec,trim = 0cm 0cm 0cm 0cm, clip]{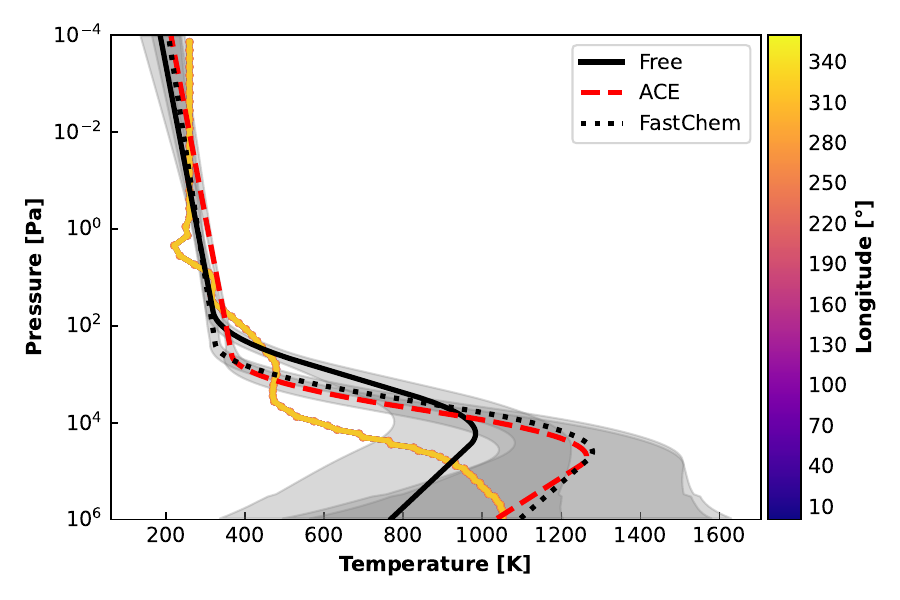}

\caption{JWST simulations. \textbf{Top to Bottom:} Temperature pressure profiles at the equator assuming 1D atmospheres for HD~189733~b, WASP-121~b and GJ~1214~b respectively. Substellar point at 180$^{\circ}$ longitude. \textbf{Left panels:} constant input chemistry. \textbf{Right:} equilibrium input chemistry. We over-plot the best TP-profiles of Free, ACE, FastChem and GGchem retrievals respectively in black solid, dashed, dotted and dashed-dotted lines. The retrieval having the highest Bayes factor is plotted in red.}
\label{fig: TP_JWST_1D}
\end{figure*}

\begin{figure*}[h!]
\centering
\includegraphics[scale=\spec,trim = 0cm 0cm 0cm 0cm, clip]{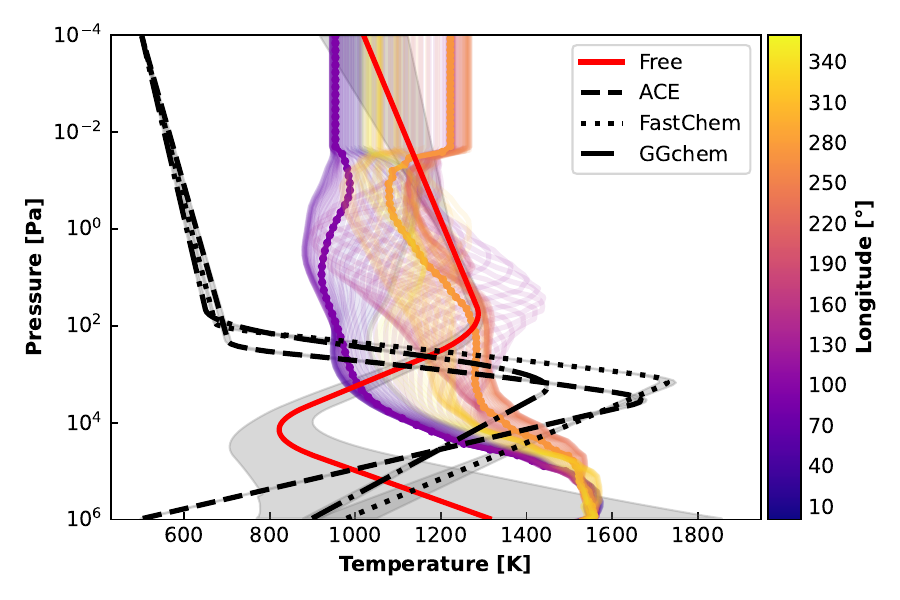}
\includegraphics[scale=\spec,trim = 0cm 0cm 0cm 0cm, clip]{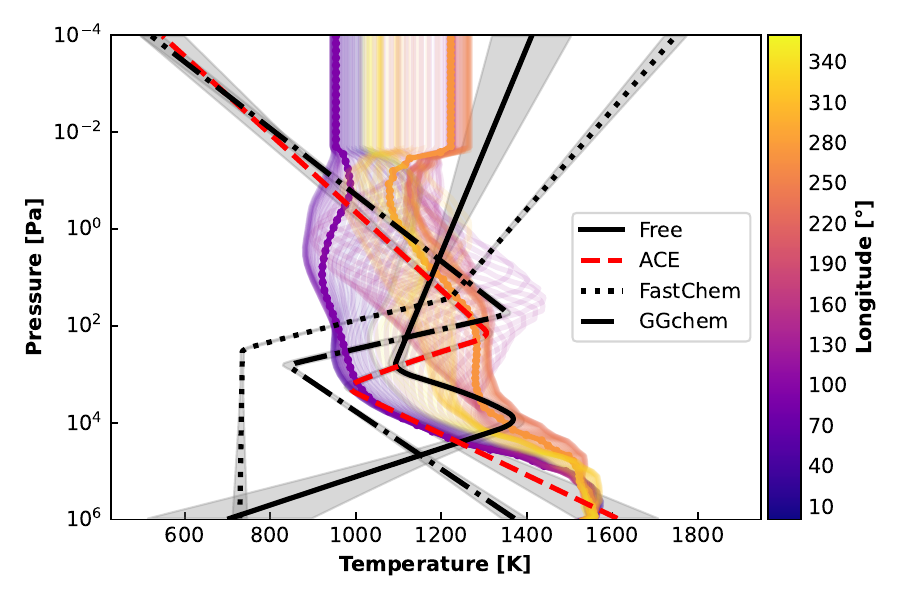}
\\
\includegraphics[scale=\spec,trim = 0cm 0cm 0cm 0cm, clip]{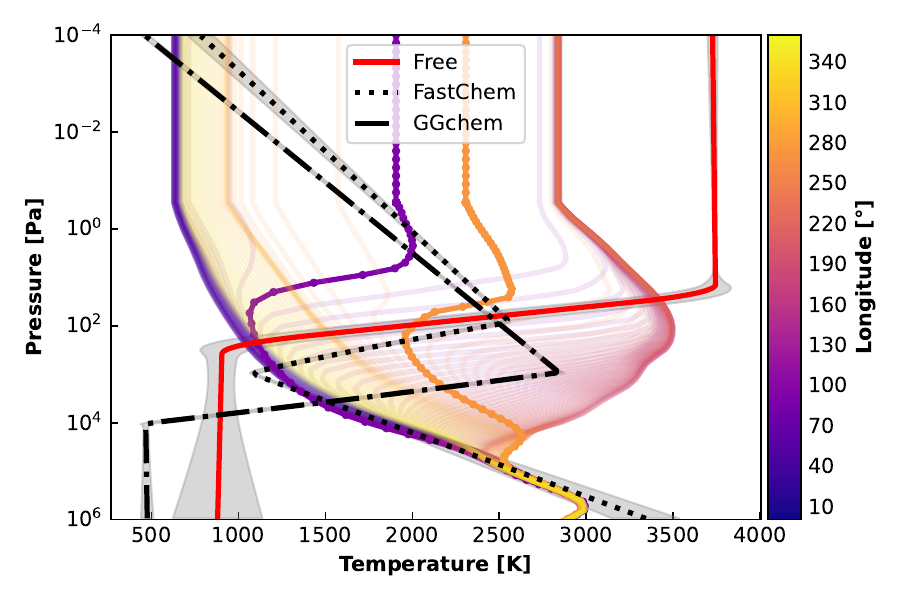}
\includegraphics[scale=\spec,trim = 0cm 0cm 0cm 0cm, clip]{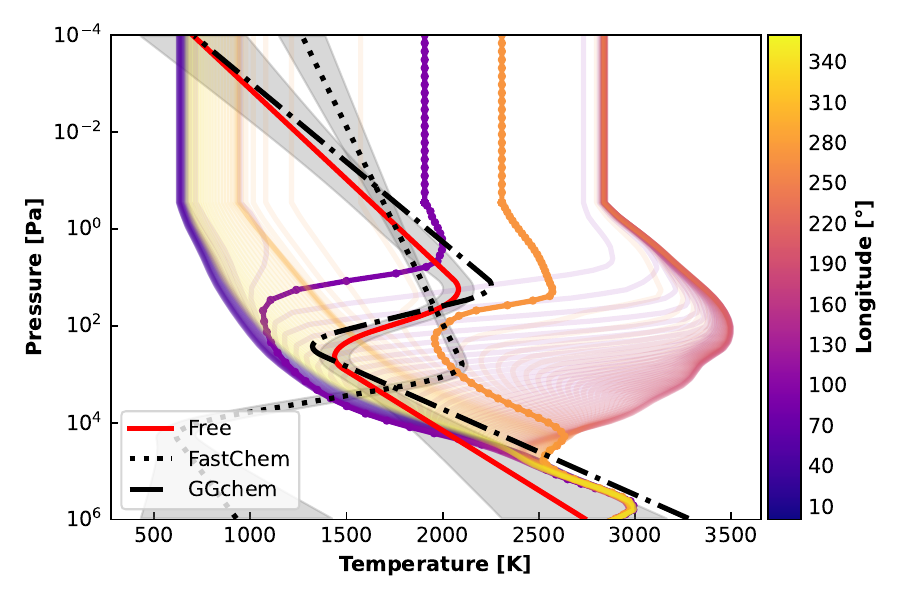}
\\
\includegraphics[scale=\spec,trim = 0cm 0cm 0cm 0cm, clip]{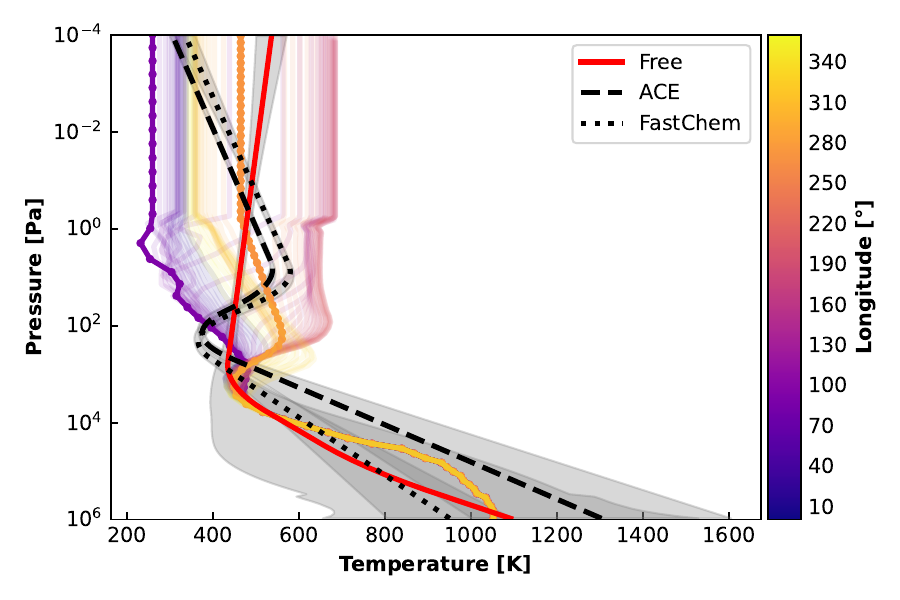}
\includegraphics[scale=\spec,trim = 0cm 0cm 0cm 0cm, clip]{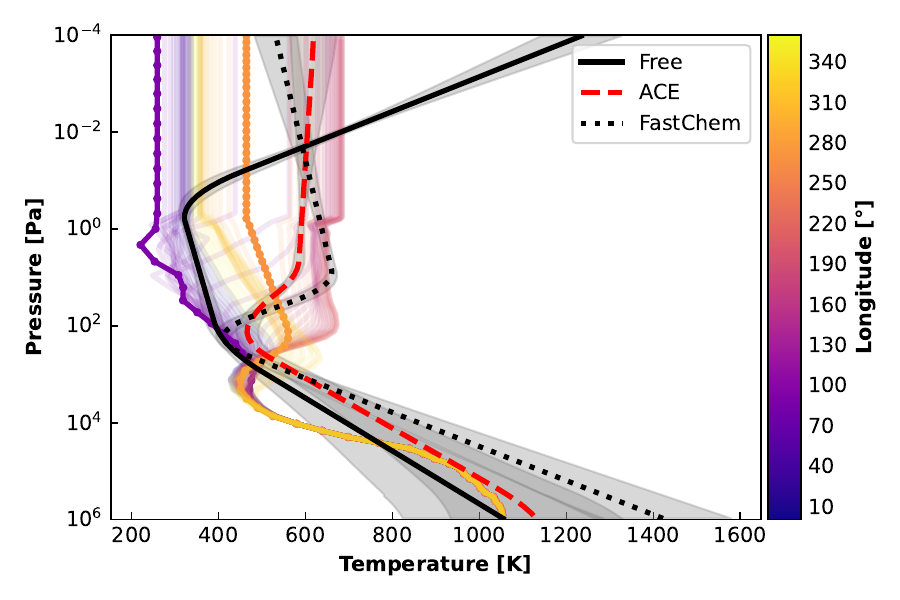}

\caption{Same as Figure \ref{fig: TP_JWST_1D} using 3D thermal structure for the atmospheres.}
\label{fig: TP_JWST_3D}
\end{figure*}

\begin{figure*}[h!]
\centering
\includegraphics[scale=\spec,trim = 0cm 0cm 0cm 0cm, clip]{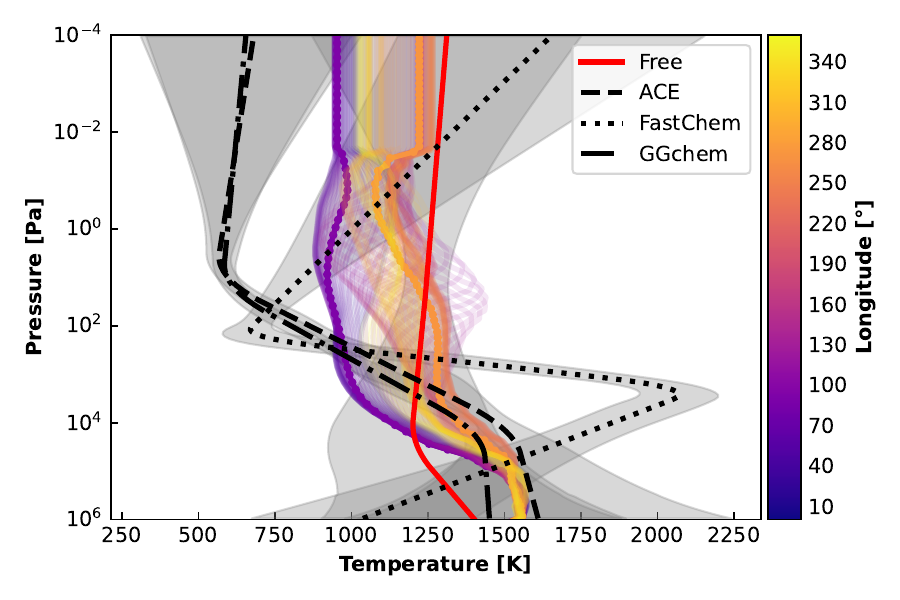}
\includegraphics[scale=\spec,trim = 0cm 0cm 0cm 0cm, clip]{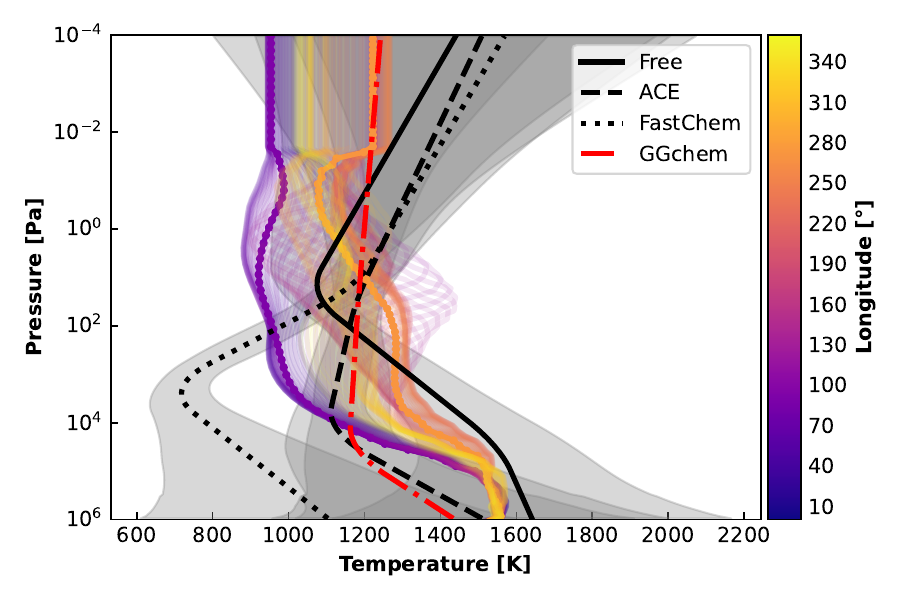}
\\
\includegraphics[scale=\spec,trim = 0cm 0cm 0cm 0cm, clip]{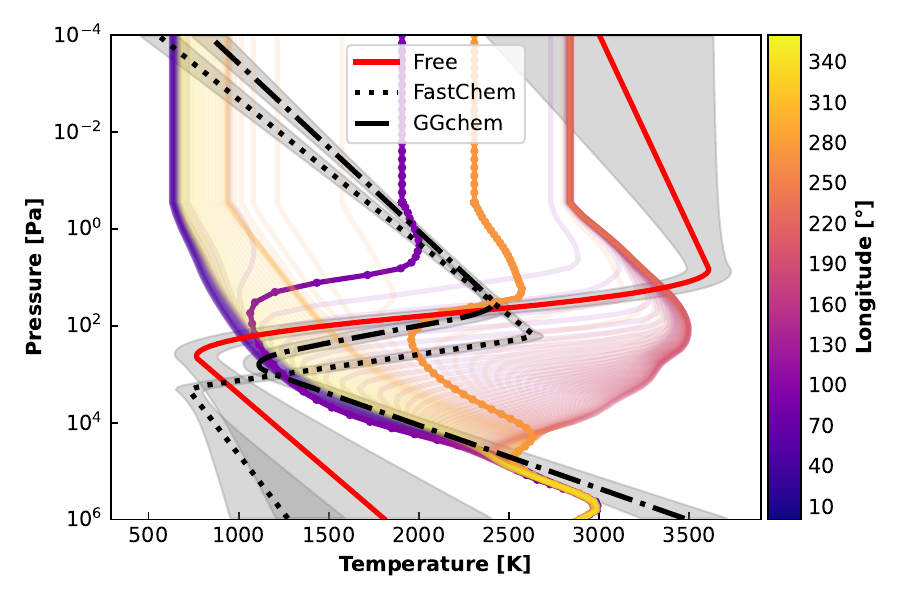}
\includegraphics[scale=\spec,trim = 0cm 0cm 0cm 0cm, clip]{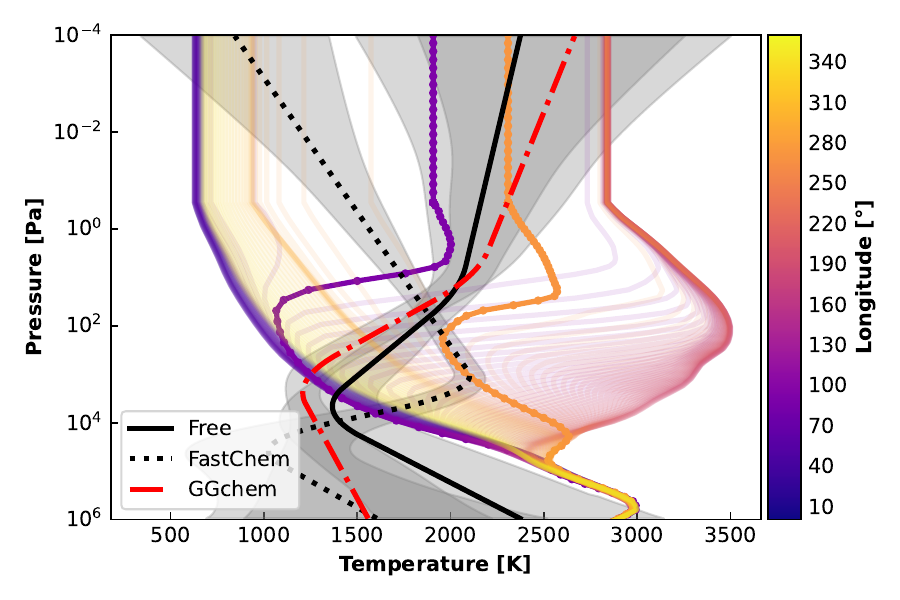}
\\
\includegraphics[scale=\spec,trim = 0cm 0cm 0cm 0cm, clip]{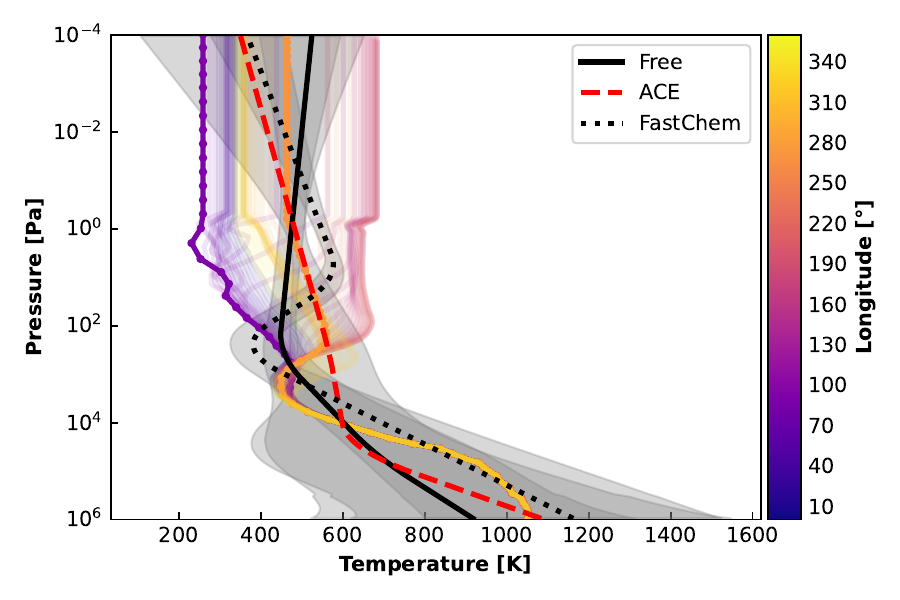}
\includegraphics[scale=\spec,trim = 0cm 0cm 0cm 0cm, clip]{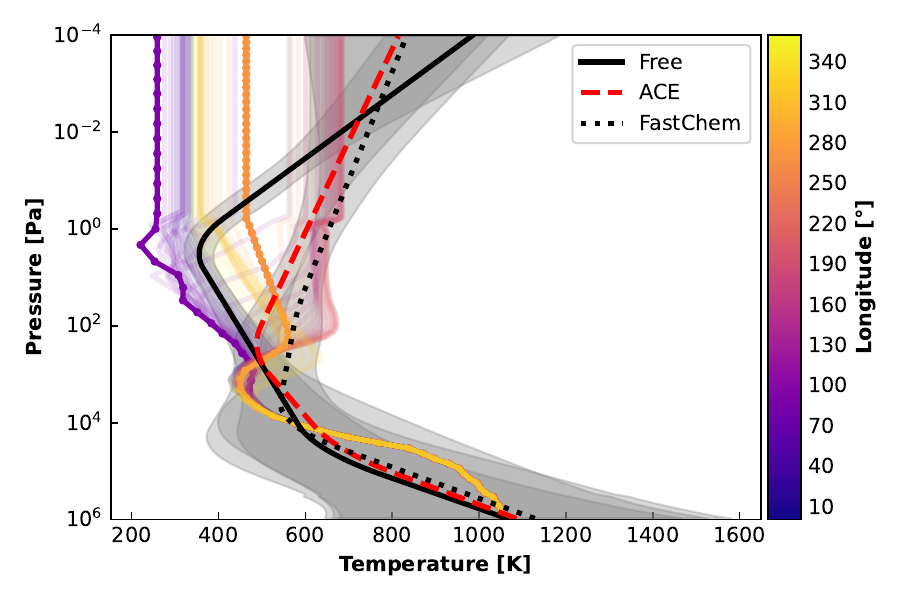}

\caption{Same as Figure \ref{fig: TP_JWST_3D} for Ariel simulations.}
\label{fig: TP_ARIEL_3D}
\end{figure*}

\clearpage

\section{Abundances profiles}

The input abundances profiles at the equator using the 1D and 3D thermal structure, over-plot with the abundances profiles of all the retrieval models are shown here. For the homogeneous 1D thermal structure model, longitude profiles are on top of each other since it has been build homogeneously.

\begin{figure*}[h!]
\centering
\includegraphics[width=\textwidth]{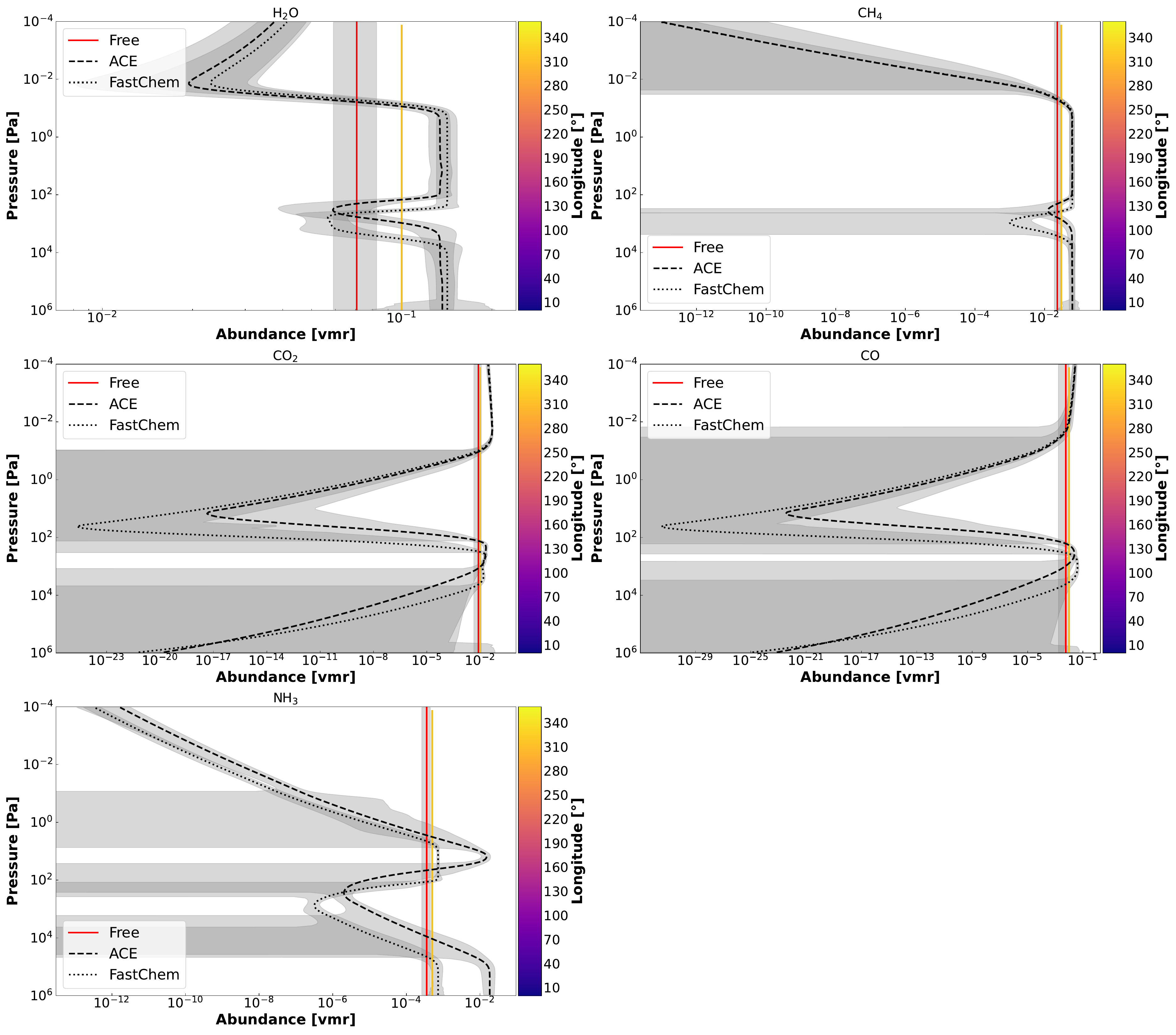}
\caption{JWST simulation with constant chemistry assuming 1D atmosphere for the thermal structure. Main absorbers volume mixing ratio (VMR) profiles at the equator on \textbf{GJ~1214~b}. Substellar point at 180$^{\circ}$ longitude. We over-plot the best species profiles of Free, ACE, FastChem and GGchem retrievals respectively in black solid, dashed, dotted and dashed-dotted lines. The retrieval having the highest Bayes factor is plotted in red.}
\label{fig: spe_JWST_1D_cst_GJ1214b}
\end{figure*}

\begin{figure*}[h!]
\centering
\includegraphics[width=\textwidth]{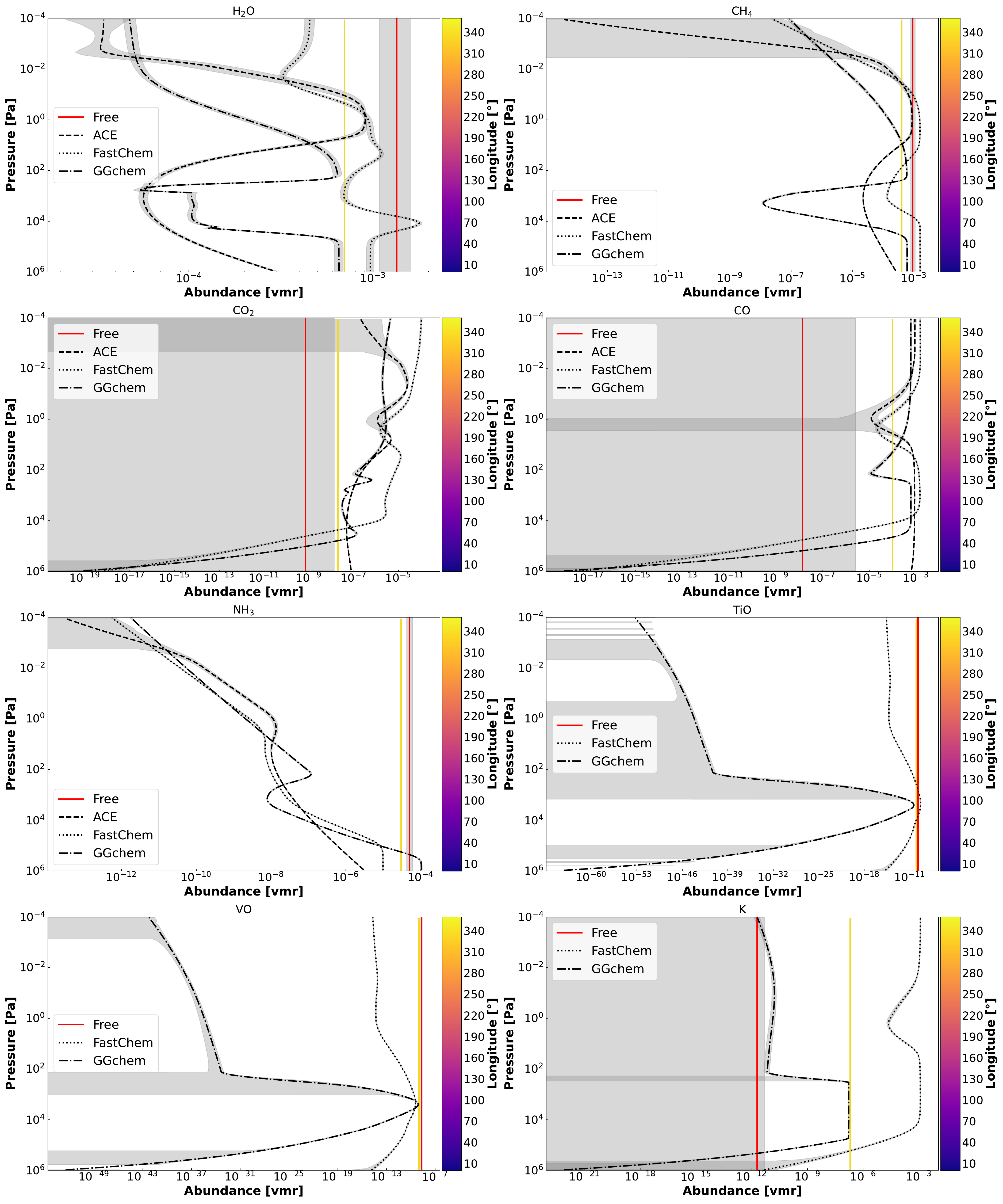}
\caption{Same as Figure \ref{fig: spe_JWST_1D_cst_GJ1214b} for \textbf{HD~189733~b}.}
\label{fig: spe_JWST_1D_cst_HD189733b}
\end{figure*}

\begin{figure*}[h!]
\centering
\includegraphics[width=\textwidth]{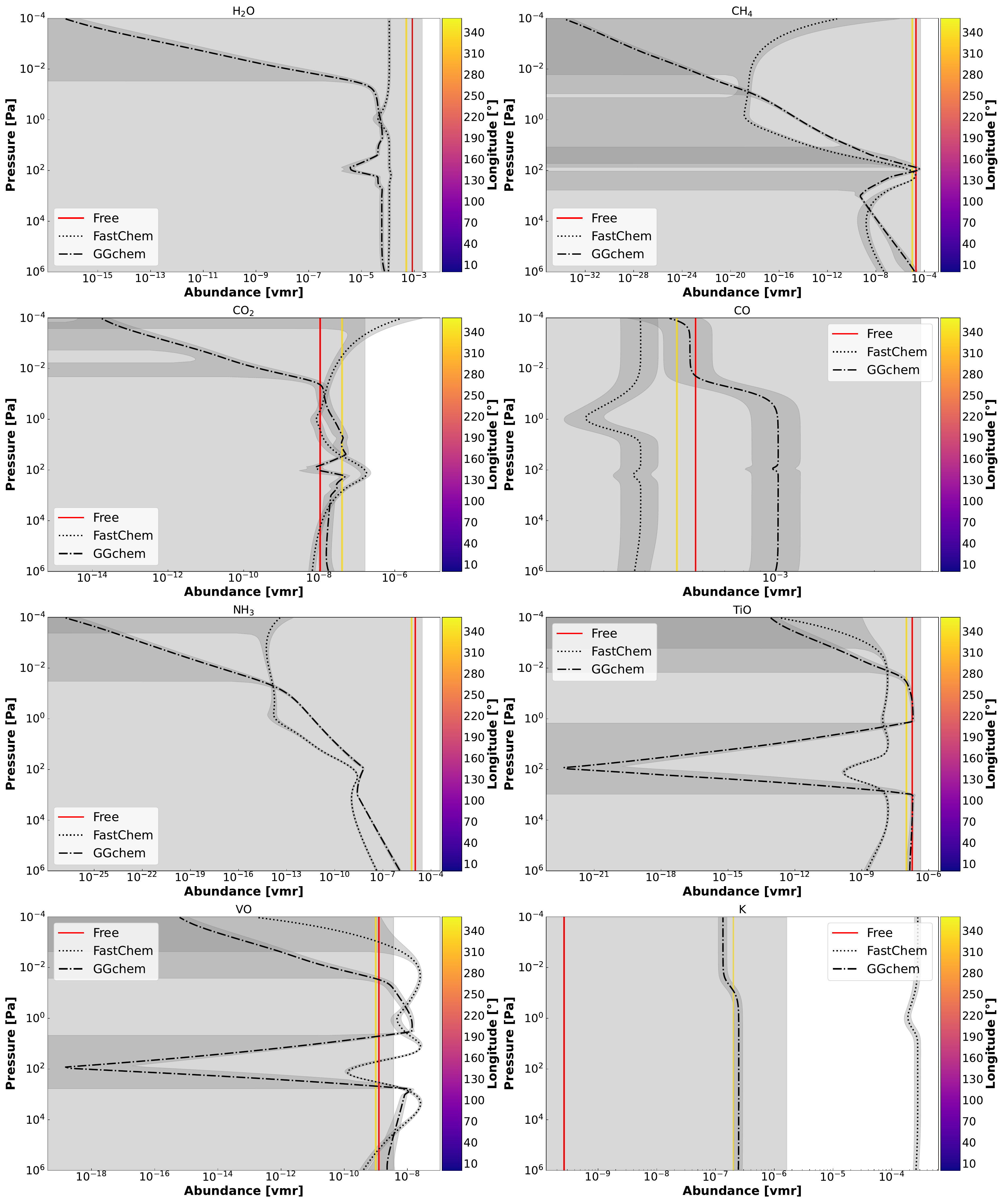}
\caption{Same as Figure \ref{fig: spe_JWST_1D_cst_GJ1214b} for \textbf{WASP-121~b}.}
\label{fig: spe_JWST_1D_cst_WASP121b}
\end{figure*}

\begin{figure*}[h!]
\centering
\includegraphics[width=\textwidth]{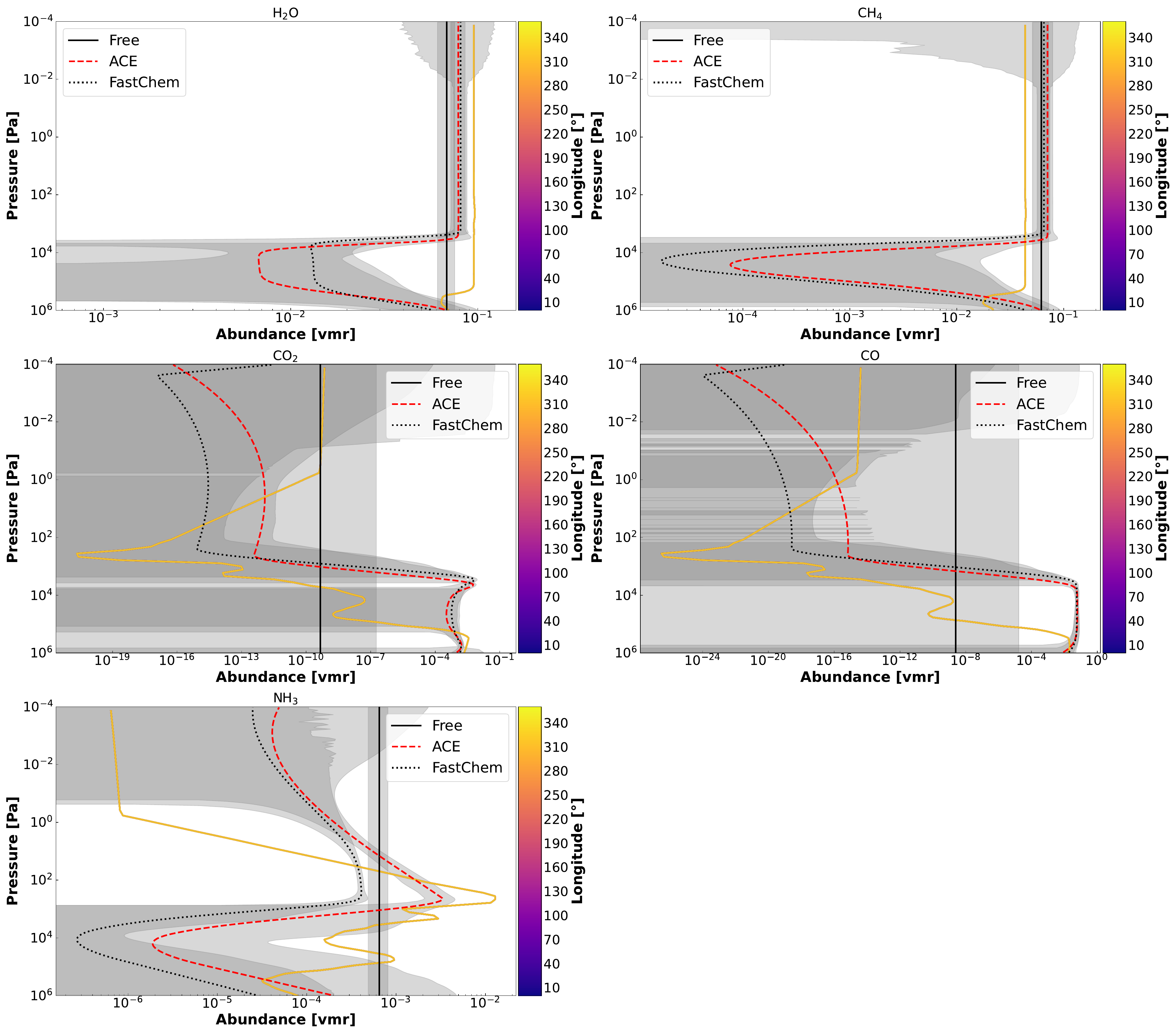}
\caption{Same as Figure \ref{fig: spe_JWST_1D_cst_GJ1214b} for \textbf{GJ~1214~b} with input equilibrium chemistry.}
\label{fig: spe_JWST_1D_eq_GJ1214b}
\end{figure*}

\begin{figure*}[h!]
\centering
\includegraphics[width=\textwidth]{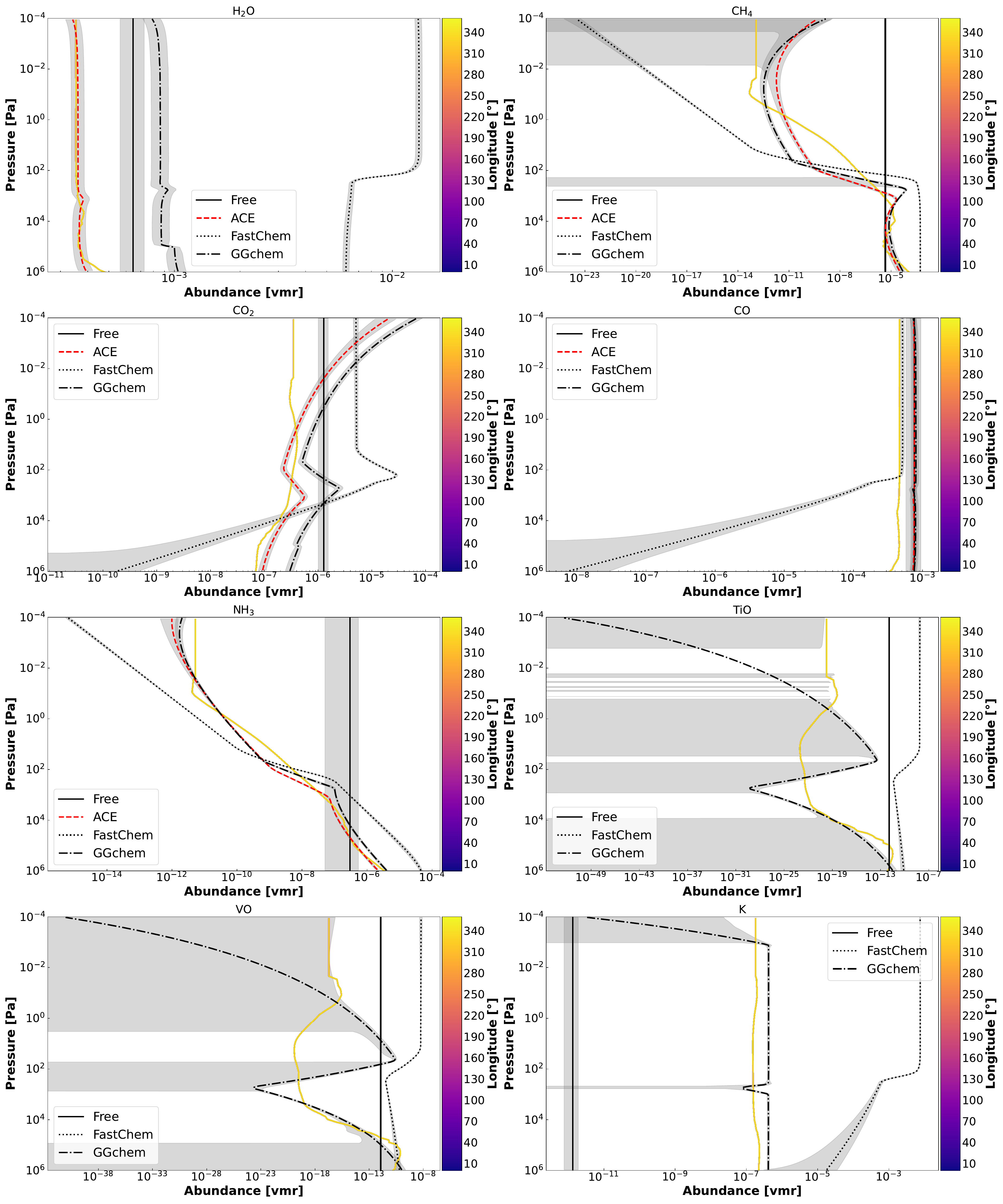}
\caption{Same as Figure \ref{fig: spe_JWST_1D_cst_GJ1214b} for \textbf{HD~189733~b} with input equilibrium chemistry.}
\label{fig: spe_JWST_1D_eq_HD189733b}
\end{figure*}

\begin{figure*}[h!]
\centering
\includegraphics[width=\textwidth]{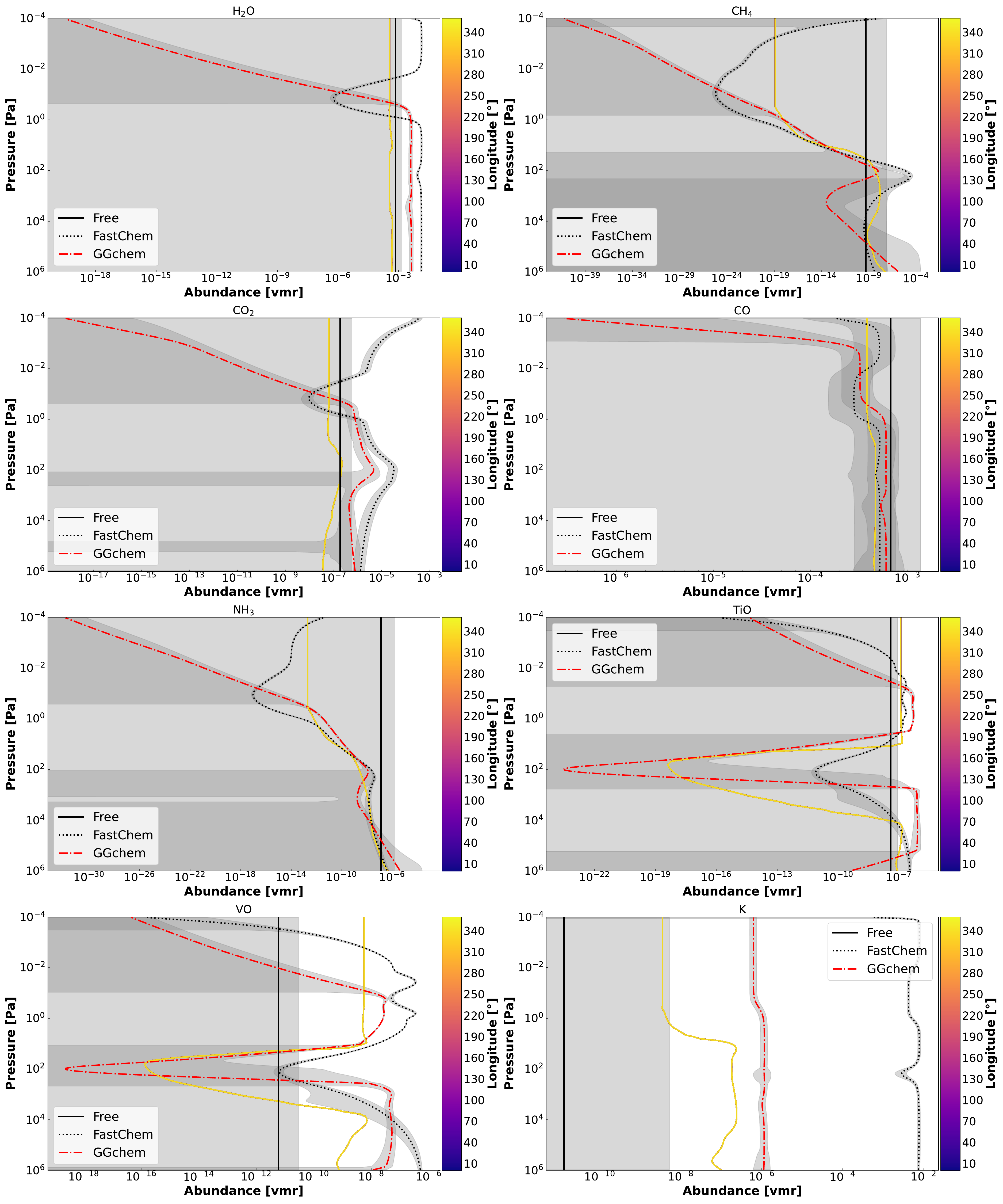}
\caption{Same as Figure \ref{fig: spe_JWST_1D_cst_GJ1214b} for \textbf{WASP-121~b} with input equilibrium chemistry.}
\label{fig: spe_JWST_1D_eq_WASP121b}
\end{figure*}

\begin{figure*}[h!]
\centering
\includegraphics[width=\textwidth]{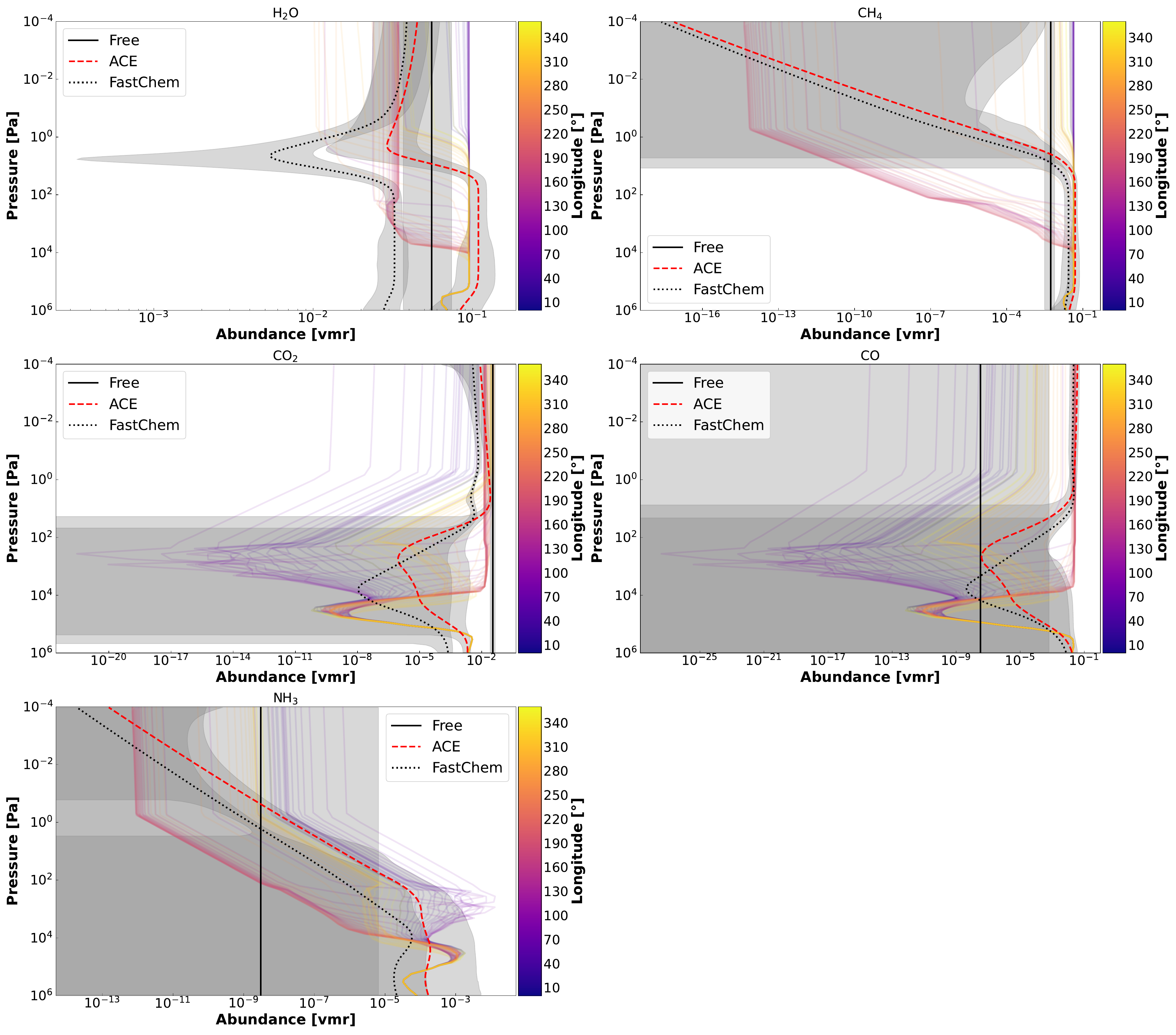}
\caption{Same as Figure \ref{fig: spe_JWST_1D_cst_GJ1214b} for \textbf{GJ~1214~b} with input equilibrium chemistry and with Ariel simulation assuming 3D atmosphere for the thermal structure.}
\label{fig: spe_ARIEL_3D_eq_GJ1214b}
\end{figure*}

\begin{figure*}[h!]
\centering
\includegraphics[width=\textwidth]{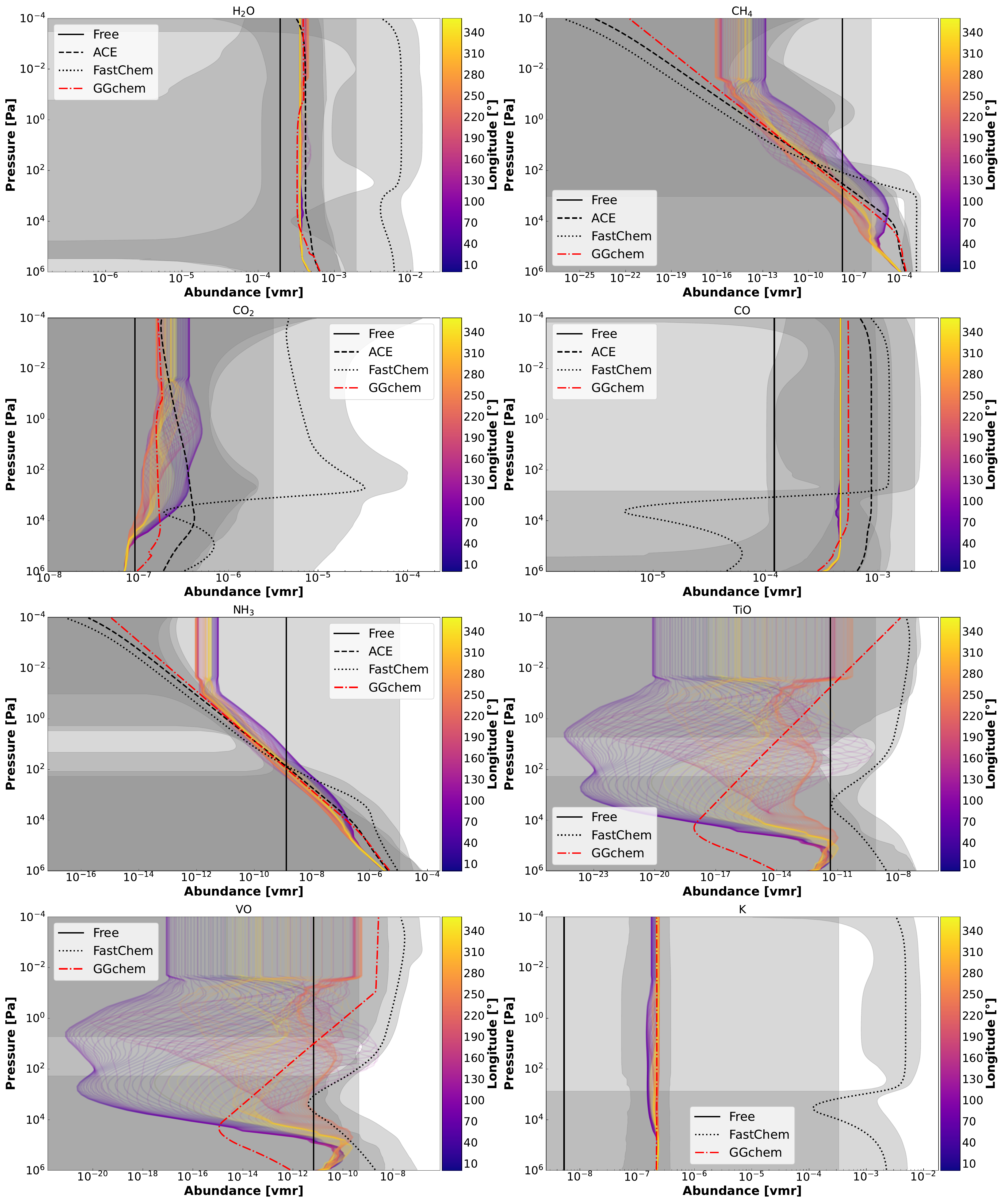}
\caption{Same as Figure \ref{fig: spe_JWST_1D_cst_GJ1214b} for \textbf{HD~189733~b} with input equilibrium chemistry and with Ariel simulation assuming 3D atmosphere for the thermal structure.}
\label{fig: spe_ARIEL_3D_eq_HD189733b}
\end{figure*}

\begin{figure*}[h!]
\centering
\includegraphics[width=\textwidth]{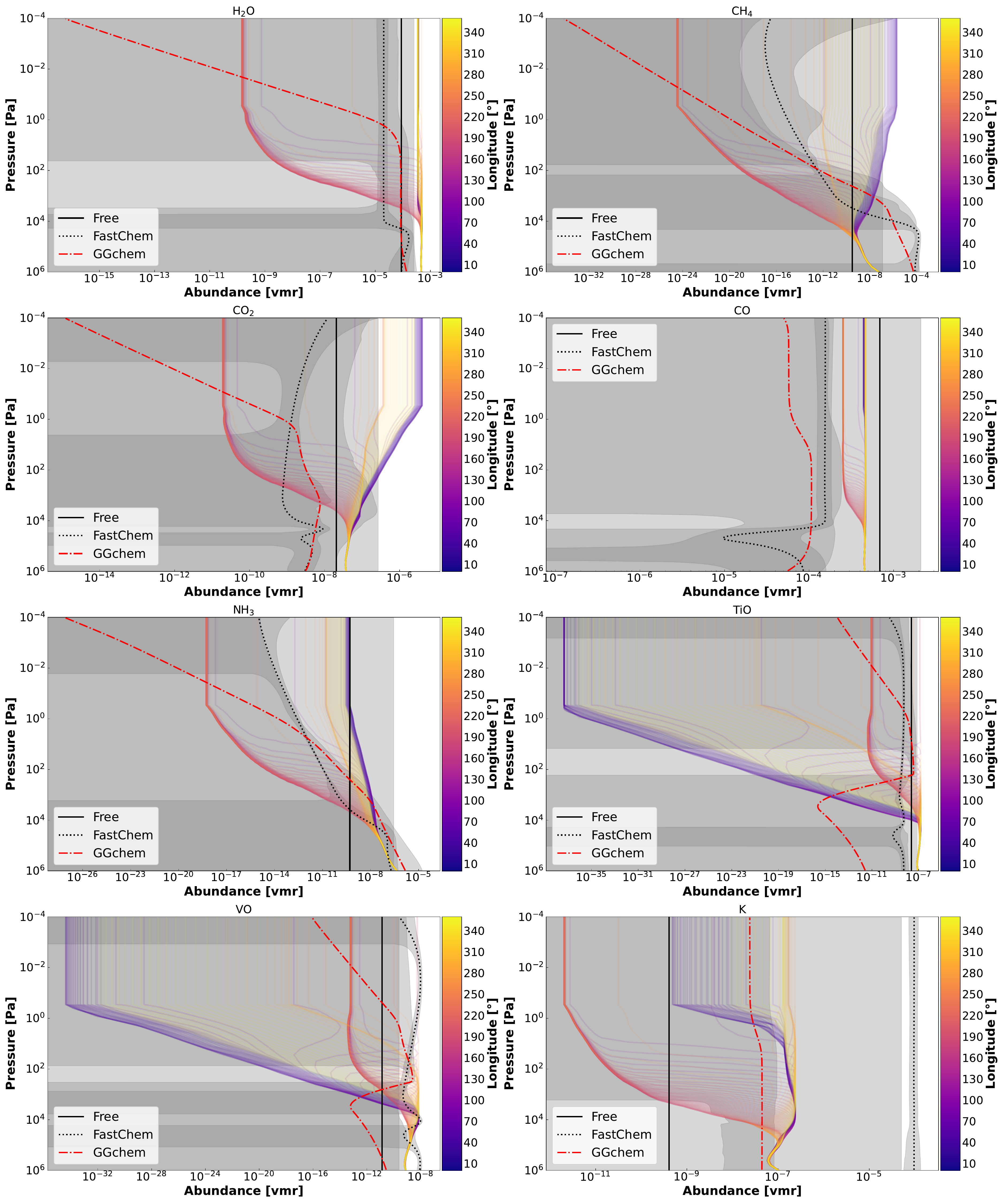}
\caption{Same as Figure \ref{fig: spe_JWST_1D_cst_GJ1214b} for \textbf{WASP-121~b} with input equilibrium chemistry and with Ariel simulation assuming 3D atmosphere for the thermal structure.}
\label{fig: spe_ARIEL_3D_eq_WASP121b}
\end{figure*}

\clearpage

\section{Corner plots}

The corner plot of the retrievals, over-plot with the input C/O ratio, metallicity (Z) and radius of the planet (R$_p$) for equilibrium chemistry are shown here.

\begin{figure*}[h!]
\centering
\includegraphics[width=\textwidth]{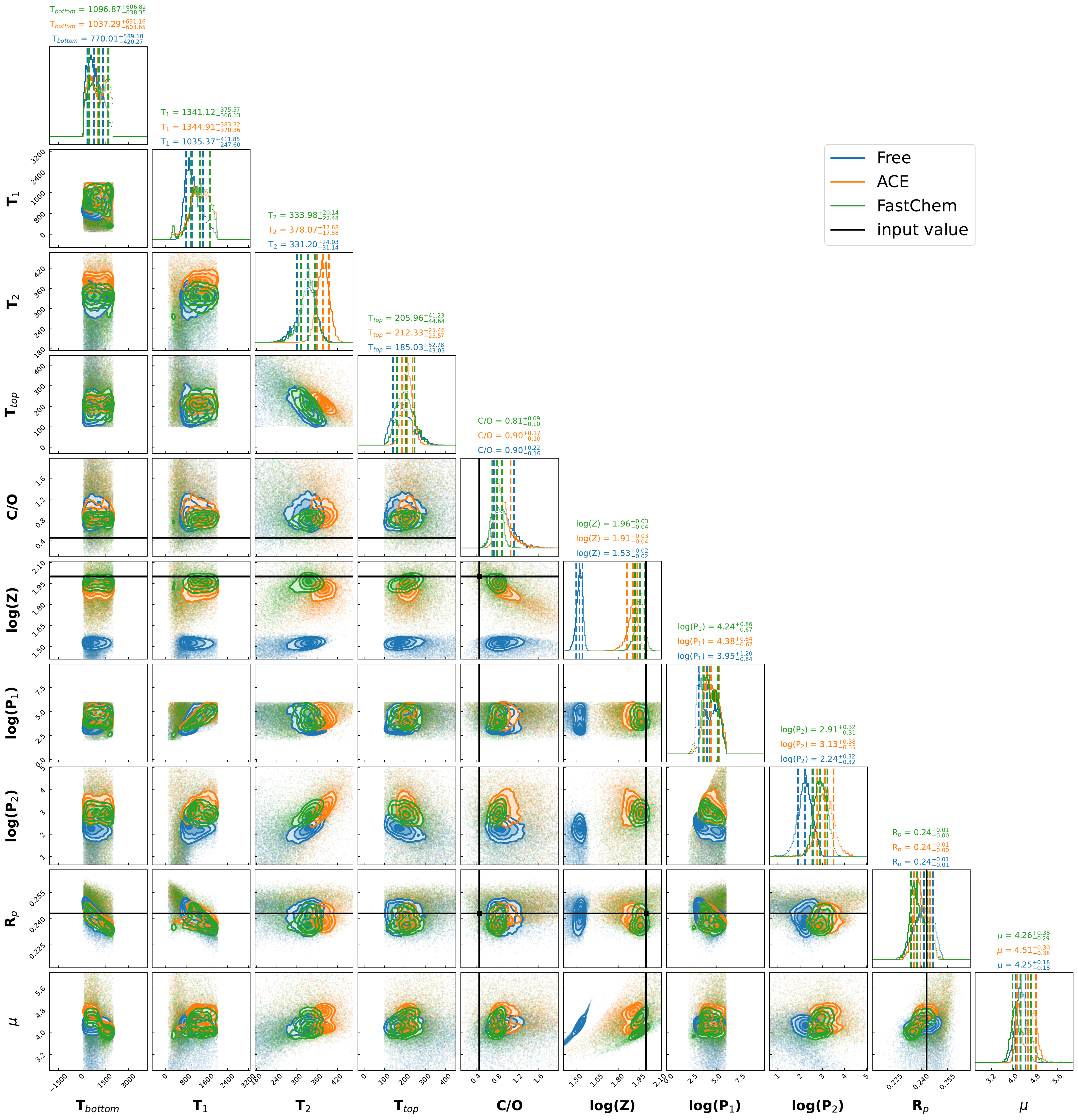}
\caption{JWST simulation with equilibrium chemistry on \textbf{GJ~1214~b} assuming 1D atmosphere for the thermal structure. We over-plot Free, ACE, FastChem and/or GGchem retrievals respectively in blue, orange, green and red. The input value is indicated by the black line for the input C/O ratio, metallicity (Z) and radius of the planet (R$_p$) values.}
\label{fig: corner_JWST_1D_eq_GJ1214b}
\end{figure*}

\begin{figure*}[h!]
\centering
\includegraphics[width=\textwidth]{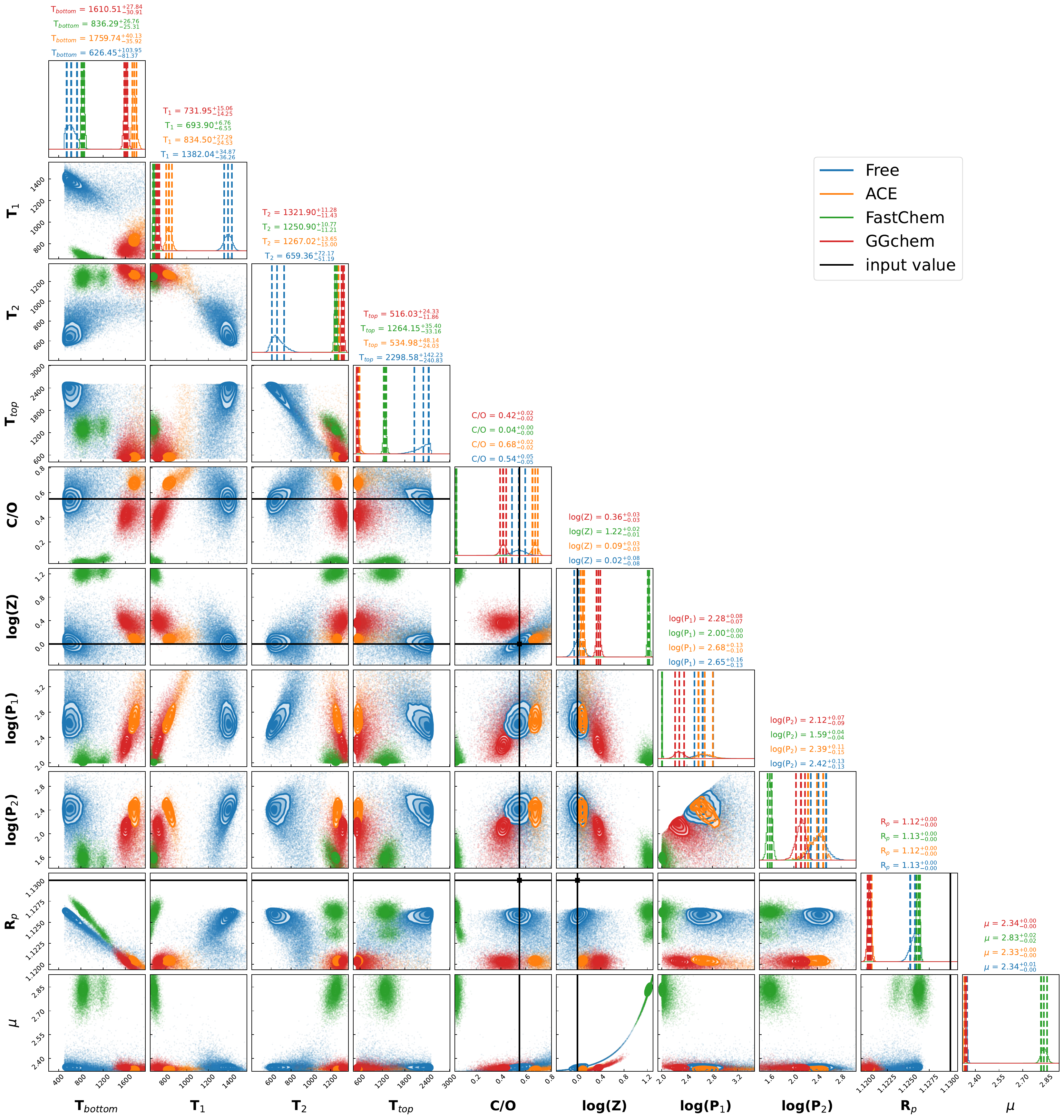}
\caption{Same as Figure \ref{fig: corner_JWST_1D_eq_GJ1214b} for \textbf{HD~189733~b}.}
\label{fig: corner_JWST_1D_eq_HD189733b}
\end{figure*}

\begin{figure*}[h!]
\centering
\includegraphics[width=\textwidth]{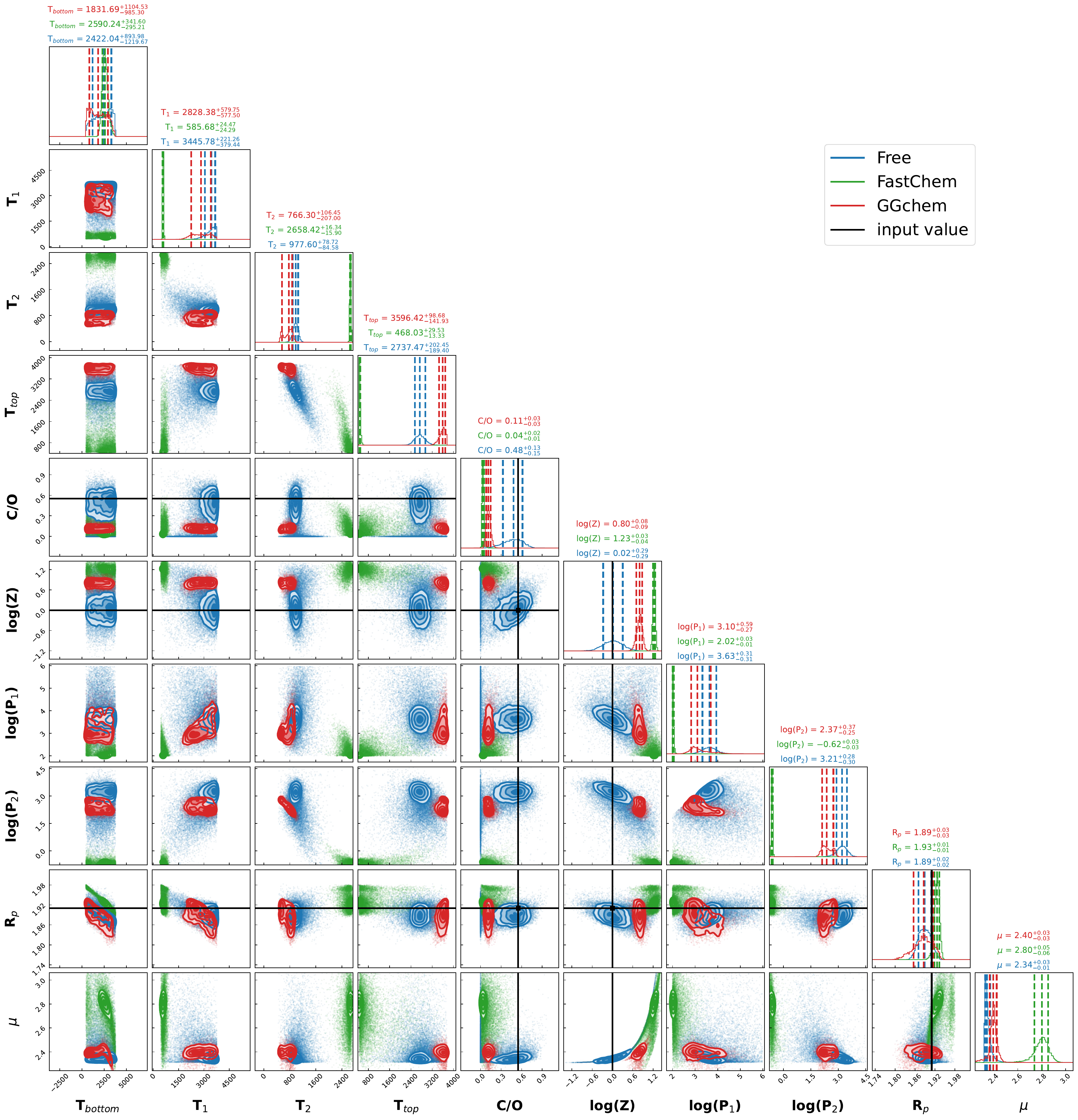}
\caption{Same as Figure \ref{fig: corner_JWST_1D_eq_GJ1214b} for \textbf{WASP121~b}.}
\label{fig: corner_JWST_1D_eq_WASP121b}
\end{figure*}

\begin{figure*}[h!]
\centering
\includegraphics[width=\textwidth]{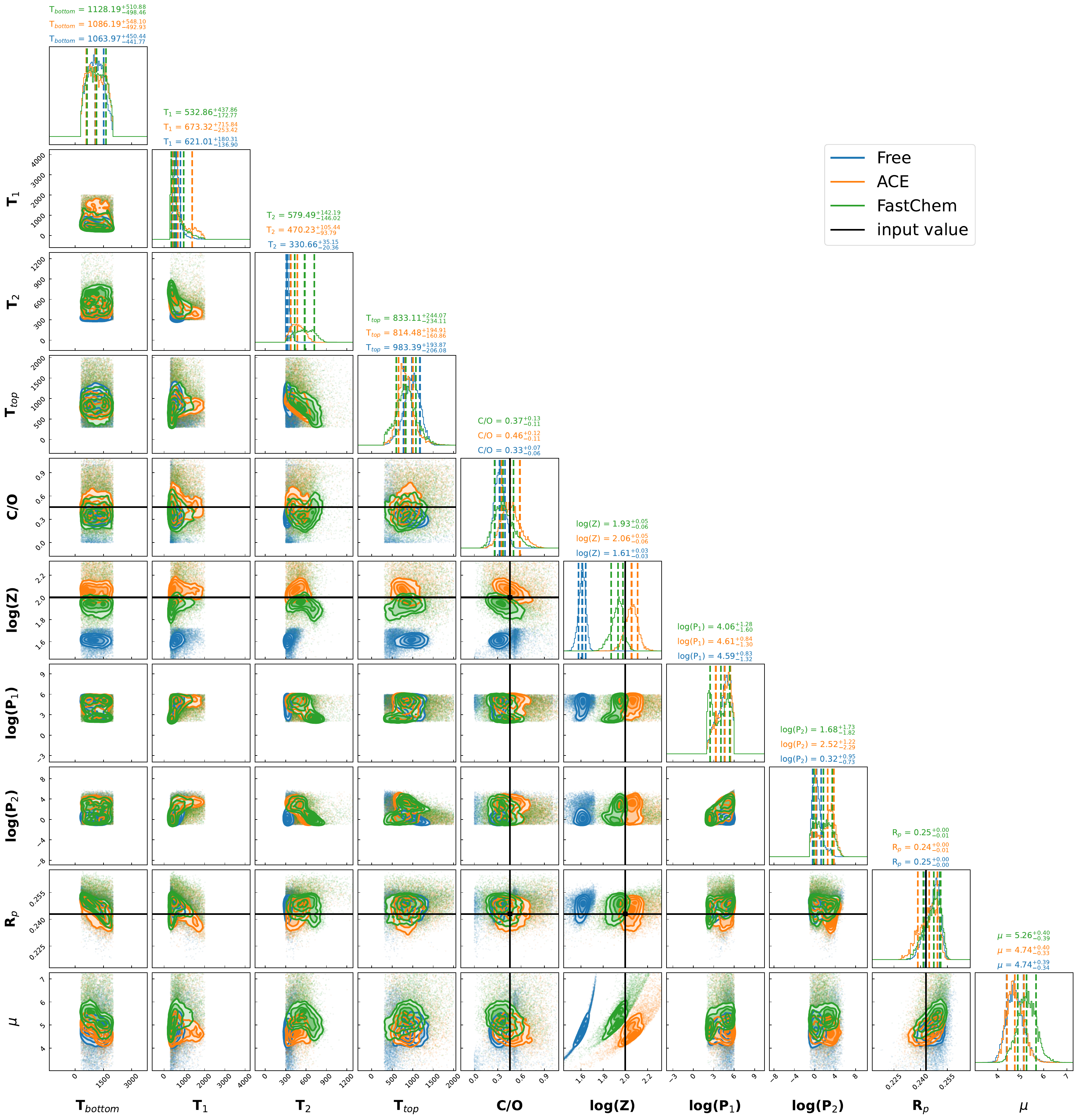}
\caption{Same as Figure \ref{fig: corner_JWST_1D_eq_GJ1214b} for \textbf{GJ~1214~b} with Ariel simulation assuming 3D atmosphere for the thermal structure.}
\label{fig: corner_ARIEL_3D_eq_GJ1214b}
\end{figure*}

\begin{figure*}[h!]
\centering
\includegraphics[width=\textwidth]{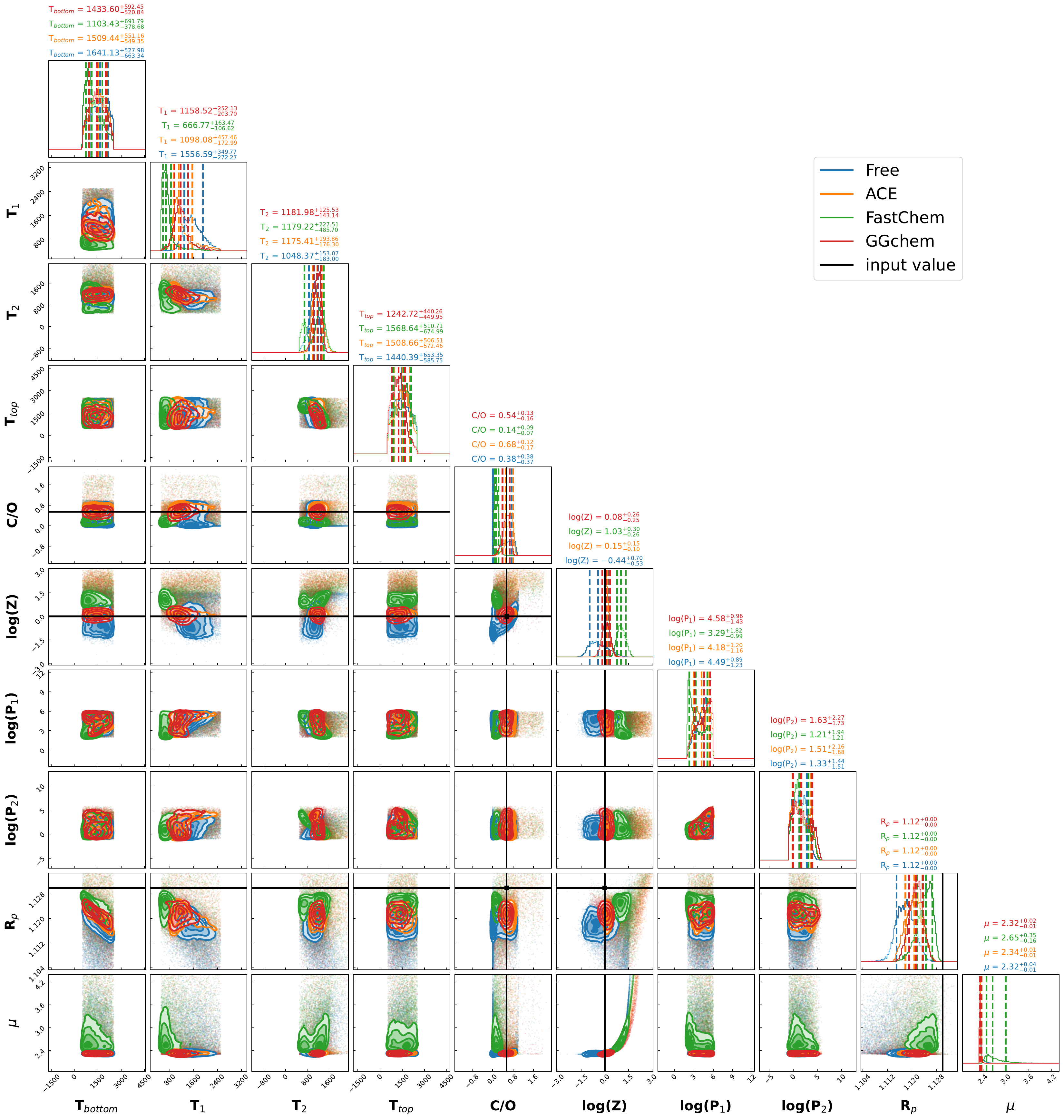}
\caption{Same as Figure \ref{fig: corner_JWST_1D_eq_GJ1214b} for \textbf{HD~189733~b} with Ariel simulation assuming 3D atmosphere for the thermal structure.}
\label{fig: corner_ARIEL_3D_eq_HD189733b}
\end{figure*}

\begin{figure*}[h!]
\centering
\includegraphics[width=\textwidth]{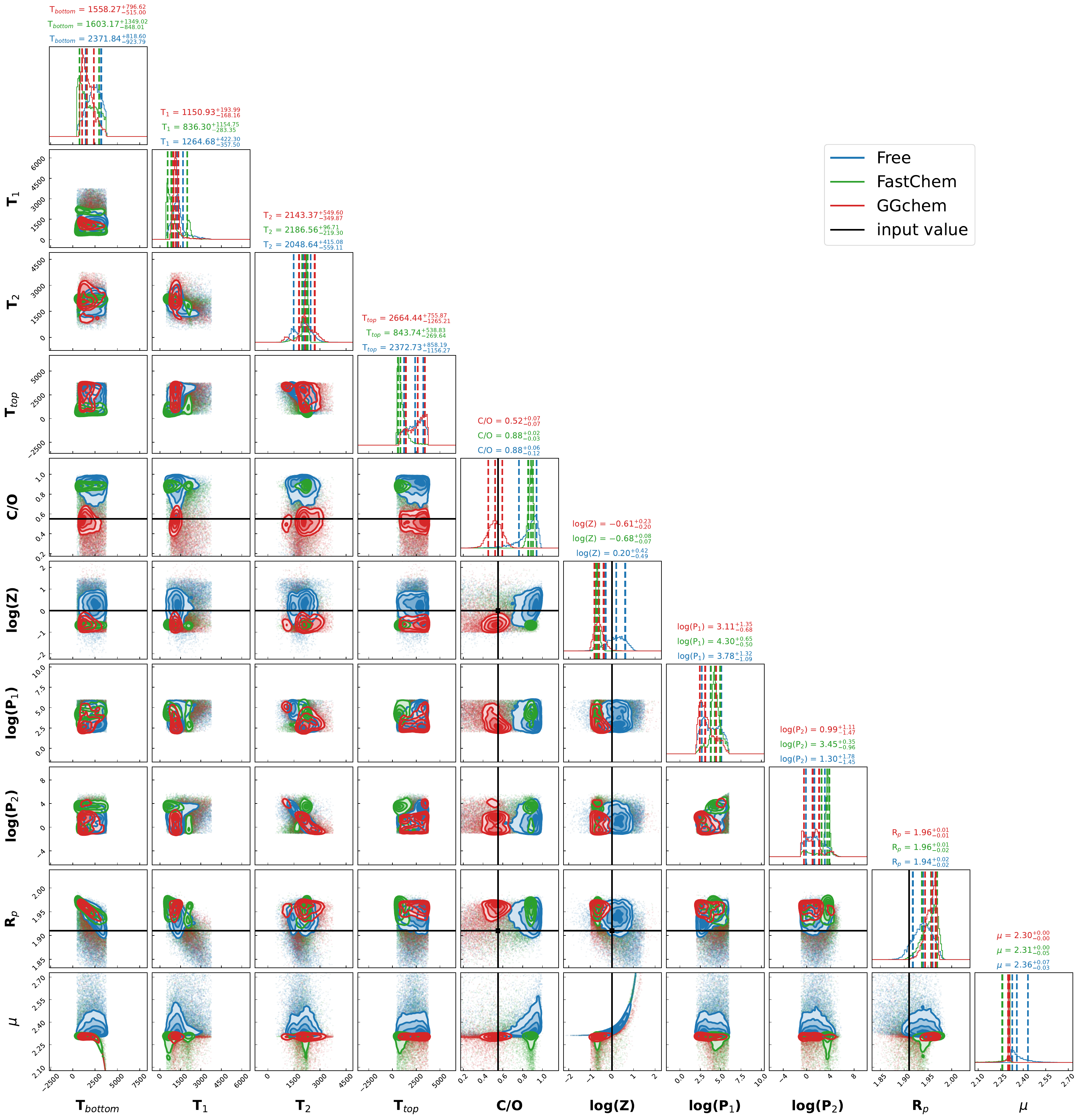}
\caption{Same as Figure \ref{fig: corner_JWST_1D_eq_GJ1214b} for \textbf{WASP121~b} with Ariel simulation assuming 3D atmosphere for the thermal structure.}
\label{fig: corner_ARIEL_3D_eq_WASP121b}
\end{figure*}

\clearpage

\end{appendix}
%
%

\end{document}